\documentclass[aps,reprint,10pt,superscriptaddress,showpacs]{revtex4-1}
\usepackage[T1]{fontenc}
\usepackage[utf8]{inputenc}
\usepackage{color}
\usepackage{amsmath}
\usepackage{amssymb}
\usepackage{graphicx}
\usepackage[unicode=true,bookmarks=true,bookmarksnumbered=true,bookmarksopen=false,breaklinks=true,pdfborder={0 0 0},backref=false,colorlinks=true]{hyperref}

\usepackage{dcolumn}
\usepackage{bm}
\usepackage[none]{hyphenat} 
\usepackage{bbm}
\usepackage{braket}
\usepackage{tabularx}

\def\ie{{\it i.e.},\ }
\def\eg{{\it e.g.}\ }

\input{epsf}

\newcommand{\dd}{\mathbf d}

\begin{document}

\title{All  ``Magic Angles'' Are  ``Stable'' Topological}

\author{Zhida Song}
\thanks{These authors contributed to this work equally.}
\affiliation{Beijing National Research Center for Condensed Matter Physics, and Institute of Physics, Chinese Academy of Sciences, Beijing 100190, China}
\affiliation{University of Chinese Academy of Sciences, Beijing 100049, China}
\author{Zhijun Wang}
\thanks{These authors contributed to this work equally.}
\affiliation{Department of Physics, Princeton University, Princeton, New Jersey 08544, USA}
\author{Wujun Shi}
\affiliation{School of Physical Science and Technology, ShanghaiTech University, Shanghai 200031, China}
\affiliation{Max Planck Institute for Chemical Physics of Solids, D-01187 Dresden, Germany}
\author{Gang Li}
\affiliation{School of Physical Science and Technology, ShanghaiTech University, Shanghai 200031, China}
\author{Chen Fang}
\email{cfang@iphy.ac.cn}
\affiliation{Beijing National Research Center for Condensed Matter Physics, and Institute of Physics, Chinese Academy of Sciences, Beijing 100190, China}
\affiliation{CAS Center for Excellence in Topological Quantum Computation, Beijing, China}
\author{B. Andrei Bernevig}
\email{bernevig@princeton.edu}
\affiliation{Department of Physics, Princeton University, Princeton, New Jersey 08544, USA}
\affiliation{Physics Department, Freie Universitat Berlin, Arnimallee 14, 14195 Berlin, Germany}
\affiliation{Max Planck Institute of Microstructure Physics, 06120 Halle, Germany}

\date{\today}

\begin{abstract}
We show that the electronic structure of the low-energy bands in the small angle-twisted bilayer graphene consists of a series of semi-metallic and \textit{topological} phases. In particular we are able to prove, using an approximate low-energy particle-hole symmetry, that the gapped set of bands that exist around \textit{all} magic angles has what we conjecture to be a \textit{stable} topological \textit{index} stabilized by a magnetic symmetry and reflected in the \textit{odd} winding of the Wilson loop in the Moir{\'e} BZ. The approximate, emergent particle-hole symmetry is essential to the topology of graphene: when strongly broken, non-topological phases can appear. Our paper underpins topology as the crucial ingredient to the description of low-energy graphene. We provide a $4$-band short range tight-binding model whose $2$ lower bands have the same topology, symmetry, and flatness as those of the twisted graphene, and which can be used as an effective low-energy model. We then perform large-scale ($11000$ atoms per unit cell, 40 days per $\bf k$-point computing time) {\it ab-initio} calculations of a series of small angles, from $3^\circ$ to $1^\circ$, which show a more complex and somewhat qualitatively different evolution of the symmetry of the low-energy bands than that of the theoretical Moir{\'e} model, but which confirms the topological nature of the system. At certain angles, we find no insulating filling in graphene at $-4$ electrons per Moir{\'e} unit cell. The {\it ab-initio} evolution of gaps tends to differ from that of the continuum Moir{\'e} model. \end{abstract}

\maketitle

Twisted bilayer graphene (TBG) is an engineered material consisting of  two layers of graphene, coupled via van-der-Waals interaction and rotated relative to each other by some twist angle $\theta$. 
This material exhibits rich single and many-body physics, both in nonzero and in zero magnetic field \cite{dean2010,Young2011,maher2013,hunt2017}.
Recently, it was suggested that for $\theta_e\sim1.1^\circ$, charge gaps appear at fillings of $\pm4N_s$, where $N_s$ is the number of superlattice cells. Importantly, another charge gap at $-2N_s$ was also detected, and conjectured to be a ``Mott gap'' \cite{cao_TBG1,cao_TBG2}. The value of  $\theta$ cannot be accurately  measured directly. Hence the filling per Moir{\'e} unit cell was not experimentally obtained, but conjectured through matching of single-particle gaps. However, if true that the ``Mott'' state appears, this represents the first many-body phase in zero-field graphene. Upon gating the sample, zero resistivity was observed at low temperatures within a range of carrier density near the Mott gap. The superconductivity in TBG is \textit{conjectured} to be unconventional \cite{xu2018,roy2018,you2018,huang2018,isobe2018,wu2018}. TBG could be a new platform for the study of strong correlation physics \cite{dodaro2018,po2018,padhi2018,yuan2018,chen2018}.

The observed single-particle charge gaps, if at density $\pm4N_s$, are consistent with the prediction of an earlier theoretical model in Ref. \cite{MacDonald_M-Model}, which we call the Moir{\'e} band model (MBM). In MBM, the two valleys at $K$ and $K'$ in the graphene Brillouin zone (BZ) decouple.  In each valley, the electronic bands of TBG are obtained by coupling the two Dirac cones in the two layers offset in momentum by the angle twist. The model predicts the vanishing of the Fermi velocity at half-filling for certain twist angles called the ``magic angles'' (labeled as $\theta_{mi}$ with $i$ integers). The predicted first (largest) magic angle is $\theta_{m1}\sim1.05^\circ$, close to the experimental $\theta_e$, and therefore the experimental observation of the narrow bands and ``Mott physics'' may be related to the vanishing Fermi velocity, \ie the first ``magic angle''.

In this Letter, we show, by exhaustive analytical, numerical and {\it ab-initio} methods, that nontrivial band topology is prevalent in TBG \textit{near every magic angle}, given band gaps appearing at $\pm4N_s$. We prove that in the MBM for each valley, as long as direct band gaps exist between the middle two bands (not counting spin) and the rest, the two bands possess nontrivial band topology protected by the magnetic group symmetry $C_2T$, a composite operation of twofold rotation and time-reversal \cite{Shiozaki2014,Fang2015a,Fang2015}. The nontrivial topology is diagnosed both from irreducible representations (irreps) of magnetic groups at high-symmetry momenta as well as from the \textit{odd} winding number of Wilson loops. The proof for arbitrary $\theta$ exploits an approximate particle-hole symmetry in the original MBM; without invoking this symmetry, or if the symmetry is strongly broken, the zero energy bands do not necessarily need to be topological.  When PHS is broken (softly, in the MBM), the statement is proved for $\theta_{m1} \ge \theta\ge\theta_{m6}$ via explicit calculations. Four separate regions of $\theta$ are numerically found to host a gapped phase in which two middle bands are gapped from others. We also diagnose several higher energy bands as topological. Topology and the low-energy physics of graphene are strongly related.
We conjecture that the fragile $\mathbb{Z}$ topological index of the two middle bands collapses to a stable $\mathbb{Z}_2$ index when more bands are considered

\begin{figure}
\begin{centering}
\includegraphics[width=1\linewidth]{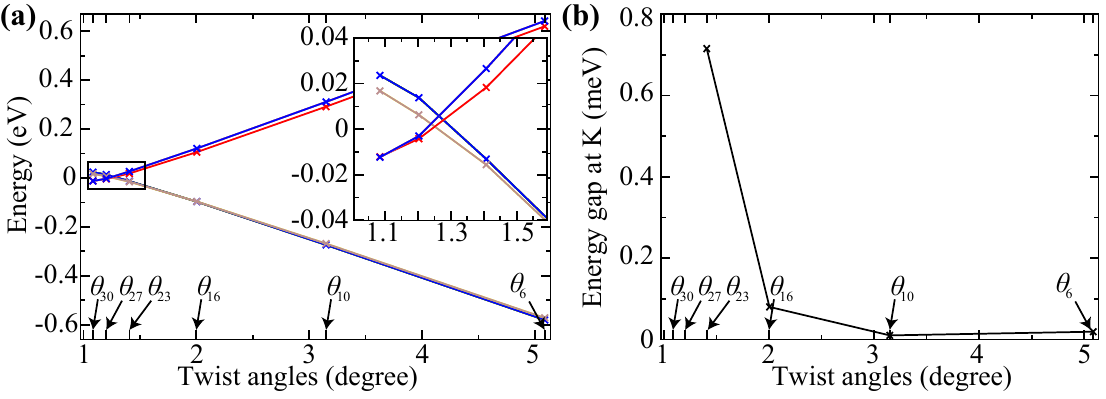}
\par\end{centering}
\caption{Electronic band structures with different commensurate twist angles $\theta_i$. (a) The {\it ab-initio} electronic band structures clearly show that there is no ``partical-hole'' symmetry. The energy bands change monotonically as decreasing $\theta$. The inset is the zoomin of the blackbox area. The two blue bands indicate two $\Gamma_5+\Gamma_6$ bands; the gray band is the $\Gamma_2+\Gamma_3$ band; the red band is the $\Gamma_1+\Gamma_4$ band. (b) Starting from a large twist angle (\eg $\theta_6$), we find that the gap is first decreasing until $\theta_{10}$, then is increasing as decreasing $\theta$ further.}
\label{dftgaps}
\end{figure}

The nontrivial band topology obstructs the building of a two-band tight-binding model for one valley with correct symmetries and finite range of hopping. In order to capture the symmetry, dispersion and topology of the Moir{\'e} bands, we - for the first time -  write down a four-band tight-binding model (dubbed TB4-1V) defined on the Moir{\'e} superlattice with short range  - up to third neighbor -  hopping. In our model, the four bands split into two energetically separate bands; the lower two bands correspond to the low-energy two bands in MBM. TB4-1V offers an anchor point for the study of correlation physics in TBG, when interactions are projected to its lower two bands. We test the validity of MBM and of the symmetry eigenvalues used to obtain our TB4-1V by performing large scale first principles calculations very close to $\theta_{m1}$. Each $k$ point we computed takes about 30-40 days for $\theta \sim \theta_{m1}$. Several qualitative differences with the MBM arise. The particle-hole symmetry breaking is larger than in the MBM (See the $\Gamma$ bands in Fig.~\ref{dftgaps}(a)). The energy gap of $+4N_s$ filling is larger than that of $-4N_s$ filling. Second, the Moir{\'e} $K$ point gap due to the coupling between $K$ and $K'$ graphene points is \textit{growing} as  $\theta$ decreases. This (along with graphene buckling), could explain the decrease in conductivity at half filling observed in experiments. In our \textit{ab-initio} calculations, a gap closing at $-4N_s$ filling happens around $\theta_{16}\approx 2.00^\circ$. Two other gap closings at $4N_s$ and $-4N_s$ fillings happen around $\theta_{30}\approx 1.08^\circ$. This gives rise to a complicated metallic phase for $\theta_{i=30}<\theta<\theta_{i=16}$, in which the middle 4 bands are not separated from others. (Here $\theta_i=\arccos \frac{3i^2+3i+1/2}{3i^2+3i+1}$, with $i=1,2,\cdots$, labels the commensurate twist angles \cite{TBGstr}, among which $\theta_{31}\approx 1.05^\circ$ is closest to $\theta_{m1}$.)

We now analyze the symmetries of the one-valley MBM to find the character of the lowest energy bands at  $\theta > \theta_{m1}$. Due to the vanishing of the $K$ to $K'$  inter-valley coupling,  any symmetry that relates $K$ to $K'$ in the original layer is hence not present in the one-valley model: time reversal, $C_{6z}$  and $C_{2y}$ are absent, but $C_{3z}$ remains. We find that the correct symmetry group of the one-valley MBM is magnetic space group (MSG) $P6'2'2$ (\#177.151 in the BNS notation~\cite{Bilbao-MSG}), which contains $C_{6z}T$, $C_{2y}T$ and $C_{2x}$ symmetries, satisfying $(C_{6z}T)^2=C_{3z}$ and $(C_{6z}T)^3=C_{2z}T$. Spin-orbital coupling (SOC) is neglected in the MBM. One should be aware that the band connections in the MBM  differ from those in the \textit{ab-initio} bands: in the latter for $\theta\gtrsim\theta_{m1}$ the middle four bands (two bands per valley) are connected to lower bands due to a level crossing shown in Fig. \ref{dftgaps}.

\begin{figure}
\begin{centering}
\includegraphics[width=1\linewidth]{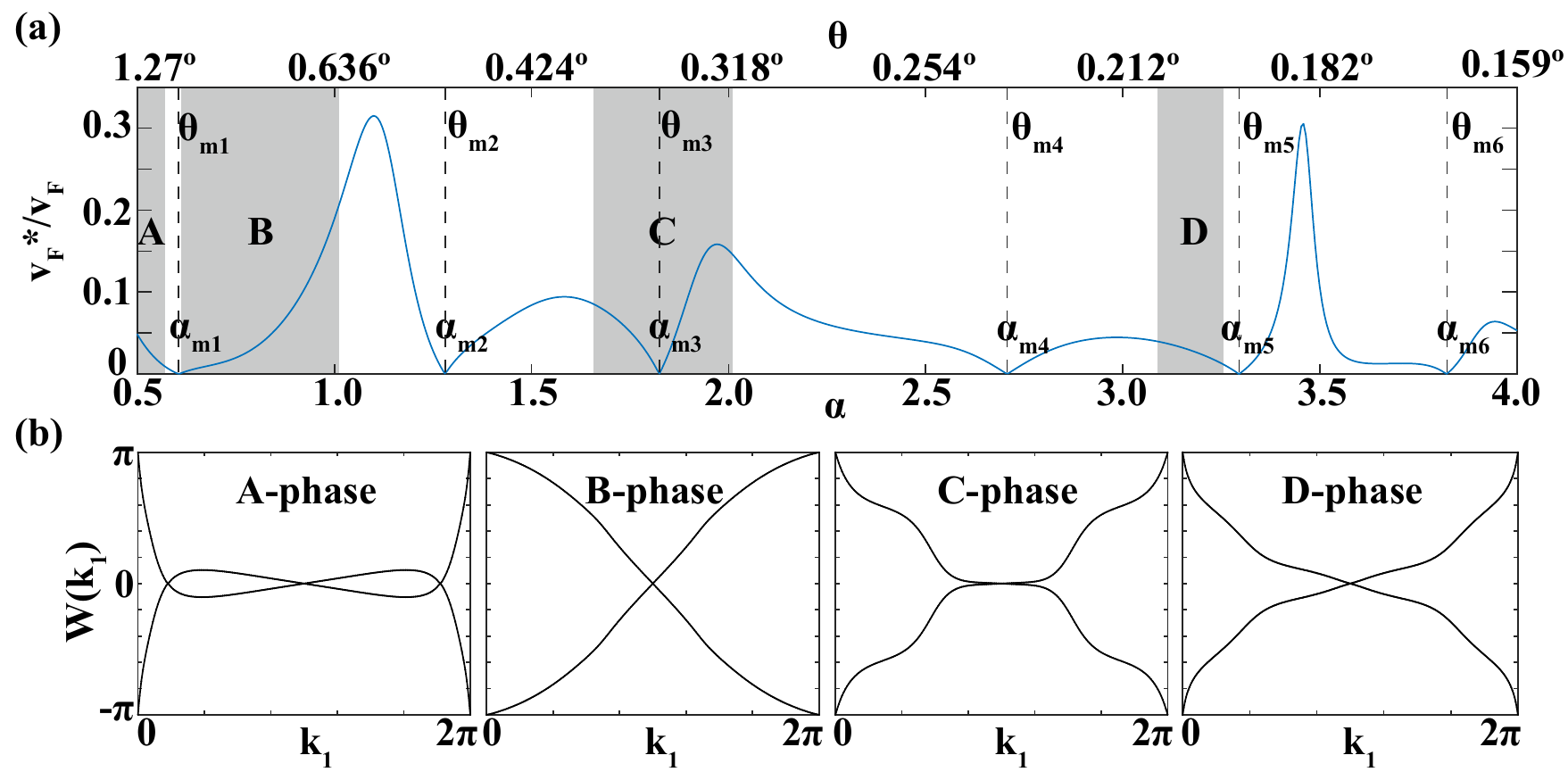}
\par\end{centering}
\caption{\label{fig:MBM-phase}(a) Fermi velocity at the charge neutrality point of the MBM plotted as a function of the twist angle. The dimensionless parameter $\alpha=w/(2v_F |\mathbf{K}| \sin\frac{\theta}{2})$  uniquely determines the band structure of the MBM (up to a scaling); $w$ is the inter-layer coupling, $v_F$ is the Fermi velocity of single layer graphene, $|\mathbf{K}|$ is the distance between $\Gamma$ and $\mathrm{K}$, and $\theta$ is the twist angle. (See Ref. \cite{MacDonald_M-Model} or appendix \ref{sub:M-1V} for more details.) In this plot we set $w=110\mathrm{meV}$, $v_F|\mathbf{K}|=19.81\mathrm{eV}$.  The gapped regions, $\Theta_a$, where the two bands near charge neutrality point are fully disconnected from all other bands are shadowed with grey, and labeled as A, B, C, D.  The magic angles, where the Fermi velocity vanishes, are marked by dashed lines. By numerical calculation we find that $\alpha_{m1}\approx 0.605$ ($\theta_{m1}\approx 1.05^\circ$), $\alpha_{m2}\approx 1.28$ ($\theta_{m1}\approx  0.497^\circ$), $\alpha_{m3}\approx 1.83$ ($\theta_{m3}\approx  0.348^\circ$), $\alpha_{m4}\approx 2.71$ ($\theta_{m4}\approx 0.235^\circ$), $\alpha_{m5}\approx 3.30$ ($\theta_{m5}\approx 0.193^\circ$), $\alpha_{m6}\approx 3.82$ ($\theta_{m6}\approx 0.167^\circ$). (b) The Wilson loops of the four gapped phases have the same winding number, $1$.  }
\end{figure}

We implement the one-valley MBM and compute the characters of the irreps of the bands at all high-symmetry points (see appendixes \ref{sub:Sym} and \ref{sub:evolution} for symmetries and irreps, respectively). 
For the phase at $\theta\gtrsim\theta_{m1}$, where the two bands near charge neutrality point are disconnected from other bands (A-phase in Fig. \ref{fig:MBM-phase}(a)), under the symmetry of MSG $P6^\prime2^\prime2$, the irreps of the two bands are calculated to be $\Gamma_1+\Gamma_2$, $M_1+M_2$ and $K_2+K_3$ at $\Gamma$, $M$ and $K$ points, respectively (see table \ref{tab:irreps-MSG} for the definitions of irreps \cite{BCS-MSG}). 
Compared to the elementary band representations \cite{Bernevig_TQC,Vergniory2017,cano2017} (EBR's), which represent the atomic limits of MSG $P6^\prime2^\prime2$, listed in table \ref{tab:EBR-magSG}, we conclude that these two bands (dubbed 2B-1V for ``two-band -- one valley'') cannot be topologically trivial: they cannot be decomposed into any linear combination of EBR's with positive coefficients. Hence 2B-1V has to be topological. However, if we allow coefficients to be negative, we obtain this decomposition: 2B-1V$=G^{2c}_{A_1}+G^{1a}_{A_2}-G^{1a}_{A_1}$ ($G^{site}_{irrep}$ are EBR notations \cite{EBRnotation} of MSG $P6^\prime2^\prime2$, given in table \ref{tab:EBR-magSG}). The negative integer indicates that the two bands \textit{at least} host a ``fragile'' topological phase~\cite{Ashvin_fragile,Bernevig_TQC,cano2017}; namely, one {\it cannot} construct exponentially localized Wannier functions unless they are coupled to some other atomic bands.
However, the topology in TBG is much richer. 
Using an approximate symmetry of the MBM, we make four statements:  
{\bf 1.} The EBR decomposition shows that 2B-1V for any gapped region of twist angles has \textit{at least} fragile topology. 
{\bf 2.} The Wilson loop winding that characterizes the topology is \textit{odd}, equal to $1$.  
{\bf 3.} In appendix \ref{sec:homotopy}, we use the \textit{odd} winding of Wilson loop to argue that 2B-1V has a \textit{stable} $\mathbb{Z}_2$ topological index protected by $C_{2z}T$, in the sense that upon adding trivial bands, i.e., bands with trivial $\mathbb{Z}_2$ index, the system remains as nontrivial element in the $\mathbb{Z}_2$ homotopy group of the Hamiltonian. 
In general, we define stable index as topological number that are well defined for arbitrary number of bands.
It should be emphasized that, different from the ``strong topological index'', the stable topological index does not imply non-Wannierizability.
A well known example having stable topological index is the SSH model \cite{SSH1979}, in which both the trivial phase and the nontrivial phase are Wannierizable. 
{\bf 4.} Breaking the approximate symmetry used to prove 1 and 2, we numerically show that, for $\theta>\theta_{m6}$, our results do not change 
\footnote{In appendix \ref{sub:PHbreaking}, four sets of parameters from fitting \textit{ab-initio} results at different momenta are used to confirm this statement. Three out of the four angles selected from the four gapped regions are still ``gapped'' angles for all four different sets; for the fourth angle $\alpha=3.1$, using set-(ii) parameters, 2B-1V is connected to the other bands.} 
(See appendix \ref{sub:PHbreaking}).

\begin{table}
\begin{tabular}{lrrr|lrr|lrr}
\hline 
& $\Gamma_{1}$ & $\Gamma_{2}$ & $\Gamma_{3}$ &  & $M_1$ & $M_2$ &  & $K_1$ & $K_2K_3$ \tabularnewline
\hline 
$E$ & 1 & 1 & 2 &        $E$ & 1 & 1 &        $E$ & 1 & 2\tabularnewline
$2C_{3}$ & 1 & 1 & -1 &  $C_{2}^\prime$ & 1 & -1 & $C_{3}$ & 1 & -1 \tabularnewline
$3C_{2}^\prime$ & 1 & -1 & 0 &     &   &   &       $C_{3}^{-1}$ & 1 & -1 \tabularnewline
\hline
\end{tabular}
\protect\caption{\label{tab:irreps-MSG} Character table of irreps at high symmetry momenta in magnetic space group $P6^\prime2^\prime2$ (\#177.151 in BNS settings) \cite{BCS-MSG}. The definitions of high symmetry momenta are given in table \ref{tab:EBR-magSG}. For the little group of $\Gamma$, $E$, $C_3$, and $C_2^\prime$ represent the conjugation classes generated from identity, $C_{3z}$, and $C_{2x}$, respectively. The number before each conjugate class represents the number of operations in this class. Conjugate class symbols at $M$ and $K$ are defined in similar ways. }
\end{table}

\begin{table}
\footnotesize
\begin{tabular}{lrrrrrr|lrrrr|lrrr}
\hline 
& $\Gamma_{1}$ & $\Gamma_{2}$ & $\Gamma_{3}$ & $\Gamma_{4}$ & $\Gamma_{5}$ & $\Gamma_{6}$ &  & $M_{1}$ & $M_{2}$ & $M_{3}$ & $M_{4}$ & & $K_1$ & $K_2$ & $K_3$ \tabularnewline
\hline 
$E$ & 1 & 1 & 1 & 1 & 2 & 2 & $E$ & 1 & 1 & 1 & 1 & $E$ & 1 & 1 & 2\tabularnewline
$2C_{6}$ & 1 & 1 & -1 & -1 & -1 & 1 & $C_{2}$ & 1 & 1 & -1 & -1 & $C_3$ & 1 & 1 & -1\tabularnewline
$2C_{3}$ & 1 & 1 & 1 & 1 & -1 & -1 & $C_{2}^\prime$ & 1 & -1 & -1 & 1 & $3C_2^{\prime\prime}$ & 1 & -1 & 0 \tabularnewline
$C_{2}$ & 1 & 1 & -1 & -1 & 2 & -2 & $C_{2}^{\prime\prime}$ & 1 & -1 & 1 & -1\tabularnewline
$3C_{2}^\prime$ & 1 & -1 & -1 & 1 & 0 & 0 &  &   &  &  & \tabularnewline
$3C_{2}^{\prime\prime}$ & 1 & -1 & 1 & -1 & 0 & 0 &  &  &  &  & \tabularnewline
\hline 
\end{tabular}
\protect\caption{\label{tab:irrep-SG} Character table of irreps at high symmetry momenta in space group $P622$ (\#177) with time-reversal symmetry. The definitions of high symmetry momenta are given in table \ref{tab:EBR-SG}. For the little group of $\Gamma$, $E$, $2C_6$, $2C_3$, $C_2$, $3C_2^\prime$, and $3C_2^{\prime\prime}$ represent the conjugation classes generated from identity, $C_{6z}$, $C_{3z}$, $C_{2z}$, $C_{2x}$, and $C_{2y}$ respectively. The number before each conjugate class represents the number of operations in this class.
Conjugate class symbols at $M$ and $K$ are defined in similar ways, wherein the $C_2^\prime$ and/or $C_2^{\prime\prime}$ classes are always subsets of the $C_2^\prime$ and/or $C_2^{\prime\prime}$ classes at $\Gamma$.}
\end{table}

As the twist angle decreases away from $\theta_{m1}$ in the MBM, the band structure and the irreps at high-symmetry momenta experience non-monotonous changes, but ``magic angles'' -  where the Fermi velocity at half-filling vanishes - reappear periodically. In Fig. \ref{fig:MBM-phase}, we show the evolution of the Fermi velocity as function of twist angle $\theta$. We also plot the first six magic angles up to $\theta_{m6}=0.167^\circ$. As a function of $\theta$, the middle two bands are not always separated by a (direct) gap at all momenta: there are four gapped intervals, denoted as $\Theta_a$, where the middle two bands form a separate group of bands, shaded by grey in Fig.~\ref{fig:MBM-phase}. While $\theta_{m3}\in \Theta_a$,  $\theta_{m1,m2,m4,m5,m6} \notin \Theta_a$. Therefore, the topology is ill-defined for the middle two bands in the MBM at higher magic angles.  However, for any $\theta\in \Theta_a$, as well as for smaller angles, we show below that 2B-1V is topologically nontrivial.

Besides the $P6^{\prime}2^{\prime}2$ symmetry, the MBM also has an approximate particle-hole (PH) symmetry. In the limit of zero rotation of the Pauli matrices of the spin between the two graphene layers, this symmetry is exact - one of the approximations used in \cite{MacDonald_M-Model}. The symmetry is crucial in proving a theorem for TBG. We then numerically show that our results, understood in light of this symmetry, hold for the general case. It is easy to show that $H\left(\mathbf{k}\right)  =  -P^\dagger H\left(-\mathbf{k}\right)P$ where $H(\mathbf{k})$ represents the MBM and $P$ is the particle-hole operator. Both are expanded in detail in appendix \ref{sub:Sym}. We emphasize that this PH symmetry is unitary, and squares to $-1$. An important property we use is that the PH operator anti-commutes with $C_{2x}$, \ie $\{P,C_{2x}\}=0$ (see appendix \ref{sub:Sym}).

First we show that in the presence of the approximate PH symmetry, for $\theta\in\Theta_a$ 2B-1V has \textit{at least} nontrivial fragile topology. The explicit full proof is given in appendices \ref{sub:Irreps}, \ref{sub:fragile-Moire} and \ref{sub:WLfromIrrep}. The irreps at $\mathrm{K}$ and $\mathrm{K'}$ of 2B-1V are always the same as the irreps at $\mathrm{K}$ and $\mathrm{K'}$ in single layer graphene. For the irreps of 2B-1V at $\Gamma$ and M, one uses that the approximate PH symmetry matrix anti-commutes with $C_{2x}$ but commutes with the other generators of $C_{2z}T$ and $C_3$. Due to these relations, the upper and the lower irreps have opposite $C_{2x}$ number, \ie $\Gamma_1+{\Gamma_2}$. Similarly, the irreps at M are forced to be $\mathrm{M}_1+\mathrm{M}_2$.
The bands at any $\theta\in\Theta_a$ have the same irreps at all high-symmetry momenta as those for $\theta\gtrsim\theta_{m1}$ (A-phase in Fig. \ref{fig:MBM-phase}(a)). They then have the same EBR decomposition as the A-phase and we know that the 2B-1V again has fragile topology. The fragile topology can be proved from another perspective. In appendix \ref{sub:WLfromIrrep}, we prove a lemma relating the winding of the Wilson loop eigenvalues of any 2B-1V model  to the irreps at high-symmetry momenta, similar to index theorems in Ref. \cite{Hughes2011,Turner2010,Fang2012}. Applying the lemma to the irreps of 2B-1V at any $\theta\in$ any gapped interval, we find the winding to be $\pm 1$ mod 3, \ie nontrivial.

The irreps cannot, by themselves, distinguish if the winding is even (but nonzero) or odd. For the first four gapped phases, we calculate the Wilson loop of 2B-1V, and find its winding to always be $\pm1$. This suggests something stronger: we \textit{conjecture} that 2B-1V has in fact stable $\mathbb{Z}_2$ nontrivial topology protected by $C_{2z}T$. 
To see this, one realizes that the homotopy group of gapped Hamiltonians in 2D is given by $\pi_2[O(N_{occ}+N_{unocc})/O(N_{occ}\oplus{N}_{unocc})]=\mathbb{Z}_2$ \cite{Fang2015} for $N_{unocc,occ}>2$ and $=\mathbb{Z}$ for $N_{occ}=2$ where the latter is nothing but the winding number of Wilson loop eigenvalues. Adding trivial bands to the lowest two bands maps one element in $\mathbb{Z}$ to one $\mathbb{Z}_2$ following the simple rule $z_2=z$ mod 2, \ie odd windings are stable to superposition of trivial bands (see appendix \ref{sec:homotopy}). 
The collapse from $\mathbb{Z}$ classification to $\mathbb{Z}_2$ classification when more bands are considered has been discussed in Ref. \cite{Ahn2018b}, and the $\mathbb{Z}_2$ index is identified as the second Stiefel-Whitney class.
In appendix \ref{sub:higherbands}, we also analyze the topological character of higher energy bands, some, but not all, of which are nontrivial.  
 
Having proved the topological nature of the 2B-1V model, we turn to the Moir\'e model with two valleys. In the MBM, {\it no} inter-valley coupling exists and the four bands - two valleys (4B-2V) near $E_F$, without inter-valley coupling, are just two copies of the 2B-1V. The $C_{6z}$, $C_{2z}$, $C_{2y}$ and time-reversal symmetries, which send one valley to another, are recovered in the two-valley system; the symmetry group is SG $P622$ (\#177). We compute the irreps of the $4$ bands around the charge neutrality point to be $\Gamma_2+\Gamma_3+\Gamma_1+\Gamma_4$, $\mathrm{M}_2+\mathrm{M}_3+ \mathrm{M}_1+\mathrm{M}_4$, and $2\mathrm{K}_3$ at the respective high-symmetry points $\Gamma, \mathrm{M}, \mathrm{K}$. Due to the different space-group, the symbols of irreps here have different meanings than those in MSG $P6^\prime2^\prime2$ (table \ref{tab:irreps-MSG}); their definitions are given in table \ref{tab:irrep-SG}. Comparing the irreps of the $4$ lowest bands to those of the EBR's of SG $P622$ \cite{Bernevig_TQC,Bilbao-Rep,Vergniory2017}, we find that the four bands decompose into $G^{2c}_{A_1}+G^{2c}_{A_2}$ ($G^{site}_{irrep}$ are EBR notations \cite{EBRnotation} of SG $P622$, given in table \ref{tab:EBR-SG}). Hence their eigenvalues are the same as those of a sum of two atomic insulators. If we  break the valley quantum number by considering the inter-valley coupling, a small gap is opened at the $K$ point, and the model is now Wannierizable. This gap opening, impossible in the MBM, does exist in our {\it ab-initio} calculations and can be tuned by increasing the inter-layer hopping $w$ (See Fig.~\ref{fig:DFT-gap} in the appendix).

\begin{figure}
\begin{centering}
\includegraphics[width=1\linewidth]{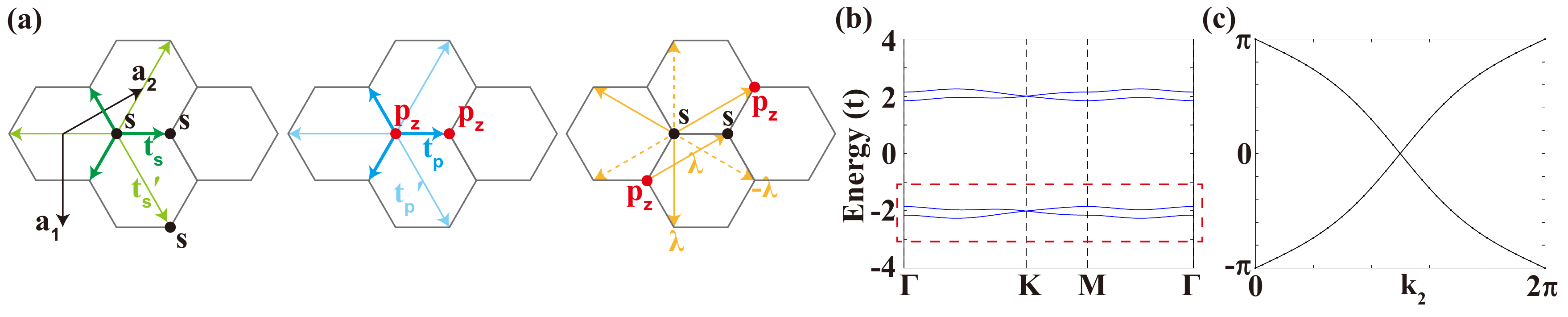}
\par\end{centering}
\protect\caption{\label{fig:TB4-1V} 
The four-band tight-binding model for one valley (TB4-1V). The \textit{lower} two bands of TB4-1V have identical irreps and Wilson loop winding of the gapped middle two bands in the one-valley MBM. (a) The tight-binding model. The energy splitting between $s$ and $p_z$ is $\Delta$. The hopping parameters $t_{s,p}$, $t_{s,p}^\prime$, $\lambda$ are all in general complex numbers. (b) The band structure. (c) The Wilson loop of the lower two bands. (b) and (c) are calculated with the parameters $t_s=t_p=t$, $t_s^\prime=t_p^\prime=-\frac{1}{3}t$, $\lambda=\frac{2}{\sqrt{27}}t$, and $\Delta=0.15t$.}
\end{figure}

We now build minimal (number of orbitals), short-range hopping models with the correct symmetries and topology, that should be used as toy-models for the two topological graphene bands. 
Due to the topological obstruction, a two-band short range model with the correct symmetry that simulates 2B-1V is excluded \cite{kang2018,po2018}. 
However, we can successfully build a tight-binding 4-band, 1 valley  (TB4-1V) lattice model with short range hopping parameters, which has 2 bands separated by a band gap from another 2 bands: either of these two bands is a model for the 2B-1V. 
Interacting physics in the 2B-1V model can be obtained by projecting to the lower $2$ bands of our TB4-1V model. We here give the strategy for building this model, and leave the details in appendix \ref{sub:TB4-1V}. We obtain the correct symmetry content of a TB4-1V model from the viewpoint of EBR's \cite{Bernevig_TQC}. We showed that the irreps ($\Gamma_1+\Gamma_2$, $\mathrm{M}_1+\mathrm{M}_2$, and $\mathrm{K}_2\mathrm{K}_3$) of the middle two bands in one-valley MBM (appendix \ref{sub:fragile-Moire}), can only be written as a difference of EBR's: $G^{2c}_{A_1}+G^{1a}_{A_1}-G^{1a}_{A_2}$. (See table \ref{tab:EBR-magSG} for the EBR's in MSG $P6^\prime 2^\prime 2$.) We now reinterpret these irreps differently: from table \ref{tab:EBR-magSG} we see that they can be thought as forming one disconnected, split, topological, branch of the composite BR formed by the sum of EBR's, $G^{2c}_{A_1}$ and $G^{2c}_{A_2}$. This understanding gives us an ansatz for building the TB4-1V model. We start with two independent EBR's,  $G^{2c}_{A_1}$ and $G^{2c}_{A_2}$, which give the irreps $2\Gamma_1$, $2\mathrm{M}_1$, $\mathrm{K}_2\mathrm{K}_3$, and $2\Gamma_2$, $2\mathrm{M}_2$, $\mathrm{K}_2\mathrm{K}_3$, respectively. These EBR's are formed by $s$ and $p_z$ orbitals (which differ from each other by $C_{2x}$ eigenvalues, $1$ and $-1$), respectively, sitting on the $2c$ Wyckoff position (the honeycomb sites).  We then mix the two EBR's, undergo a phase transition, and decompose the bands into two new branches, each of which has the irreps $\Gamma_1+\Gamma_2$, $\mathrm{M}_1+\mathrm{M}_2$, and $\mathrm{K}_2\mathrm{K}_3$ - the correct representations of 2B-1V. Each branch then has a Wilson loop winding $3n\pm1$ and is topological. By this method the TB4-1V model  reproduces the irreps of the middle two bands. What is left  is to reproduce the correct Wilson loop winding (in principle it may wind $3n\pm 1$ times.) Heuristically, since the winding of the Wilson loop is $1$ in the 2B-1V model, only one phase transition separates this phase from the phase described by the two EBR's - $G^{2c}_{A_1}$ and $G^{2c}_{A_2}$ - with a gap between them. Heuristically, the number of phase transitions gives the winding of the Wilson loop. The orbitals and hoppings are shown in Fig. \ref{fig:TB4-1V}. There are six parameters in total: $\Delta$- a real number -  the energy splitting between $s$ and $p_z$ orbitals, $t_{s,p}$ the nearest hoppings, $\lambda$ the second-nearest hopping, and $t_{s,p}^\prime$ the third-nearest hoppings. For simplicity, here we set $t_s=t_p=t$, $t_s^\prime=t_p^\prime=t^\prime$ and $\lambda$ all real numbers. If the parameters satisfy $3|t+t^\prime|>|\Delta|$ and $|3t-3t^\prime|>|\Delta|$, the lower two bands form the irreps $\Gamma_1+\Gamma_2$, $M_1+M_2$, $K_2K_3$. To make the bands flat, we further ask that the averaged energy of the lower two bands at $\Gamma$, $M$, $K$ equals each other, which requires $t^\prime=-\frac{1}{3}t$, $\lambda=\frac{2}{\sqrt{27}}t$.  
The band structure and Wilson loop with these parameters are plotted in Fig. \ref{fig:TB4-1V}(b) and (c), where $\Delta$ is set as $0.15t$. Our TB4-1V model reproduces both the correct irreps, Wilson loop winding and flat dispersion. 

The two-valley tight-binding model that preserves valley symmetry, in SG $P622$, is a stacking of the TB4-1V model and its time-reversal counterpart (see appendix \ref{sub:TB8-2V}). We refer to this model as TB8-2V. 
The lower (or upper) 4 bands of this short range hopping model give the same physics as the two-valley graphene model.  
As shown by our first principle calculations in section \ref{sec:DFT}, the inter-valley coupling of perfect TBG is small but nonzero ($\sim$0.5meV) and increasing (see section \ref{sec:DFT}) at a small twisting angle ($\theta_{23}\approx 1.41^\circ$), as shown in Fig. \ref{dftgaps}(b) (likely to be smaller in the real material due to buckling effects that diminish the gap). We simulate this in the TB8-2V model by adding valley symmetry breaking terms such as $\zeta \tau_y \mu_z \sigma_z$, where $\tau_y$, $\mu_z$, $\sigma_z$ act in valley, orbital, and sublattice degrees of freedom, respectively. These terms  open gap at the $K$ point and trivialize the Wilson loop, (See Fig. \ref{fig:TB8-2V}.)  
In the absence of valley symmetry, the Wilson loop winding is no longer protected and the Wannier obstruction for the lowest $4$ bands is removed. 
This allows us (appendix \ref{sub:TB4-2V}) to build a tight-binding model just for the lower four bands by explicitly constructing Wannier functions for the lower four bands in TB8-2V \cite{kang2018}. 
We sketch the reasoning here. 

To build a TB4-2V model, we start with a TB8-2V model \textit{preserving} the valley symmetry, \ie $\zeta=0$, but take a symmetry-breaking gauge for its Wannier functions. We follow the projection procedure of Ref. \cite{Vanderbilt_Wannier_Z2}. The lower four bands of TB8-2V form the irreps $\Gamma_1+\Gamma_2+\Gamma_3+\Gamma_4$, $\mathrm{M}_1+\mathrm{M}_2+\mathrm{M}_3+\mathrm{M}_4$, and $2 \mathrm{K}_3$, which can be induced by the EBR's $G^{2c}_{A_1}$ and $G^{2c}_{A_2}$ of SG $P622$. (See table \ref{tab:irrep-SG} and \ref{tab:EBR-SG} for the definitions of irreps and EBR's, respectively.) 
Just as in TB4-1V model, the two EBR's $G^{2c}_{A_1}$ and $G^{2c}_{A_2}$ can be realized by $s$ and $p_z$ orbitals on the $2c$ position, respectively, (of opposite $C_{2x}$ eigenvalues). We then define the four projection orbitals for Wannier functions  such that they respect the symmetry of SG $P622$ plus time-reversal but break valley symmetry. Hence they can be localized in a short range tight-binding model - all details are given in appendix \ref{sub:TB4-2V}. This model can be used for angles for which the inter-valley hybridization is observed. 

Our proofs of topology in low energy TBG are based on the simplified MBM. To give them any credibility, we now must relate our predictions to the realistic calculations of twisted bilayer graphene in SG $P622$ with negligible SOC. We took the time to perform a series of {\it ab-initio} calculations for $i\in \{6, 10, 16, 23, 30\}$  where $i$ denotes the commensurate twist angle by the formula  $\theta_i=\arccos\frac{3i^2+3i+0.5}{3i^2+3i+1}$~\cite{TBGstr}. Full band structures are given in appendix \ref{sec:DFT}. Due to the technical difficulty in these calculations with VASP, our number of $k$-points sampled is about 30, 18, 6, and 1 for i = 6, 10, 23 and 30, respectively. The computational time for one point is  6, 24, 72, and 720 hours, respectively. Our {\it ab-initio} calculations show three remarkable features different from the MBM: {\bf 1}. the PHS breaking is  larger, with effects on the metallic vs insulating phases of low energy graphene, {\bf 2}. the gap at the $K$ point of the Moir{\'e} BZ varies as the $\theta$ is changed, {\bf 3}. the representation sequence of the two lowest bands at the high symmetry points can be different and undergoes phase transitions - leading to metallic phases at the $\Gamma$ point mixed with higher bands, depending on the twist angle.

In Fig.~\ref{dftgaps}(a), we show the evolution of the \textit{ab-initio} energy bands at $\Gamma$ explicitly. The gray line stands for the  $2$-fold $\Gamma_2+\Gamma_3$ band (``23'') and the red line stands for the $2$-fold  $\Gamma_1+\Gamma_4$ band (``14''). Please see table \ref{tab:irrep-SG} for definitions of these irreps.  The two blue lines stand for two different $4$-fold $\Gamma_5+\Gamma_6$ bands (``low56'' and ``high56''). The energies at the $\Gamma$ point  are not particle-hole symmetric. 
The gap between the ``low56''  and ``23'' bands is much smaller than that between the ``14'' and ``high56'' bands. The closing of the former gap happens at $i=16$, while for the latter the closing happens at $i=30$. This results in a metallic phase for the middle 4 bands  for $16<i<30$, in which the middle 4 bands are not separated from others. This metallic phase can only occur when the PHS is broken; it can be somewhat modeled by the MBM when the $\theta$ dependence in the rotated spin matrix is not neglected. In the gapped phase (\eg i=6), the ordering of the energy bands is 4224, which denotes the degeneracy of the bands in the ascending order. In the metallic phase, it becomes 2424. We can ask if two of $4$-fold   $\Gamma_5+\Gamma_6$  bands along with the $2$-fold $\Gamma_1+\Gamma_4$ can form a separate set of $4$ bands around the charge neutrality point. The set of irreps $\{\Gamma_5+\Gamma_1+\Gamma_4, M_1+M_2+M_3+M_4, 2K_{3}\}$, (or, alternatively  $\{\Gamma_6+\Gamma_1+\Gamma_4, M_1+M_2+M_3+M_4, 2K_{3}\}$), do not satisfy the compatibility relations of SG $P622$ and hence there is no gapped set of 4 middle bands.

Lastly, in Fig.~\ref{dftgaps}(b), we calculate the Moir{\'e} $K$ point gap as a function of the twist angles ($\theta_i$). Starting for a large $\theta$ (\ie $\theta_6$), we find a decreasing $K$ gap as the twist angle changes to $\theta_{10}$; then it increases as $\theta$ decreases further. In the MBM, no energy difference is expected between two $K_3$ representations coming from different valleys. However, in {\it ab-initio} calculations of perfect TBG, this is not the case. Decreasing $\theta$, the Moir\'e supercell grows rapidly (\eg for $\theta_{30}$, there are 11000 atoms in the supercell). Due to numerical difficulties, we can only obtain the K gap in {\it ab-initio} for $i=6$, 10, 16 and 23. We conjecture that the K gap of perfect TBG is not negligible for an extremely small angle because of the inter-valley coupling. We can simulate the trends of the K gap by decreasing the distance $z$ (increasing the inter-layer hopping $w$) for a relatively larger angle $\theta_{10}$, which shows the same tendency as decreasing $\theta$~\cite{MacDonald_M-Model}. Both  the flattening of middle bands and the increase of the Moir{\'e}  $K$  gap are obtained and presented in Fig.~\ref{fig:DFT-gap} in appendix \ref{sec:DFT}. Since the {\it ab-initio} models of perfect graphene predict an observable gap at K, while the experimental data suggests that this gap is small, we are left with the conclusion that effects not introduced in our {\it ab-initio} (or in the MBM) such as lattice relaxation and/or graphene lattice warping, are important and render the gap at the K point much smaller. Understanding these issues is of utmost importance for a microscopic, non-phenomonologic theory of TBG.

In conclusion, we have performed a complete and exhaustive study of the TBG band structure and showed that the low energy bands are always topological, near \textit{all} magic angles, as long as the lowest energy bands are gapped from the rest. Moreover, we conjecture that there is a stable character of their topology, as the Wilson loop winding is $1$, which allows for the definition of a stable topological index, even though the analysis based on the high-symmetry points characters of the  Elementary Band Representations can only prove a fragile phase. This underpins topology as the fundamental property of the low energy bands of TBG. We then provided short-range toy tight-binding models for the low energy TBG, which should (with atomic orbitals matching the charge density of a triangular lattice) be used to study the effects of correlations in graphene. We then checked our results by performing a large and time-consuming set of {\it ab-initio} calculations, and presented the areas of agreement and disagreement with the simple, continuum model, as well as possible solutions to these disagreements. 

{\it Note.} During the extended period of time that passed while obtaining our {\it ab-initio} results, Ref.~\cite{po2018} has also predicted that the bands at half filling are fragile topological. The differences between our papers are: our paper contains {\it ab-initio} results essential in confirming the nature of the bands, uses the particle-hole symmetry to \textit{prove} all low-energy bands in TBG at \textit{any} angle are topological, conjectures that the bands contain a stable index. Our paper does not present conjectures or proofs about the Mott phase of TBG.
After our paper was posted, Ref. \cite{Po2018b} proved that the $C_2T$ fragile phase with nontrivial $\mathbb{Z}_2$ index became Wannierizable after adding atomic bands. 
This conclusion is not inconsistent with ours since stable index (unlike a strong index) does not necessarily implies non-Wannierizability.

After a very brief discussion with us, the authors of Ref. \cite{Po2018b} posted a version 2 of their paper which identifies a stable index (referred as the second Stiefel-Whitney index $w_2$). This index is another indication of the collapse of the homotopy group from $Z$ to $Z_2$, pointed out in our paper for TBG. After version 2 of Ref. \cite{Po2018b} appeared, Ref. \cite{Ahn2018} presented an elaborate study on the $w_2$ index and its relation with Wilson loop. Ref. \cite{Ahn2018} also worked out a relation between $w_2$ and the nested Wilson loop, which was consistent with our claim (in appendix \ref{sec:homotopy}) that the nested Wilson loop could be used to diagnose our new $\mathbb{Z}_2$ index (classification) of TBG.

Among the several changes in version 2 of Ref. \cite{Po2018b} are also added several fragile “resolutions” of the 2 band TBG model, in which atomic bands of nontrivial stable $w_2$ index are added to cancel the $w_2$ index of the $2$ fragile TBG bands. We emphasize that in the (one-valley) MBM, in presence of particle hole symmetry, one cannot trivialize the $w_2$ index of the middle two bands by adjusting parameters, even if adjusting parameters may couple additional bands to the middle two bands.
This is because if a set of bands with $w_2=1$ is moved from positive energy to Fermi level, then according to particle-hole symmetry another set of bands with $w_2=1$ is moved from negative energy to Fermi level, and thus the total $w_2$ index remains invariant.
(The statement above requires Ref. \cite{Ahn2018}. They pointed out that upon adding two sets of bands, the total $w_2$ index was in general not the sum of $w_2$ of each: the Berry phases of each band set also contributed to the total $w_2$. 
However, in presence of $C_3$ symmetry, one can easily find that the Berry phases' contribution always vanish. 
Therefore, in presence of $C_3$ symmetry, the $w_2$ index is additive.)

{\it Acknowledgments.}
 We thank Barry Bradlyn, Felix von Oppen, Mike Zaletel, Hoi Chun Po for helpful discussions.  Z.~S. and C.~F. were supported by MOST (No. 2016YFA302400, 2016YFA302600) and NSFC (No. 11674370, 11421092). Z. W., and B. A. B. were supported by the Department of Energy Grant No. DE-SC0016239, the National Science Foundation EAGER Grant No. NOA-AWD1004957, Simons Investigator Grants No. ONR-N00014-14-1-0330, No. ARO MURI W911NF-12-1-0461, and No. NSF-MRSEC DMR- 1420541, the Packard Foundation, the Schmidt Fund for Innovative Research. G. L. acknowledges the starting grant of ShanghaiTech University and Program for Professor of Special Appointment (Shanghai Eastern Scholar). The calculations were carried out at the HPC Platform of Shanghaitech University Library and Information Services, and School of Physical Science and Technology.  

\bibliographystyle{apsrev4-1}
\bibliography{ref}

\newpage

\begin{widetext}
\appendix


\section{Notations on symmetries and representations} \label{sec:notation}

We first specify the notations about symmetry and representations. 
If the Hamiltonian is symmetric under a group operation, $g$, we have 
\begin{equation}
H(g\mathbf{k}) = D_\mathbf{k}(g)H(\mathbf{k})D_\mathbf{k}(g^{-1}), \label{eq:symm-D}
\end{equation}
where $D_\mathbf{k}$, the representation matrix of symmetry operator, in general depends on $\mathbf{k}$. In general $H(\mathbf{k})$ is not periodic in $\mathbf k$: it may change through a unitary transformation after a translation of reciprocal lattice, \ie 
\begin{equation}
H(\mathbf{k+G})=V^\mathbf{G}H(\mathbf{k})V^{\mathbf{G}\dagger}. \label{eq:embedding}
\end{equation}
Here $V^\mathbf{G}$ is the ``embedding matrix'' \cite{alexandradinata2016} and it satisfies $V^{\mathbf{G}_1}V^{\mathbf{G}_2}=V^{\mathbf{G}_1+\mathbf{G}_2}$ and $V^0=\mathbb{I}$. 
For \textit{symmorphic} space groups, we can always make $D_\mathbf{k}$ independent on $\mathbf{k}$ by properly choosing the embedding matrices.
To be specific, for tight-binding models we can define the Hamiltonian in momentum space as $H_{\alpha s,\beta s^\prime}(\mathbf{k}) = \langle \phi_{\alpha s \mathbf{k}}|\hat{H}|\phi_{\beta s^\prime \mathbf{k}}\rangle$, where the Bl\"och bases are $|\phi_{\alpha s \mathbf{k}}\rangle = \frac{1}{\sqrt{N}}\sum_{\mathbf{R}\alpha s} e^{i\mathbf{k}\cdot(\mathbf{R+t}_s)}|\alpha \mathbf{R}+\mathbf{t}_s \rangle$, such that the embedding matrices are $V_{\alpha s, \beta s^\prime}^\mathbf{G} = \delta_{\alpha \beta}\delta_{s,s^\prime}e^{-i\mathbf{G}\cdot\mathbf{t}_s}$ and the $D_\mathbf{k}$'s are independent on $\mathbf{k}$.
Here $\alpha$ is the orbital label, $\mathbf{R}$ is the lattice vector, $\mathbf{t}_s$ is the sublattice vector (the position of the sites in the unit cell), and $N$ is the number of unit cells in real space.
In this gauge, adopted in the rest of the paper, we sometimes omit the notation $D_\mathbf{k}(g)$ and replace it directly with $g$. When we write $g$ acting on Hamiltonian and wave function, \ie $gH(\mathbf{k})g^{-1}$ and $g|u\rangle$, we mean it shorthand  for $D_\mathbf{k}(g)H(\mathbf{k}) D_\mathbf{k}(g)^{-1}$ and $D_\mathbf{k}(g) |u\rangle$, respectively.

Substituting Eq. (\ref{eq:embedding}) to Eq. (\ref{eq:symm-D}), we get the following the identity
\begin{equation}
    g V^{\mathbf{G}} g^{-1} = V^{g\mathbf{G}}. \label{eq:embedding-id}
\end{equation}
For a high symmetry momentum $\bf k$ on the Brillouin Zone (BZ) boundary, which satisfies $g\mathbf{k} = \mathbf{k+G}$ with $\mathbf{G}$ some reciprocal lattice, we have
\begin{equation}
    g H(\mathbf{k}) g^{-1} = V^{g\mathbf{k}-\mathbf{k}} H(\mathbf{k}) V^{g\mathbf{k}-\mathbf{k} \dagger}.
\end{equation}
Using identity (\ref{eq:embedding-id}), it is direct to prove that the matrices $V^{\mathbf{k}-g\mathbf{k}}g$, with $g\mathbf{k}=\mathbf{k}$ (modulo a reciprocal lattice), form a group commutting with $H(\mathbf{k})$. 
Therefore, when we say the wave functions at $\mathbf{k}$, $|u_{n\mathbf{k}}\rangle$, form a representation of the symmetry $g$, we actually mean that
\begin{equation}
    V^{\mathbf{k}-g\mathbf{k}} g|u_{n\mathbf{k}}\rangle = \sum_m |u_{m\mathbf{k}}\rangle S_{mn}(g).
\end{equation}
Here $S_{mn}(g)$ is the corresponding representation matrix.

\section{Moir{\'e} band model (MBM)}

\subsection{One-valley Moir{\'e} band model (MBM-1V)} \label{sub:M-1V}

We here present a detailed derivation of the band structure of the Moir{\'e} pattern in twisted bilayer graphene, with emphasis on the symmetries of the system. When the twisting angle is small, a Moir{\'e} pattern is formed by the interference of lattices from the two layers. The Moir{\'e} pattern has a very large length
scale (or unit cell, if commensurate).  The low energy, close to half-filling band structure is formed only from the electron states around the Dirac cones in each layer. A Moir{\'e} band theory describing such a
state is built by the authors of Ref. \cite{MacDonald_M-Model}. In the following, we refer to it as
one-valley Moir{\'e} model (MBM-1V). Part of our analytical study, especially the topological
study, is mainly based on this model. Thus in this section, we give
a review of the one-valley Moir{\'e} model.

We define the Bl{\"o}ch bases in the top  (not rotated) layer as
\begin{equation}
\left|\phi_{\mathbf{p}\alpha}^{\left(T\right)}\right\rangle =\frac{1}{\sqrt{N}}\sum_{\mathbf{R}}e^{i\left(\mathbf{R}+\mathbf{t}_{\alpha}\right)\cdot\mathbf{p}}\left|\mathbf{R}+\mathbf{t}_{\alpha}\right\rangle \label{eq:phi-T}
\end{equation}
where $\alpha=1,2$ is the sublattice index, $\mathbf{t}_{\alpha}$
is the sublattice vector, $\mathbf{R}$ is lattice vector, and
$\left|\mathbf{R}+\mathbf{t}_{\alpha}\right\rangle $ is the Wannier
state at site $\mathbf{t}_{\alpha}$ in lattice $\mathbf{R}$. The
Bl{\"o}ch bases in bottom layer can be obtained by rotating and shifting the Bl{\"o}ch
basis in the top layer 
\begin{eqnarray}
\left|\phi_{\mathbf{p}\beta}^{\left(B\right)}\right\rangle  & = & \hat{P}_{\left\{ M_{\theta}|\mathbf{d}\right\} }\left|\phi_{\left(M_{\theta}^{-1}\mathbf{p}\right)\beta}^{\left(T\right)}\right\rangle =\frac{1}{\sqrt{N}}\sum_{\mathbf{R}^{\prime}}e^{i\left(\mathbf{R}^{\prime}+\mathbf{t}_{\beta}^{\prime}\right)\cdot\mathbf{p}}\left|\mathbf{R}^{\prime}+\mathbf{t}_{\beta}^{\prime}\right\rangle 
\end{eqnarray}
Here $M_{\theta}$ is a rotation (by an angle $\theta$) along the $z$ axis, $\mathbf{d}$ is the translation
from top layer to bottom layer, $\hat{P}_{\left\{ M_{\theta}|\mathbf{d}\right\} }$
is the corresponding rotation operator on wave-functions, $\mathbf{R}^{\prime}$
is the lattice vector in the  bottom layer, and $\mathbf{t}_{\beta}^{\prime}$ is
the sublattice vector in the bottom layer. It is easy to show that the
intra-layer  single-particle first quantized Hamiltonians of the top and bottom layers are related
by:
\begin{eqnarray}
H_{\alpha\beta}^{\left(BB\right)}\left(\mathbf{p}\right) & = & H_{\alpha\beta}^{\left(TT\right)}\left(M_{\theta}^{-1}\mathbf{p}\right)\label{eq:HBB-HTT}
\end{eqnarray}
Here $H^{\left(TT\right)}$ is the top layer Hamiltonian and $H^{\left(BB\right)}$
is the bottom layer Hamiltonian. 

We now derive the inter-layer coupling. First we Fourier transform the inter-layer hopping to real space:
\begin{eqnarray}
H_{\alpha\beta}^{\left(TB\right)}\left(\mathbf{p},\mathbf{p}^\prime\right) & = & \left\langle \phi_{\mathbf{p}\alpha}^{\left(T\right)}\right|\hat{H}\left|\phi_{\mathbf{p}^\prime\beta}^{\left(B\right)}\right\rangle \nonumber \\
 & = & \frac{1}{N}\sum_{\mathbf{R}\mathbf{R}^{\prime}}e^{-i\left(\mathbf{R}+\mathbf{t}_{\alpha}\right)\cdot\mathbf{p}+i\left(\mathbf{R}^{\prime}+\mathbf{t}_{\beta}^{\prime}\right)\cdot\mathbf{p}^\prime}\left\langle \mathbf{R}+\mathbf{t}_{\alpha}\right|\hat{H}\left|\mathbf{R}^{\prime}+\mathbf{t}_{\beta}^{\prime}\right\rangle \label{eq:H(TB)0}
\end{eqnarray}
We follow Ref. \cite{MacDonald_M-Model} to adopt the tight-binding, two-center approximation,
for the inter-layer hopping, \ie
\begin{eqnarray}
\left\langle \mathbf{R}+\mathbf{t}_{\alpha}\right|\hat{H}\left|\mathbf{R}^{\prime}+\mathbf{t}_{\beta}^{\prime}\right\rangle  & = & t\left(\mathbf{R}+\mathbf{t}_{\alpha}-\mathbf{R}^{\prime}-\mathbf{t}_{\beta}^{\prime}\right)\nonumber \\
 & = & \frac{1}{N\Omega}\sum_{\mathbf{q}}\sum_{\mathbf{G}}t_{\mathbf{q}+\mathbf{G}}e^{i\left(\mathbf{q}+\mathbf{G}\right)\cdot\left(\mathbf{R}+\mathbf{t}_{\alpha}-\mathbf{R}^{\prime}-\mathbf{t}_{\beta}^{\prime}\right)}\label{eq:T-matrix}
\end{eqnarray}
Here $\mathbf{q}$ sums over all momenta in the top layer BZ, $\mathbf{G}$ sums over all top layer reciprocal lattices,
and $t_{\mathbf{q}+\mathbf{G}}$ is the Fourier transformation of
$t\left(\mathbf{r}\right)$. Substituting this back into Eq. (\ref{eq:H(TB)0}),
we get
\begin{eqnarray}
H_{\alpha\beta}^{\left(TB\right)}\left(\mathbf{p},\mathbf{p}^\prime\right) & = & \sum_{\mathbf{G}_{1}\mathbf{G}_{2}}\frac{t_{\mathbf{p}+\mathbf{G}_{1}}}{\Omega}e^{-i\mathbf{t}_{\beta}\cdot\mathbf{G}_{2}+i\mathbf{G}_{1}\cdot\mathbf{t}_{\alpha}}\delta_{\mathbf{p}+\mathbf{G}_{1},\mathbf{p}^\prime+M_{\theta}\mathbf{G}_{2}}\label{eq:H(TB)}
\end{eqnarray}
where both $\mathbf{G}_{1}$ and $\mathbf{G}_{2}$ are reciprocal
lattice vectors in the top layer. 

In the MBM-1V, only the low energy electron states
around Dirac cones are  considered. Thus we approximate
the intra-layer Hamiltonian as
\begin{equation}
H^{\left(TT\right)}\left(\mathbf{K}+\delta\mathbf{p}\right)\approx v_{F}\delta\mathbf{p}\cdot\boldsymbol{\sigma}
\end{equation}
\begin{equation}
H^{\left(BB\right)}\left(M_{\theta}\mathbf{K}+\delta\mathbf{p}\right)\approx v_{F}\left(M_{\theta}^{-1}\delta\mathbf{p}\right)\cdot\boldsymbol{\sigma}\approx v_{F}\delta\mathbf{p}\cdot\boldsymbol{\sigma}\label{eq:H(BB)}
\end{equation}
Here $\delta\mathbf{p}$ is a small momentum deviation from the $K$ point. In the bottom layer Hamiltonian, we perform a second approximation and neglect the $\theta$-dependence
of $H^{\left(BB\right)}$. We leave the discussion of the effect of this approximation 
for appendix \ref{sub:PHbreaking}.

The computations and approximations involved in $H^{\left(TB\right)}$ are not so direct and need more discussion. First,  in Eq. (\ref{eq:H(TB)}), for small $\theta$,
we only need to consider the electron states around the Dirac cone in each layer,
\ie states with momenta $\mathbf{p}=\mathbf{K}+\delta\mathbf{p}$
and $\mathbf{p}^\prime=M_{\theta}\mathbf{K}+\delta\mathbf{p}^\prime$ (modulo a
reciprocal lattice vector). On the other hand, $t_{\mathbf{q}}$ depends
only on the magnitude of $\mathbf{q}$ and decays exponentially when $\left|\mathbf{q}\right|$ becomes larger than $1/d_\perp$: $t_\mathbf{q}\propto \exp({-\alpha(|\mathbf{q}|d_\perp)^\gamma})$, here $d_\perp$ is the distance between two layers,. Fitting to data gives $\alpha\approx0.13$, and $\gamma\approx1.25$ \cite{Bistritzer2010}.
Therefore we keep the three largest relevant $t_{\mathbf{k}+\mathbf{G}_{1}}$
terms as $t_{\mathbf{K}+\delta\mathbf{p}}$, $t_{C_{3z}\mathbf{K}+\delta\mathbf{p}}$,
and $t_{C_{3z}^{2}\mathbf{K}+\delta\mathbf{p}}$, corresponding in Eq. (\ref{eq:H(TB)}) to $\bf{p}= \mathbf{K}+\delta\mathbf{p}$ and 
$\mathbf{G}_{1}=0$, $\mathbf{G}_{1}=C_{3z}\mathbf{K}-\mathbf{K}$,
and $\mathbf{G}_{1}=C_{3z}^{2}\mathbf{K}-\mathbf{K}$ in Eq. (\ref{eq:H(TB)}), respectively.
With $C_3$ symmetry, all these three terms are equal to $w\Omega$,
and the inter-layer coupling can be reformulated as
\begin{align}
H_{\alpha\beta}^{\left(TB\right)}\left(\mathbf{K}+\delta\mathbf{p},M_{\theta}\mathbf{K}+\delta\mathbf{p}^\prime\right) \approx & w\sum_{\mathbf{G}_{2}}\bigg\{\delta_{\mathbf{K}+\delta\mathbf{p},\ M_{\theta}(\mathbf{K}+\mathbf{G}_{2})+\delta\mathbf{p}^\prime}e^{-i\mathbf{t}_{\beta}\cdot\mathbf{G}_{2}}\nonumber \\
  + & \delta_{C_{3z}\mathbf{K}+\delta\mathbf{p},\ M_{\theta}(\mathbf{K}+\mathbf{G}_{2})+\delta\mathbf{p}^\prime}e^{i\mathbf{t}_{\alpha}\cdot\left(C_{3z}\mathbf{K}-\mathbf{K}\right)-i\mathbf{t}_{\beta}\cdot\mathbf{G}_{2}}\nonumber \\
  + & \delta_{C_{3z}^{2}\mathbf{K}+\delta\mathbf{p},\ M_{\theta}(\mathbf{K}+\mathbf{G}_{2})+\delta\mathbf{p}^\prime}e^{i\mathbf{t}_{\alpha}\cdot\left(C_{3z}^{2}\mathbf{K}-\mathbf{K}\right)-i\mathbf{t}_{\beta}\cdot\mathbf{G}_{2}}\bigg\}
\end{align}
Since $\delta\mathbf{p}$, $\delta\mathbf{p}^\prime$, and $M_{\theta}\mathbf{K}-\mathbf{K}$
are all small quantities compared with reciprocal lattice vector, only the $\mathbf{G}_{2}=0$, $\mathbf{G}_2=C_{3z}\mathbf{K}-\mathbf{K}$, and $\mathbf{G}_2=C_{3z}^2\mathbf{K}-\mathbf{K}$ terms are considered in the above three lines, respectively.
Therefore the inter-layer Hamiltonian can be finally written as
\begin{equation}
H_{\alpha\beta}^{\left(TB\right)}\approx w\sum_{j=1}^{3}\delta_{\delta\mathbf{p},\delta\mathbf{p}^\prime+\mathbf{q}_{j}}T_{\alpha\beta}^{j}
\end{equation}
where $\mathbf{q}_{1}=M_{\theta}\mathbf{K}-\mathbf{K}$, $\mathbf{q}_{2}=C_{3z}\mathbf{q}_{1}$,
$\mathbf{q}_{3}=C_{3z}\mathbf{q}_{2}$ (Fig. \ref{fig:M-model}),
and 
\begin{equation}
T_{1}=\sigma_{0}+\sigma_{x}
\end{equation}
\begin{equation}
T_{2}=\sigma_{0}+\cos\left(\frac{2\pi}{3}\right)\sigma_{x}+\sin\left(\frac{2\pi}{3}\right)\sigma_{y}
\end{equation}
\begin{equation}
T_{3}=\sigma_{0}+\cos\left(\frac{2\pi}{3}\right)\sigma_{x}-\sin\left(\frac{2\pi}{3}\right)\sigma_{y}
\end{equation}

\begin{figure}
\begin{centering}
\includegraphics[width=1\linewidth]{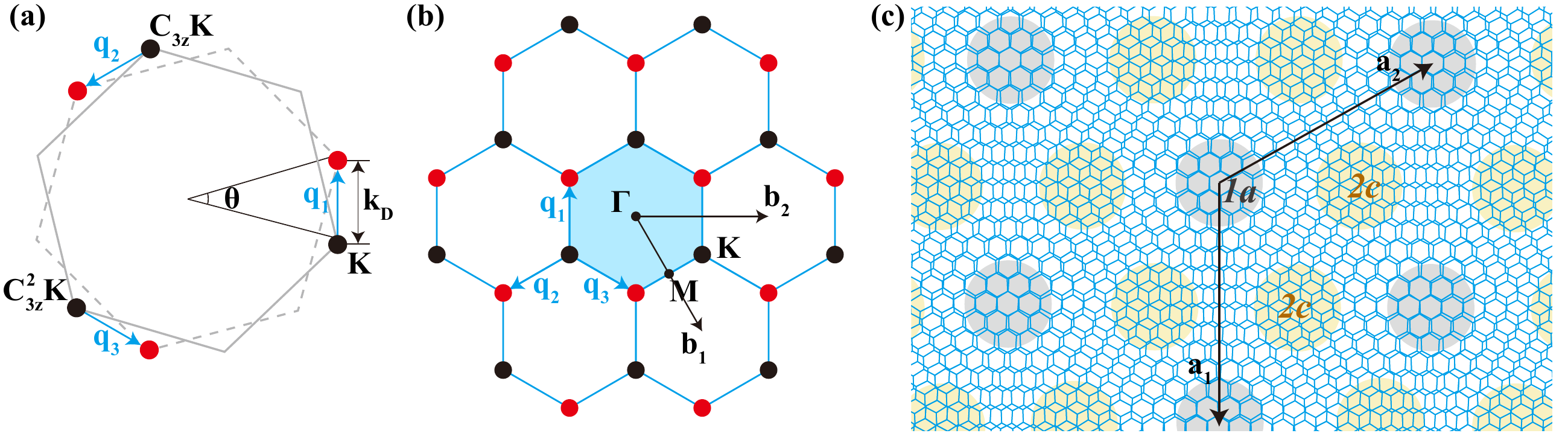}
\par\end{centering}

\protect\caption{\label{fig:M-model}One-valley Moir{\'e} band model (MBM-1V). (a) The Brillouin Zones of
two Graphene layers. The grey solid line and black circles represent
the BZ and Dirac cones of top layer, and the grey dashed
line and red circles represent the BZ and Dirac cones
of the bottom layer. (b) The lattice formed by adding $\mathbf{q}_{1,2,3}$
iteratively. Black and red circles represent Dirac cones from top
layer and bottom layer respectively. The Moir{\'e} Hamiltonian has
periodicity  $\mathbf{b}_{1}=\mathbf{q}_{1}-\mathbf{q}_{3}$ and
$\mathbf{b}_{2}=\mathbf{q}_{2}-\mathbf{q}_{3}$, and the corresponding
BZ (Moir{\'e} BZ) is represented by the shadowed
area. (c) The Moir{\'e} pattern in real space. The AA-stacking area
(shadowed with grey) is identified as $1a$ position in the magnetic
space group $P6^{\prime}2^{\prime}2$ and/or $P6221^\prime$,
and the AB-stacking area (shadowed with yellow) is identified as $2c$
position in the same groups.}.
\end{figure}

To write the Hamiltonian in more compact form, hereafter we re-label
$\delta\mathbf{p}$ as $\mathbf{k}-\mathbf{Q}$, and $\delta\mathbf{p}^\prime$
as $\mathbf{k}-\mathbf{Q}^{\prime}$. 
$\mathbf{Q}$ and $\mathbf{Q}^{\prime}$ take value in the hexagonal lattice formed by adding
$\mathbf{q}_{1,2,3}$ iteratively (Fig. \ref{fig:M-model}(b)) with integer coefficients. There
are two types of $\mathbf{Q}$'s, denoted as black and red circles respectively,
one for the top layer states and the other for the bottom layer states.
Then, the Moir{\'e} Hamiltonian can be written as
\begin{eqnarray}
H^{(MBM-1V)}_{\mathbf{Q},\mathbf{Q}^{\prime}}\left(\mathbf{k}\right) & = & \delta_{\mathbf{Q}\mathbf{Q}^{\prime}}v_{F}\left(\mathbf{k}-\mathbf{Q}\right)\cdot\boldsymbol{\sigma}+w\sum_{j=1}^{3}\left(\delta_{\mathbf{Q}^{\prime}-\mathbf{Q},\mathbf{q}_{j}}+\delta_{\mathbf{Q}-\mathbf{Q}^{\prime},\mathbf{q}_{j}}\right)T^{j}\label{eq:M-model}
\end{eqnarray}
Such a Hamiltonian, when an infinite number of $\mathbf{Q}, \mathbf{Q}'$ are included, is periodic in momentum space: it keeps invariant (up to a unitary transformation) under a translation of $\mathbf{b}_{1}=\mathbf{q}_{1}-\mathbf{q}_{3}$ or $\mathbf{b}_{2}=\mathbf{q}_{2}-\mathbf{q}_{3}$:
\begin{equation}
    H^{MBM-1V}(\mathbf{k}+\mathbf{b}_i) = V^{\mathbf{b}_i}H^{MBM-1V}(\mathbf{k})V^{\mathbf{b}_i\dagger},
\end{equation}
where $i=1,2$, and $V^\mathbf{G}_{\mathbf{Q},\mathbf{Q}^\prime} = \delta_{\mathbf{Q}^\prime,\mathbf{Q}+\mathbf{G}}$ is  the embedding matrix.
Thus $\mathbf{b}_{i=1,2}$ can be thought as Moir{\'e} reciprocal bases, and a Moir{\'e} BZ can then be defined, as shown Fig. \ref{fig:M-model}(b). 
An important property of this Hamiltonian is that, up to a scaling constant,
it depends on a single parameter 
\begin{eqnarray}
H^{(MBM-1V)}_{\mathbf{Q},\mathbf{Q}^{\prime}}\left(\mathbf{k}\right) & = & v_{F}k_{D}\bigg\{\delta_{\mathbf{Q}\mathbf{Q}^{\prime}}\left(\bar{\mathbf{k}}-\bar{\mathbf{Q}}\right)\cdot\boldsymbol{\sigma}+\alpha\sum_{j=1}^{3}\left(\delta_{\bar{\mathbf{Q}}^{\prime}-\bar{\mathbf{Q}},\bar{\mathbf{q}}_{j}}+\delta_{\bar{\mathbf{Q}}-\bar{\mathbf{Q}}^{\prime},\bar{\mathbf{q}}_{j}}\right)T^{j}\bigg\}\label{eq:M-model-1}
\end{eqnarray}
Here $k_{D}=\left|M_{\theta}\mathbf{K}-\mathbf{K}\right|=2\left|\mathbf{K}\right|\sin\frac{\theta}{2}$
is the distance between the top layer Dirac cone and the bottom layer Dirac
cone (Fig. \ref{fig:M-model}(a)), $\bar{\mathbf{k}}=\mathbf{k}/k_{D}$,
$\bar{\mathbf{Q}}=\mathbf{Q}/k_{D}$, $\bar{\mathbf{q}}_{j}=\mathbf{q}_{j}/k_{D}$
are dimensionless momenta, and $\alpha=\frac{w}{v_{F}k_{D}}$ is the
single parameter that the Hamiltonian depends on.

\subsection{Symmetries and elementary band representations \label{sub:Sym}}

In order to build the Moir{\'e} band structure, the MBM-1V Hamiltonian in Eq. (\ref{eq:M-model-1}) uses the states of Dirac Hamiltonian around the $\mathbf{K}$ point in both the top and bottom layers of the bilayer Graphene. The Dirac cone at $\mathbf{K'}=-\mathbf{K}$, is missing in this construction. 
The electronic energy states from the cone  $\mathbf{K'}$  form their own, time-reversed, MBM-1V. Thus MBM-1V breaks time-reversal symmetry. Nevertheless, we find that the MBM-1V respects a combined anti-unitary operation: time-reversal followed by $C_{2z}$. 
Therefore the symmetry group of MBM-1V is a magnetic space group, which turns out to be $P6^{\prime}2^{\prime}2$ (\#177.151
in the BNS setting \cite{Bilbao-MSG}). The generators of this group are:
\begin{enumerate}
\item The $C_{3z}$ symmetry 
\begin{equation}
H^{(MBM-1V)}\left(\mathbf{k}\right)=D^{\dagger}\left(C_{3z}\right)H^{(MBM-1V)}\left(C_{3z}\mathbf{k}\right)D\left(C_{3z}\right)
\end{equation}
where $D_{\mathbf{Q}^{\prime},\mathbf{Q}}\left(C_{3z}\right)=e^{i\frac{2\pi}{3}\sigma_{z}}\delta_{\mathbf{Q}^{\prime},C_{3z}\mathbf{Q}}$,
and $C_{3z}\mathbf{q}_{j}=\mathbf{q}_{j+1}$.
\item The $C_{2x}$ symmetry
\begin{equation}
H^{(MBM-1V)}\left(\mathbf{k}\right)=D^{\dagger}\left(C_{2x}\right)H^{(MBM-1V)}\left(C_{2x}\mathbf{k}\right)D\left(C_{2x}\right)
\end{equation}
where $D_{\mathbf{Q}^{\prime},\mathbf{Q}}\left(C_{2x}\right)=\sigma_{x}\delta_{\mathbf{Q}^{\prime},C_{2x}\mathbf{Q}}$,
and $C_{2x}\mathbf{q}_{1}=-\mathbf{q}_{1}$, $C_{2x}\mathbf{q}_{2}=-\mathbf{q}_{3}$,
$C_{2x}\mathbf{q}_{3}=-\mathbf{q}_{2}$.
\item The $C_{2z}T$ symmetry
\begin{equation}
H^{(MBM-1V)}\left(\mathbf{k}\right)=D\left(C_{2z}T\right)H_{\mathbf{Q},\mathbf{Q}^{\prime}}^{(MBM-1V)*}\left(\mathbf{k}\right)D^{T}\left(C_{2z}T\right) \label{eq:D-C2T}
\end{equation}
where $D_{\mathbf{Q}^{\prime},\mathbf{Q}}\left(C_{2z}T\right)=\sigma_{x}\delta_{\mathbf{Q}^{\prime},\mathbf{Q}}$.
\end{enumerate}
It should be noticed that all  rotations of momenta here are with respect to the $\Gamma$ point of the Moir{\'e} BZ. (Fig. \ref{fig:M-model}(b)).

\begin{table*}
\begin{centering}
\begin{tabular}{|c|c|c|c|c|c|c|c|c|}
\hline 
Wyckoff pos. & \multicolumn{3}{c|}{$1a$ $\left(000\right)$} & \multicolumn{3}{c|}{$2c$ $\left(\frac{1}{3}\frac{2}{3}0\right)$, $\left(\frac{2}{3}\frac{1}{3}0\right)$} & \multicolumn{2}{c|}{$3f$ $(\frac1200)$, $(0\frac120)$, $(\frac12\frac120)$}\tabularnewline
\hline 
Site sym. & \multicolumn{3}{c|}{$6^{\prime}22^{\prime}$, $32$} & \multicolumn{3}{c|}{$32$, $32$} & \multicolumn{2}{c|}{$2^\prime2^\prime2$, $2$}\tabularnewline
\hline 
BR & $G^{1a}_{A_{1}}(1)$ & $G^{1a}_{A_{2}}(1)$ & $G^{1a}_{E}(2)$ & $G^{2c}_{A_1}(2)$ & $G^{2c}_{A_{2}}(2)$ & $G^{2c}_E(4)$ & {$G^{3f}_{A}(3)$} & $G^{3f}_B(3)$ \tabularnewline
\hline 
$\Gamma\left(000\right)$ & $\Gamma_{1}\left(1\right)$ & $\Gamma_{2}\left(1\right)$ & $\Gamma_{3}\left(2\right)$ & $2\Gamma_{1}\left(1\right)$ & $2\Gamma_{2}\left(1\right)$ & $2\Gamma_{3}\left(2\right)$ & {$\Gamma_1(1)+\Gamma_3(2)$} & $\Gamma_2(1)+\Gamma_3(2)$ \tabularnewline
\hline 
$K\left(\frac{1}{3}\frac{1}{3}0\right)$ & $K_{1}\left(1\right)$ & $K_{1}\left(1\right)$ & $K_{2}K_{3}\left(2\right)$ & $K_{2}K_{3}\left(2\right)$ & $K_{2}K_{3}\left(2\right)$ & $2K_{1}\left(1\right)+K_{2}K_{3}\left(2\right)$ & {$K_1(1)+K_2K_3(2)$} & $K_1(1)+K_2K_3(2)$ \tabularnewline
\hline 
$M\left(\frac{1}{2}00\right)$ & $M_{1}\left(1\right)$  & $M_{2}\left(1\right)$  & $M_{1}\left(1\right)+M_{2}\left(1\right)$  & $2M_{1}\left(1\right)$  & $2M_{2}\left(1\right)$  & $2M_{1}\left(1\right)+2M_{2}\left(1\right)$ & {$2M_1(1)+M_2(1)$}  & $M_1(1)+2M_2(1)$ \tabularnewline
\hline
\end{tabular}
\par\end{centering}
\protect\caption{\label{tab:EBR-magSG} Elementary band representations of the magnetic
space group $P6^{\prime}2^{\prime}2$ (\#177.151 in the BNS setting \cite{Bilbao-MSG}). Since twisted bilayer graphene is two-dimensional, here we only list the $z=0$ Wyckoff positions and $k_{z}=0$ high symmetry momenta.
In the first row are the symbols and coordinates of Wyckoff positions. 
In the second row are the corresponding site symmetry groups (magnetic point groups) and their unitary subgroups
(point groups). The symbol $G_\rho^{w}$ in the third row means the BR generated from the co-rep $\rho$ of the site symmetry group at Wyckoff position $w$, and the number in parentheses represents the dimension of this BR.
The co-reps $\rho$ are induced from the irreps of the unitary subgroups\cite{SG-book}, and the latter can be read from the Bilbao server \cite{Bilbao-Rep}. 
The fourth row to sixth row give the co-reps at high symmetry momenta of each BR, and the numbers in parentheses represent the dimensions of these co-reps. 
All co-reps at high symmetry momenta are induced from the irreps at corresponding high symmetry momenta of the unitary sub space group $P312$ (\#149) \cite{SG-book}, and these irreps can be read from the Bilbao server \cite{Bilbao-Rep}.
In table \ref{tab:irreps-MSG} we list definitions of these co-reps in momentum space.  }
\end{table*}

\begin{table*}
\begin{centering}
\begin{tabular}{|c|c|c|c|c|c|c|}
\hline 
Wyckoff pos. & \multicolumn{6}{c|}{$1a$ $\left(000\right)$}\tabularnewline
\hline 
Site sym. & \multicolumn{6}{c|}{$622$}\tabularnewline
\hline 
BR & $G^{1a}_{A_1}(1)$ & $G^{1a}_{A_2}(1)$ & $G^{1a}_{B_1}(1)$ & $G^{1a}_{B_2}(1))$ & $G^{1a}_{E_1}(2)$ & $G^{1a}_{E_2}(2)$\tabularnewline
\hline 
$\Gamma\left(000\right)$ & $\Gamma_{1}\left(1\right)$ & $\Gamma_{2}\left(1\right)$ & $\Gamma_{4}\left(1\right)$ & $\Gamma_{3}\left(1\right)$ & $\Gamma_{6}\left(2\right)$ & $\Gamma_{5}\left(2\right)$\tabularnewline
\hline 
$K\left(\frac{1}{3}\frac{1}{3}0\right)$ & $K_{1}\left(1\right)$ & $K_{2}\left(1\right)$ & $K_{2}\left(1\right)$ & $K_{1}\left(1\right)$ & $K_{3}\left(2\right)$ & $K_{3}\left(2\right)$\tabularnewline
\hline 
$M\left(\frac{1}{2}00\right)$ & $M_{1}\left(1\right)$  & $M_{2}\left(1\right)$  & $M_{4}\left(1\right)$  & $M_{3}\left(1\right)$  & $M_{3}\left(1\right)+M_{4}\left(1\right)$  & $M_{1}\left(1\right)+M_{2}\left(1\right)$ \tabularnewline
\hline 
Wyckoff pos. & \multicolumn{6}{c|}{$2c$ $\left(\frac{1}{3}\frac{2}{3}0\right)$, $\left(\frac{2}{3}\frac{1}{3}0\right)$}\tabularnewline
\hline 
Site sym. & \multicolumn{6}{c|}{$32$}\tabularnewline
\hline 
BR & \multicolumn{2}{c|}{$G^{1a}_{A_1}(2)$} & \multicolumn{2}{c|}{$G^{1a}_{A_2}(2)$}& \multicolumn{2}{c|}{$G^{1a}_{E}(4)$}\tabularnewline
\hline 
$\Gamma\left(000\right)$ & \multicolumn{2}{c|}{$\Gamma_{1}\left(1\right)+\Gamma_{4}\left(1\right)$} & \multicolumn{2}{c|}{$\Gamma_{2}\left(1\right)+\Gamma_{3}\left(1\right)$} & \multicolumn{2}{c|}{$\Gamma_{5}\left(2\right)+\Gamma_{6}\left(2\right)$}\tabularnewline
\hline 
$K\left(\frac{1}{3}\frac{1}{3}0\right)$ & \multicolumn{2}{c|}{$K_{3}\left(2\right)$} & \multicolumn{2}{c|}{$K_{3}\left(2\right)$} & \multicolumn{2}{c|}{$K_{1}\left(1\right)+K_{2}\left(1\right)+K_{3}\left(2\right)$}\tabularnewline
\hline 
$M\left(\frac{1}{2}00\right)$ & \multicolumn{2}{c|}{$M_{1}\left(1\right)+M_{4}\left(1\right)$}  & \multicolumn{2}{c|}{$M_{2}\left(1\right)+M_{3}\left(1\right)$}  & \multicolumn{2}{c|}{$M_{1}\left(1\right)+M_{2}\left(1\right)+M_{3}\left(1\right)+M_{4}\left(1\right)$ }\tabularnewline
\hline
Wyckoff pos. & \multicolumn{6}{c|}{$3f$ $(\frac1200)$, $(0\frac120)$, $(\frac12\frac120)$} \\
\hline
Site sym. & \multicolumn{6}{c|}{$222$} \\
\hline
BR & \multicolumn{2}{c|}{$G^{3f}_{A_1}(3)$} & \multicolumn{2}{c|}{$G^{3f}_{B_1}(3)$} & {$G^{3f}_{B_2}(3)$} & {$G^{3f}_{B_3}(3)$} \\
\hline
$\Gamma(000)$ & \multicolumn{2}{c|}{$\Gamma_1(1)+\Gamma_5(2)$} & \multicolumn{2}{c|}{$\Gamma_2(1)+\Gamma_5(2)$} & $\Gamma_4(1)+\Gamma_6(2)$ & $\Gamma_3(1)+\Gamma_6(2)$\\
\hline
$K(\frac13\frac130)$ & \multicolumn{2}{c|}{$K_1(1)+K_3(2)$} & \multicolumn{2}{c|}{$K_2(1)+K_3(2)$} & $K_2(1)+K_3(2)$ & $K_1(1)+K_3(2)$\\
\hline
$M(\frac1200)$ & \multicolumn{2}{c|}{$M_1(1)+M_3(1)+M_4(1)$} & \multicolumn{2}{c|}{$M_2(1)+M_3(1)+M_4(1)$} & $M_1(1)+M_2(1)+M_4(1)$ & $M_1(1)+M_2(1)+M_3(1)$\\
\hline 
\end{tabular}
\par\end{centering}

\protect\caption{\label{tab:EBR-SG} Elementary band representations of the the space group $P622$ (\#177) with time-reversal symmetry. 
Since the twisted bilayer graphene is two-dimensional, here we only list the $z=0$
Wyckoff positions and $k_{z}=0$ high symmetry momenta.
In the first row we give the symbols and coordinates of Wyckoff positions. In the second row we give the corresponding site symmetry groups. The symbol $G^w_\rho$ in the third row
means the BR generated from the irrep $\rho$ of the site symmetry
group at Wyckoff position $w$, and the number in parentheses represents the dimension of this
BR. The irrep $\rho$ can be read from the Bilbao server \cite{Bilbao-Rep}.
The left rows give the irreps at high symmetry momenta of each BR,
and the numbers in parentheses represent the dimensions of these irreps.
These irreps can be read from the Bilbao server \cite{Bilbao-Rep}.}
\end{table*}

Band representation (BR) theory was recently developed as a powerful tool
to diagnose band topology from symmetry eigenvalues \cite{Bernevig_TQC,Ashvin_indicator_2017}.
In this work, we take advantage of this tool to analyze the band topology
of the Moir{\'e} model. We follow the terminology of Topological Quantum Chemistry
\cite{Bernevig_TQC}. A key concept in this theory is the elementary
BR (EBR), which is defined as minimal set of bands that can be generated
from atomic insulators. Topological Quantum Chemistry has, for the first time, tabulated all the EBR's existent in all the 230 non-magnetic symmetry groups in nature \cite{Bernevig_TQC,Vergniory2017,Bilbao-Rep}. Any BR is a sum of a \textit{positive} number
of EBR's. Given a group of isolated bands, if it is \textit{neither}
a BR \textit{nor} a difference of two BRs it must possess some robust
topology; whereas if it is \textit{not} a BR but it \textit{is}
a difference between two BRs, it should have a fragile topology \cite{Ashvin_fragile, cano2017}. 

We tabulate the EBR's of magnetic space group $P6^{\prime}2^{\prime}2$
in table \ref{tab:EBR-magSG}. Since the twisted bilayer graphene is two dimensional,
only the $z=0$ Wyckoff positions and $k_{z}=0$ high symmetry momenta
are listed. All the co-representations (co-reps) in table \ref{tab:EBR-magSG} are defined in table \ref{tab:irreps-MSG} and can be induced from irreps of the unitary sub space group $P312$ (\#149)
\cite{SG-book}. To be concrete, let us take the $K$ point as an example.
In $P312$ the $K$ point has three one-dimensional irreps $K_{1}$, $K_{2}$, and $K_{3}$, wherein $K_{1}$ is a real irrep whereas $K_2$ and $K_3$ are complex conjugate with each other. Under the $C_{2z}T$ operation, $K_{1}$ transforms to itself and $K_{2}$ and $K_{3}$ transform to each other due to the complex conjugation.
Therefore, in $P6^{\prime}2^{\prime}2$ $K_{1}$ itself form a co-rep, and $K_{2}$ and $K_{3}$ together form a co-rep denoted as $K_{2}K_{3}$.   We emphasize that in the Moir{\'e} pattern/unit cell each Wyckoff position corresponds to a specific area. For example, the $1a$ and $2c$ positions correspond to the AA-stacking area and AB-stacking area of the Moir{\'e} model, respectively (Fig. \ref{fig:M-model}(c)).

\subsection{Two-valley Moir{\'e} band model (MBM-2V)}

In the previous section we showed how electron states around the Dirac cones at $K$ in the lower and upper layer interfere with each other in the Moir{\'e} pattern to form the MBM-1V. The MBM-1V is half of the TBG system, with similar physics taking place in the electron states around $K'$.  In this subsection we build a two-valley Moir{\'e} band model (MBM-2V) by stacking the MBM-1V and its time-reversal counterpart
together, and analyze its symmetries. Such a model can be written as
\begin{align}
H^{(MBM-2V)}_{\mathbf{Q},\mathbf{Q}^{\prime}}\left(\mathbf{k}\right) & =\delta_{\mathbf{Q}\mathbf{Q}^{\prime}}v_{F}\left(\mathbf{k}-\mathbf{Q}\right)\cdot\boldsymbol{\sigma}\otimes\tau_{z}+w\sum_{j=1}^{3}\left(\delta_{\mathbf{Q}^{\prime}-\mathbf{Q},\mathbf{q}_{j}}+\delta_{\mathbf{Q}-\mathbf{Q}^{\prime},\mathbf{q}_{j}}\right)T^{j}\otimes\tau_{0}\label{eq:Ham-Moire-2V}
\end{align}
Here $\tau_{z}$ and $\tau_{0}$ are the Pauli and identity matrix
representing the valley degree of freedom. 
All the symmetries summarized in appendix \ref{sub:Sym} are preserved in this model.
The time-reversal symmetry and the $C_{6z}$ symmetry are recovered.
Therefore the symmetry group is now $P622$ (\#177) plus time-reversal, or the grey magnetic space group $P6221^\prime$ (\#177.150 in the BNS setting). In table \ref{tab:EBR-SG} we list the corresponding EBR's.
Here we list the generators
\begin{enumerate}
\item The $C_{6z}$ symmetry 
\begin{equation}
    H^{(MBM-2V)}\left(\mathbf{k}\right)=D^{\dagger}\left(C_{6z}\right)H^{(MBM-2V)}\left(C_{6z}\mathbf{k}\right)D\left(C_{6z}\right)
\end{equation}
where $D_{\mathbf{Q}^{\prime},\mathbf{Q}}\left(C_{6z}\right)=e^{-i\frac{2\pi}{3}\sigma_{z}}\otimes\tau_x\delta_{\mathbf{Q}^{\prime},C_{6z}\mathbf{Q}}$, and $C_{6z}\mathbf{q}_j=-\mathbf{q}_{j-1}$.
\item The $C_{2x}$ symmetry 
\begin{equation}
    H^{(MBM-2V)}\left(\mathbf{k}\right)=D^{\dagger}\left(C_{2x}\right)H^{(MBM-2V)}\left(C_{2x}\mathbf{k}\right)D\left(C_{2x}\right)
\end{equation}
where $D_{\mathbf{Q}^{\prime},\mathbf{Q}}\left(C_{2x}\right)=\sigma_{x}\otimes\tau_0\delta_{\mathbf{Q}^{\prime},C_{2x}\mathbf{Q}}$, and $C_{2x}\mathbf{q}_{1}=-\mathbf{q}_{1}$, $C_{2x}\mathbf{q}_{2}=-\mathbf{q}_{3}$,
$C_{2x}\mathbf{q}_{3}=-\mathbf{q}_{2}$.
\item The time-reversal symmetry
\begin{equation}
    H^{(MBM-2V)}\left(\mathbf{k}\right)=D\left(T\right)H_{\mathbf{Q},\mathbf{Q}^{\prime}}^{(MBM-2V)*}\left(-\mathbf{k}\right)D^{T}\left(T\right)
\end{equation}
where $D_{\mathbf{Q}^{\prime},\mathbf{Q}}\left(T\right)=\sigma_{x}\otimes\tau_x\delta_{\mathbf{Q}^{\prime},-\mathbf{Q}}$.
\end{enumerate}
It should be noticed that all  rotations of momenta here are with respect to the $\Gamma$ point of the Moir{\'e} BZ. (Fig. \ref{fig:M-model}(b)).

In Eq. (\ref{eq:Ham-Moire-2V}) we have neglected the inter-valley coupling. 
In such an approximation, the valley is a good quantum number: $\left[H^{(MBM-2V)}\left(\mathbf{k}\right),\mathcal{V}\right]=0$, where $\mathcal{V}_{\mathbf{Q}^{\prime},\mathbf{Q}}=\tau_{z}\delta_{\mathbf{Q}^{\prime}\mathbf{Q}}$.
The presence of inter-valley couplings would break this symmetry: these couplings involve $\tau_x$ or $\tau_y$ matrices, both of which anti-commutes with $\mathcal{V}$.

As discussed in section \ref{sec:DFT}, first principle calculations show that such couplings do exist even for small angles - and in fact they become larger as the angle becomes smaller. 

\subsection{Particle-hole symmetry\label{sub:PH-sym}}

Besides the $P6^{\prime}2^{\prime}2$ symmetry, MBM-1V also has an important particle-hole (PH) symmetry, stemming from the PH symmetry of the low-energy Dirac node in the original graphene lattice. 
This symmetry is subtle because it can be broken by either the $\theta$-dependence in $H^{(BB)}$ or by $\mathbf{k}^2$ terms in $H^{(TT)}$ and $H^{(BB)}$ (the latter breaking corresponding to breaking the PH of the Dirac node in the original graphene lattice).
Nevertheless, it is helpful for understanding the band topology. Moreover, we numerically show that our results, understood in light of this symmetry, hold for the general case.
It is direct to show that 
\begin{eqnarray}
H\left(\mathbf{k}\right) & = & -D^{\dagger}\left(P\right)H\left(-\mathbf{k}\right)D\left(P\right) \label{PHsymm1}
\end{eqnarray}
Here $H(\mathbf{k})$ represents MBM-1V, $D_{\mathbf{Q}^{\prime},\mathbf{Q}}\left(P\right)=\delta_{\mathbf{Q}^{\prime},-\mathbf{Q}}\zeta_{\mathbf{Q}}$, and $\zeta_{\mathbf{Q}}=1$ for top layer $\mathbf{Q}$'s (black circles in Fig. \ref{fig:M-model}(b)) and $\zeta_{\mathbf{Q}}=-1$ for the bottom layer $\mathbf{Q}$'s (red circles in Fig. \ref{fig:M-model}(b)), respectively. 
We emphasize that this PH symmetry is unitary, and squares to $-1$.
An important property used later in analizing the band topology is that the PH operator anti-commutes with $C_{2x}$, \ie $\{P,C_{2x}\}=0$. 
This is because sign of $D\left(P\right)$ depends the type of $\mathbf{Q}$ while $C_{2x}$ sends one type to another. 

Since MBM-2V in Eq. (\ref{eq:Ham-Moire-2V}) consists of two uncoupled one-valley models, \ie MBM-1V and its time-reversal counterpart preserves the PH symmetry in Eq. (\ref{PHsymm1}). However, we emphasize  that in general, inter-valley coupling can break the PH symmetry.

\section{Band evolution with the twist angle}

\subsection{MBM Energy level evolution and phases} \label{sub:evolution}

\begin{figure}
\begin{centering}
\includegraphics[width=1\linewidth]{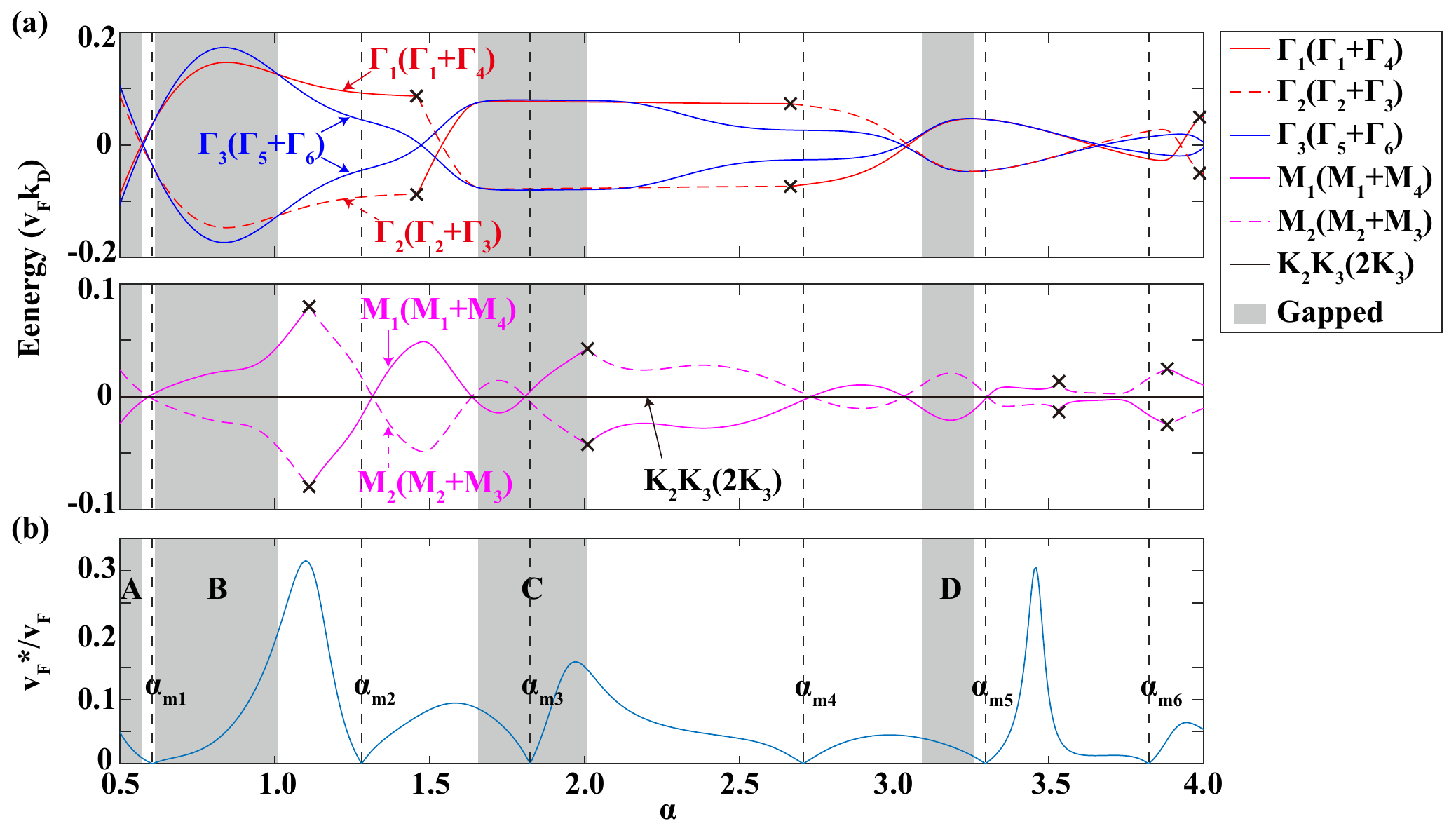}
\par\end{centering}
\protect\caption{\label{fig:evolution}Band evolution with varying twist angle. Here $\alpha=w/\left(2v_{F}\left|\mathbf{K}\right|\sin\frac{\theta}{2}\right)$, and $\theta$ is the twisting angle. (a) The \textit{four} (in the M-1v) energy bands (of different degeneracy) closest to the charge neutrality point at $\Gamma$ (upper plot), and $M$, and $K$ (middle plot) are plotted as functions of $\alpha$ (the $K_2 K_3$ rep at the $K$ point is fixed and flat at zero energy by PH symmetry). These are the levels +2,+1,-1,-2 in appendix \ref{sub:evolution}. The colored lines represent irreps $\Gamma_{1}$, $\Gamma_{2}$, $\Gamma_{3}$, $M_{1}$, $M_{2}$, $K_{2}K_{3}$ for MBM-1V (magnetic space group $P6^{\prime}2^{\prime}2$), and $\Gamma_{1}+\Gamma_{4}$, $\Gamma_{2}+\Gamma_{3}$, $\Gamma_{5}+\Gamma_{6}$, $M_{1}+M_{4}$, $K_{3}$ for MBM-2V (grey magnetic space group $P6221^\prime$). The additional degeneracy in MBM-2V is due to the valley symmetry as we neglect inter-valley coupling. 
The cross symbols represent level exchanges between the plotted levels and upper or lower levels not plotted here. (At these points, the representation can change). The gapped region, where the middle \textit{two} bands in MBM-1V (four bands in MBM-2V) are disconnected from the upper and lower bands in the whole BZ, are marked by grey shadow. (b) The velocity at $K$ is plotted as a function of $\alpha$. The magic $\alpha$'s are defined as the $\alpha$'s giving zero velocity, as denoted by the vertical dashed lines. Some magic angles (such as $\alpha_{m1}$) live in a gapless region between the +1, +2 bands  (and between -1,-2 bands) }
\end{figure}
\begin{figure}
\begin{centering}
\includegraphics[width=0.8\linewidth]{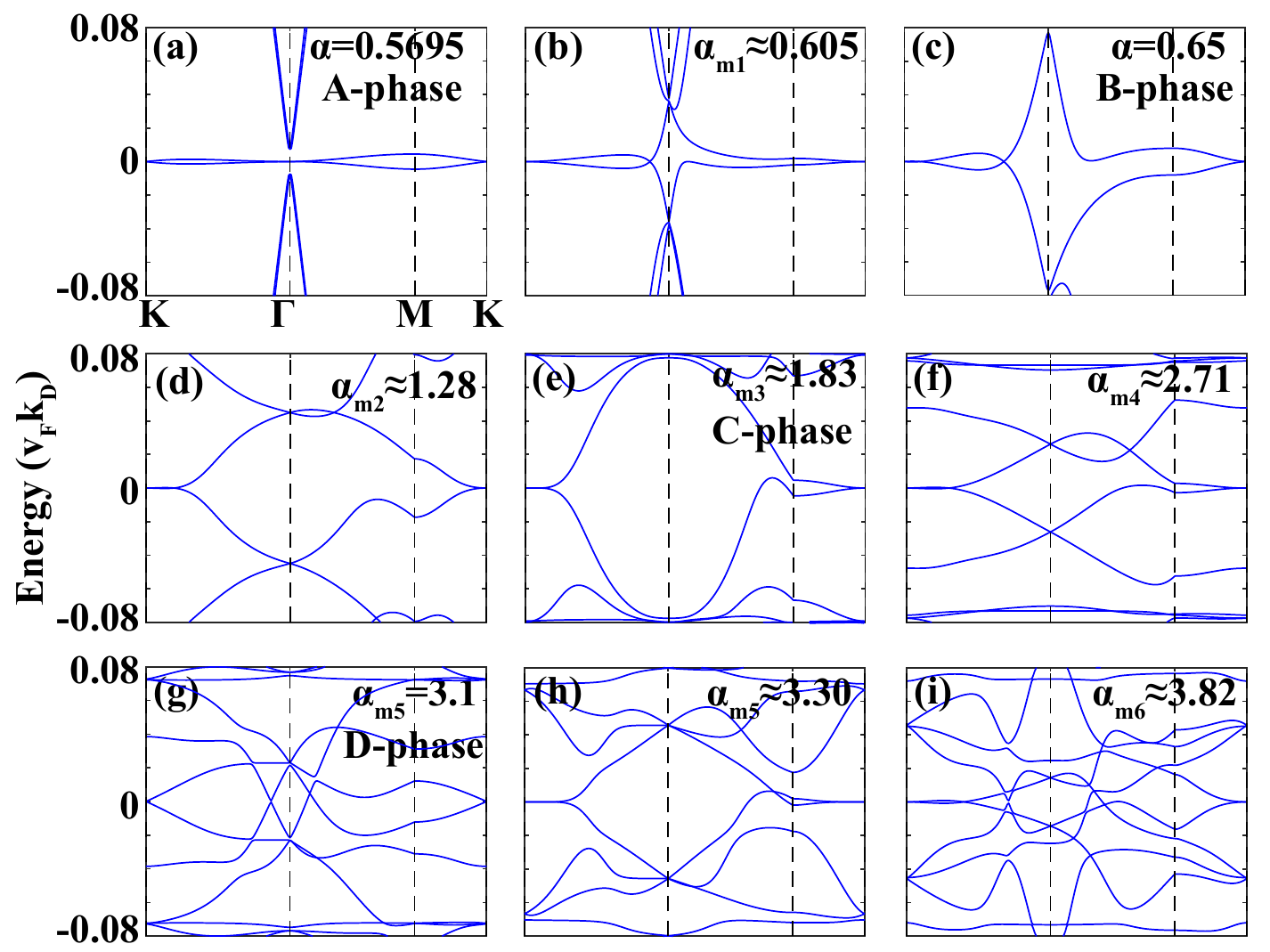}
\par\end{centering}
\protect\caption{\label{fig:Bands}Band structure at the first six magic $\alpha$'s and
    the four gapped phases.}
\end{figure}

In this section we discuss the band evolution with decreasing twisting
angle. Specifically, we are interested in (i) how the irreps of
the bands at high symmetry points close to charge neutrality point evolve, (ii) in which parameter regions
the middle two bands are disconnected from upper and lower bands, (iii)
how the magic angle emerges, and (iv) what happens if the PH symmetry in Eq. (\ref{PHsymm1}) is
broken. 

Since MBM-1V and MBM-2V depend on a single parameter $\alpha=w/\left(2v_{F}\left|\mathbf{K}\right|\sin\frac{\theta}{2}\right)$
(Eq. (\ref{eq:M-model-1})), where $\theta$ is the twisting angle,
in the following we study the band evolution with $\alpha$. For convenience, in the following we refer the first band above zero
energy as +1 band, the second as +2 band, etc., and the first band
below zero energy as -1 band, the second as -2 band, etc. The main
result is summarized in Fig. \ref{fig:evolution} and \ref{fig:Bands}.
In Fig. \ref{fig:evolution}(a) the evolutions of levels closest to the
charge neutrality point are plotted, and the gapped regions where the
+1 band is disconnected from the +2 band in the  \textit{whole} BZ are shadowed. The gapped regions are denoted as the A-, B-, C-, and
D-phases, respectively. 
In these phases the -1 and -2 bands are also disconnected due to the PH symmetry (Eq. (\ref{PHsymm1}), and  thus the middle two bands ($\pm1$ bands) are gapped from all other bands in the MBM-1V. 
In Fig. \ref{fig:evolution}(b) the velocity of +1 band at the $K$ point is plotted as a function
of $\alpha$, and the magic $\alpha$'s where the velocity reduces to zero are
marked by dashed lines. In Fig. \ref{fig:Bands} the band structures
at the first six magic $\alpha$'s and the four gapped phases around these angles are plotted. 

\textbf{Remark}. An interesting observation is that the band structure
in Fig. \ref{fig:Bands}(a), where the $\Gamma_{1}$ and $\Gamma_{2}$
levels move close to zero energy, is extremely flat, even flatter than the
band structures at magic $\alpha$'s. Our prediction is that, upon STM studies, it will be shown that the superconductivity in graphene happens at an angle different than the ``magic'' angle for which the Dirac velocity vanishes.

\subsection{Gap closing and reopening\label{sub:Gap-closing}}

Two neighboring gapped phases - defined as phases where the +2 and +1 bands do not touch at any momentum (by PH symmetry neither do -2 and -1) are obviously separated, by definition, by a gapless phase.
Here we look at the gapless phase between A- and B-phases, \ie the gapless phase near the first magic angle.
As shown in Fig. \ref{fig:Gamma}, there are several level crossings around $\Gamma$ in the process of gap closing and reopening.
For $\alpha <\alpha_{m1}$, the middle two bands form irreps $\Gamma_1+\Gamma_2$ and are gapped from other bands. 
The lowest (highest) irrep above (below) the middle two bands is the two-dimensional $\Gamma_3$ ($\Gamma_3$).
As $\alpha$ increases, the two $\Gamma_3$ irreps move toward zero enerngy.
\begin{figure}
\begin{centering}
\includegraphics[width=1\linewidth]{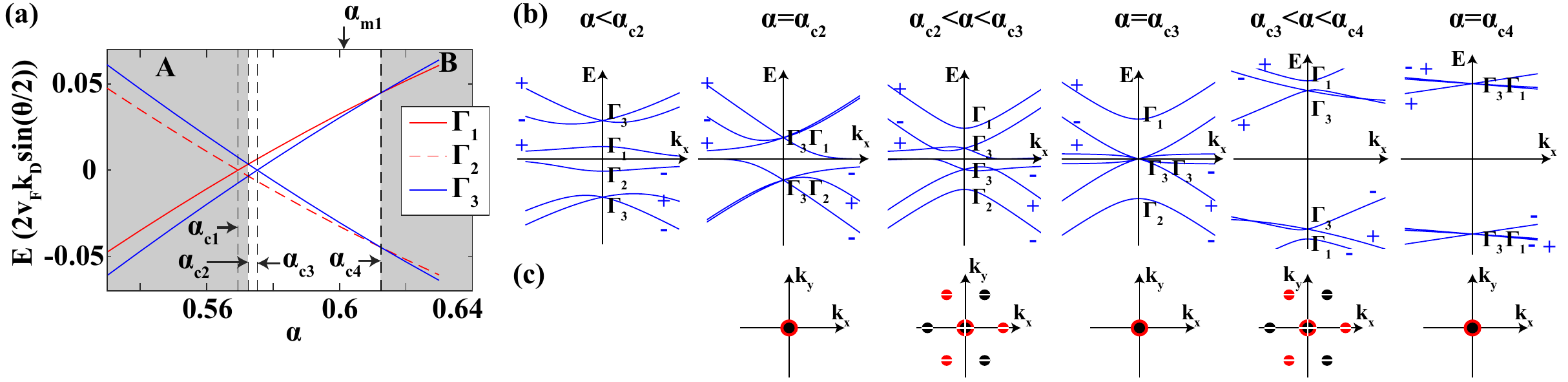}
\par\end{centering}    
\protect\caption{\label{fig:Gamma}Band evolution at $\Gamma$ around the first magic $\alpha$. Here $\alpha=w/\left(2v_{F}\left|\mathbf{K}\right|\sin\frac{\theta}{2}\right)$, and $\theta$ is the twisting angle. (a) Energy levels at $\Gamma$ are plotted as functions of $\alpha$. The colored lines represent irreps $\Gamma_{1}$, $\Gamma_{2}$, and $\Gamma_{3}$ in MBM-1V. (b) Band structures at various $\alpha$'s. The blue ``+'' and ``$-$'' symbols represent the $C_{2x}$ eigenvalues of the corresponding bands. (c) The evolutions of emerged gapless nodes. Gapless nodes at positive and negative energies are represented by black (red) circles (and are related by PH symmetry). When the gapless nodes are two-band, \ie the gapless nodes are Dirac nodes, we use white ``+'' and ``$-$'' symbols to represent the corresponding vorticities. }
\end{figure}  
At some critical $\alpha_{c2}$, the two $\Gamma_3$'s cross with $\Gamma_1$ and $\Gamma_2$, creating two three-fold accidental degenerate levels at a positive and a negative energies (due to PH symmetry), respectively.
As $\alpha$ increases further, the three-fold gapless point at positive energy splits to four Dirac nodes between +1 and +2 bands, wherein one stays at $\Gamma$ and the other three move along $C_2$-invariant lines (Fig. \ref{fig:Gamma}(c)).
The Dirac node at $\Gamma$ is just the $\Gamma_3$ level itself, whereas the other three Dirac nodes are formed by crossings of bands having $C_2$ eigenvalue $+1$ and bands having $C_2$ eigenvalue $-1$ (Fig. \ref{fig:Gamma}(b)).
As shown in Fig. \ref{fig:Gamma}(b), the bands having $C_2$ eigenvalue $+1$ are previously ($\alpha<\alpha_{c2}$) connected to $\Gamma_1$, but are now ($\alpha>\alpha_{c2}$) connected to $\Gamma_3$. 
By the k$\cdot$p model below, we show that the three Dirac nodes at $C_2$-invariant lines have the opposite vorticity with the Dirac node at $\Gamma$.
Taking the operators as $C_{3z}=e^{i\frac{2\pi}{3}\sigma_y}$, $C_{2x}=\sigma_x$, and $C_{2z}T=K$, then the k$\cdotp$p model of bands having a $\Gamma_3$ irrep at the $\Gamma$ point can be easily seen to be 
\begin{equation}
    H(\mathbf{k}) = \epsilon(\mathbf{k})\sigma_0 + (ak_x + b(k_x^2-k_y^2))\sigma_x + (ak_y - 2 bk_xk_y)\sigma_z, \label{eq:GM3}
\end{equation}
where $\epsilon(\mathbf{k})$ is an identity term irrelevant with band crossing.
Solving this model we get one Dirac node with the vorticity $+1$ at $\Gamma$, and three Dirac nodes with the vorticity $-1$ at $(-\frac{a}{b},0)$, $(\frac{a}{2b},\pm\frac{\sqrt{3}a}{2b})$.
By particle-hole symmetry, the three-fold gapless point at negative energy splits to another four Dirac nodes between -1 and -2 bands (Fig. \ref{fig:Gamma}(c)).
At some $\alpha$ between $\alpha_{c2}$ and $\alpha_{c3}$, the six Dirac nodes, three between +1 and +2 bands and three between -1 and -2 bands, reach their maximal distance from $\Gamma$ and then turn back toward $\Gamma$.
At some other $\alpha=\alpha_{c3}$, the six Dirac nodes reach $\Gamma$, and at the same time the two $\Gamma_{3}$ irreps, reach zero energy and cross each other.
In general two identical irreps can not cross each other due to level repulsion, however, as analyzed in next paragraph, the particle-hole symmetry kills any possible coupling between the two $\Gamma_3$ irreps.
As $\alpha$  increases further, the two $\Gamma_{3}$ irreps move away from zero energy, generating again four Dirac nodes between +1 and +2 bands and four Dirac nodes between -1 and -2 bands (Fig. \ref{fig:Gamma}(c)).
The four Dirac nodes between +1 and +2 bands are again described by Eq. (\ref{eq:GM3}): the three moving along $C_2$-invariant lines have the opposite vorticity with the one at $\Gamma$.  Due to PH symmetry, the four Dirac nodes between -1 and -2 bands have similar behavior.
At some $\alpha$ between $\alpha_{c3}$ and another critical $\alpha_{c4}$, the six Dirac nodes moving along $C_2$-invariant lines reach their maximal distance from $\Gamma$ and then turn back toward $\Gamma$, and finally reach $\Gamma$ at $\alpha_{c4}$  (Fig. \ref{fig:Gamma}(c)). This is a movement reminiscent of that of the region  $\alpha$ between $\alpha_{c2}$ and $\alpha_{c3}$. Right at $\alpha_{c4}$, $\Gamma_{3}$ levels cross with the $\Gamma_{1}$
and $\Gamma_{2}$ levels.
For $\alpha>\alpha_{c4}$ the middle two bands are no longer connected to $\Gamma_3$'s, and thus there are no longer gapless points between the +1(-1) and +2(-2) bands, and we have entered the gapped region B in Fig. \ref{fig:evolution}.

Now we prove that the crossing between the two $\Gamma_3$ irreps at $\alpha_{c3}$ is allowed by the particle-hole symmetry.
For the four bands forming the two $\Gamma_3$ irreps (of different energy) at $\Gamma$, we take the magnetic space group operators as $C_{3z}=\tau_0e^{i\frac{2\pi}{3}\sigma_z}$, $C_{2x}=\tau_0\sigma_x$, and $C_{2z}T=K$.
Then the partical-hole operator, $P$, which satisfies $P^2=-1$, $[P,C_{3z}]=0$, $[P,C_{2z}T]=0$, $\{P,C_{2x}\}$ (appendix \ref{sub:PH-sym}), can be chosen as either $P=i\tau_z\sigma_y$ or $P=i\tau_x\sigma_y$.
Since the two choices can be transformed to each other by a unitary transformation, in the following we proceed with $P=i\tau_x\sigma_y$.
A minimal Hamiltonian at $\Gamma$ is given by $H(\Gamma)=\Delta\tau_z\sigma_0$. We show that this is the only term allowed, and hence level crossing can happen by varying one parameter $\Delta$.
Whether the level crossing between two $\Gamma_3$'s can happen depends on whether we can add an additional symmetry allowed term, $m\gamma$, such that $\{\gamma,P\}=0$ and $\{\gamma,\tau_z\sigma_0\}=0$.
If no such term exists, at $\Delta=0$ the two $\Gamma_3$ irreps cross with each other.
The only $C_{3z}$- and $C_{2x}$-allowed terms that anti-commuting with $\tau_z\sigma_0$ are $\tau_x\sigma_0$ and $\tau_y\sigma_0$, whereas the first breaks particle-hole symmetry and the second breaks time-reversal symmetry.
Therefore, the crossing between two $\Gamma_3$'s are protected.

\subsection{Irreps of gapped phases\label{sub:Irreps}}

From our plots of the phase diagram in Fig. \ref{fig:evolution} we see that in all the gapped
phases the middle two bands form the irreps $\Gamma_{1}+\Gamma_{2}$,
$M_{1}+M_{2}$, and $K_{2}K_{3}$. In this subsection we prove that
this must be true for arbitrary $\alpha$  if both the magnetic
space group ($P6^{\prime}2^{\prime}2$)  \textit{and} the PH are symmetries of the problem.

We first prove that the $K_{2}K_{3}$ irrep is pinned at zero energy.
The $K$ point has the symmetries $C_{3z}$, $C_{2z}T$, and $PC_{2x}$, wherein the last is $C_{2x}$ followed by PH, and these symmetries satisfy $[C_{3z},C_{2z}T]=0$, $(PC_{2x})C_{3z}(PC_{2x})^{-1}=C_{3z}^{-1}$, $[C_{2z}T, PC_{2x}]=0$ (see appendix \ref{sub:Sym} for definitions of these symmetries).  
From the character table \ref{tab:irreps-MSG} we see that we can choose the basis  of $K_{2}K_{3}$ as the $C_{3z}$
eigenstates: $C_{3z}\left|1\right\rangle =e^{i\frac{2\pi}{3}}\left|1\right\rangle $,
$C_{3z}\left|\bar{1}\right\rangle =e^{-i\frac{2\pi}{3}}\left|\bar{1}\right\rangle $. We now look at the effect of the other symmetries. On one hand, $PC_{2x}$ flips the $C_{3z}$ eigenvalue 
\begin{equation}
C_{3z}PC_{2x}\left|1\right\rangle =PC_{2x}C_{3z}^{-1}\left|1\right\rangle =e^{-i\frac{2\pi}{3}}PC_{2x}\left|1\right\rangle ,
\end{equation} \ie $PC_{2x}\left|1\right\rangle $ has $C_{3z}$ eigenvalue $e^{-i\frac{2\pi}{3}}$.
On the other hand, $PC_{2x}$ anti-commutes with Hamiltonian so $PC_{2x}\left|1\right\rangle $
has the opposite energy eigenvalue as $\left|1\right\rangle $. If
$H\left(K\right)\left|1\right\rangle =\epsilon\left|1\right\rangle $
then
\begin{equation}
H\left(K\right)PC_{2x}\left|1\right\rangle =-PC_{2x}H\left(K\right)\left|1\right\rangle =-\epsilon PC_{2x}\left|1\right\rangle 
\end{equation}
Thus there are two possibilities (i) $\epsilon=0$ and $\left|\bar{1}\right\rangle =PC_{2x}\left|1\right\rangle $,
and (ii) $\epsilon\neq0$, and $\left|\bar{1}\right\rangle \neq PC_{2x}\left|1\right\rangle $.
In the second case $PC_{2x}\left|1\right\rangle $ and its $C_{2z}T$-partner form another $K_{2}K_{3}$ irrep at the energy $-\epsilon$. Therefore, the $K_{2}K_{3}$ irrep either appears in pair with opposite
energies or appears alone at zero energy. 
For MBM-1V we now prove that the zero energy $K_{2}K_{3}$ irrep must exist. 
In the $\alpha=0$ limit (Eq. (\ref{eq:M-model-1})) the zero energy Dirac node forms a $K_{2}K_{3}$ irrep, and increasing $\alpha$, which preserves the PH symmetry, can not move a single irrep away from zero energy in a PH-symmetric way. Hence a single $K_{2}K_{3}$ irrep can exist only at zero energy. 

In the second step, we prove that the middle two bands, when gapped, must form $\Gamma_{1}+\Gamma_{2}$ irreps at $\Gamma$. 
\textit{First}, as $\Gamma_{3}$ is a two-dimensional irrep, the middle two bands can, in principle, form
one $\Gamma_{3}$, if they are degenerate. However we show that can not happen in MBM-1V with PH symmetry. 
$\Gamma$ has symmetries $C_{3z}$, $C_{2x}$, $C_{2z}T$, and $P$, and these symmetries satisfy relations $C_{2x}C_{3z}C_{2x}^{-1}=C_{3z}^{-1}$, $[C_{3z},C_{2z}T]=0$, $[C_{3z},P]=0$, $[C_{2x},C_{2z}T]=0$, $\{C_{2x},P\}=0$, and $[C_{2z}T,P]=0$.
Since $\Gamma_3$ has same $C_{3z}$ eigenvalues with $K_2K_3$, the analysis above for $K_2K_3$ also applies here such that, similar with $K_2K_3$, $\Gamma_3$ also appears either in pair with opposite energies or alone at zero energy.

In MBM-1V, $\Gamma_{3}$ can only appear in pairs. This is because the $\alpha=0$ Hamiltonian
(Eq. (\ref{eq:M-model-1})) does not have zero mode at $\Gamma$ and
any PH symmetry-preserving process cannot create a single zero mode.
Thus the middle two bands, particle-hole conjugate to each other, can only form $\Gamma_{1}$ and/or $\Gamma_{2}$
irreps. \textit{Second}, due to PH symmetry, we  now show that the two bands always have opposite energy eigenvalues. Suppose $\left|\Gamma_{1}\right\rangle $
has $C_{2x}$ eigenvalue $+1$ and energy eigenvalue $\epsilon$,
then according to the anti-commutation relation $\left\{ C_{2x},P\right\} =0$
and $\left\{ H\left(\Gamma\right),P\right\} =0$, $P\left|\Gamma_{1}\right\rangle $
has $C_{2x}$ eigenvalue $-1$ and energy eigenvalue $-\epsilon$.
Thus $P\left|\Gamma_{1}\right\rangle $ forms a $\Gamma_{2}$ irrep
at energy $-\epsilon$. Therefore, we conclude that the gapped middle
two bands always form irreps $\Gamma_{1}+\Gamma_{2}$ at $\Gamma$.

In the third step, we prove that the gapped middle two bands must form $M_{1}+M_{2}$ irreps at $M$. 
The $M$ point has symmetries $C_{2(100)}=C_{3z}C_{2x}C_{3z}^{-1}$, $C_{2z}T$, and $P$, and these symmetries satisfy $[C_{2(100)},C_{2z}T]=0$, $\{C_{2(100)},P\}=0$, and $[C_{2z}T,P]=0$. 
The only two irreps at $M$, \ie $M_{1}$ and $M_{2}$, have $C_{2x}$ eigenvalues $+1$ and $-1$, respectively. Thus,  like $\Gamma_1$ and $\Gamma_2$, due to the anti-commutation relation $\{C_{2(100)},P\}=0$ $M_1$ and $M_2$ also always appear in pairs.

\subsection{Generation of magic angle}\label{sub:magic_angle_generation}

The magic angles are defined as the angles at which the velocity of the
+1 (or, identically -1) band at $K$ vanishes. In Ref \cite{cao_TBG1}, the generation
of magic angle is briefly discussed: three more Dirac nodes, which
have opposite vorticities with the Dirac node at $K$, move to $K$
point to cause a quadratic touching such that the velocity vanishes.
Here we give a complete and detailed picture of the generation
of the first magic angle ($\alpha$), as shown in Fig. \ref{fig:magic}. We checked that the generation of the other magic angles follows a similar pattern.

In Fig. \ref{fig:magic}(a), where $\alpha=0.53$, we have two Dirac nodes at the points $K$ and $K^\prime$ of the Moir{\'e} BZ with the same vorticity $+1$.
With an increasing $\alpha$, an accidental crossing between the +1 and -1 bands arises along the $K\Gamma$ line, which then creates two Dirac nodes with opposite vorticities (Fig. \ref{fig:magic}(h)) along that line. 
Due to to the $C_{3z}$ and $C_{2x}$ symmetries, there are in total 12 Dirac nodes created, six having $+1$ vorticity and six having $-1$ vorticity, as marked as black and red in Fig. \ref{fig:magic}, respectively.
When $\alpha$ is further increased, all these 12 Dirac nodes move toward $\Gamma$, and at nearly the same time all of them reach $\Gamma$ (Fig. \ref{fig:magic}(c) and (i)).
Then the Dirac nodes having vorticity $-1$ pass through $\Gamma$ and move on to the $\Gamma M$ line, leading to the interchange of $\Gamma_{1}$ and $\Gamma_{2}$ levels, whereas the Dirac nodes having vorticity  $+1$ move back along $\Gamma K$.
Before Fig. \ref{fig:magic}(d), the $\Gamma_1$ and $\Gamma_2$ levels cross with two $\Gamma_3$ levels, as reflected in Fig. \ref{fig:evolution} and \ref{fig:Gamma}, such that in Fig. \ref{fig:magic}(d) the +1 and -1 bands are connected to $\Gamma_3$ levels.
At the mean time, as shown in Fig. \ref{fig:magic}(j), the Dirac nodes having vorticity  $-1$ move on towards $M$, and the Dirac nodes having vorticity  $+1$ move slowly toward $K$.
During the remaining process, the Dirac nodes between the $+1$ and $-1$ bands, having vorticity  $+1$ do not move significantly.
At Fig. \ref{fig:magic} (e) and (k) the six Dirac nodes having vorticity  $-1$ reach $M$, causing an interchange of $M_{1}$ and $M_{2}$ levels, as reflected in Fig. \ref{fig:evolution}.  
Then these six Dirac nodes move towards $K$ along $MK$, and finally reach $K$ causing a quadratic touching and a velocity inversion, as shown in Fig. \ref{fig:magic} (f) and (l). 
Comparing the symmetry eigenvalues at $\Gamma$ and $M$ in the $+1$ (or $-1$) band before and after the magic angle, we see that they are completely inverted.

\begin{figure}
\begin{centering}
\includegraphics[width=1\linewidth]{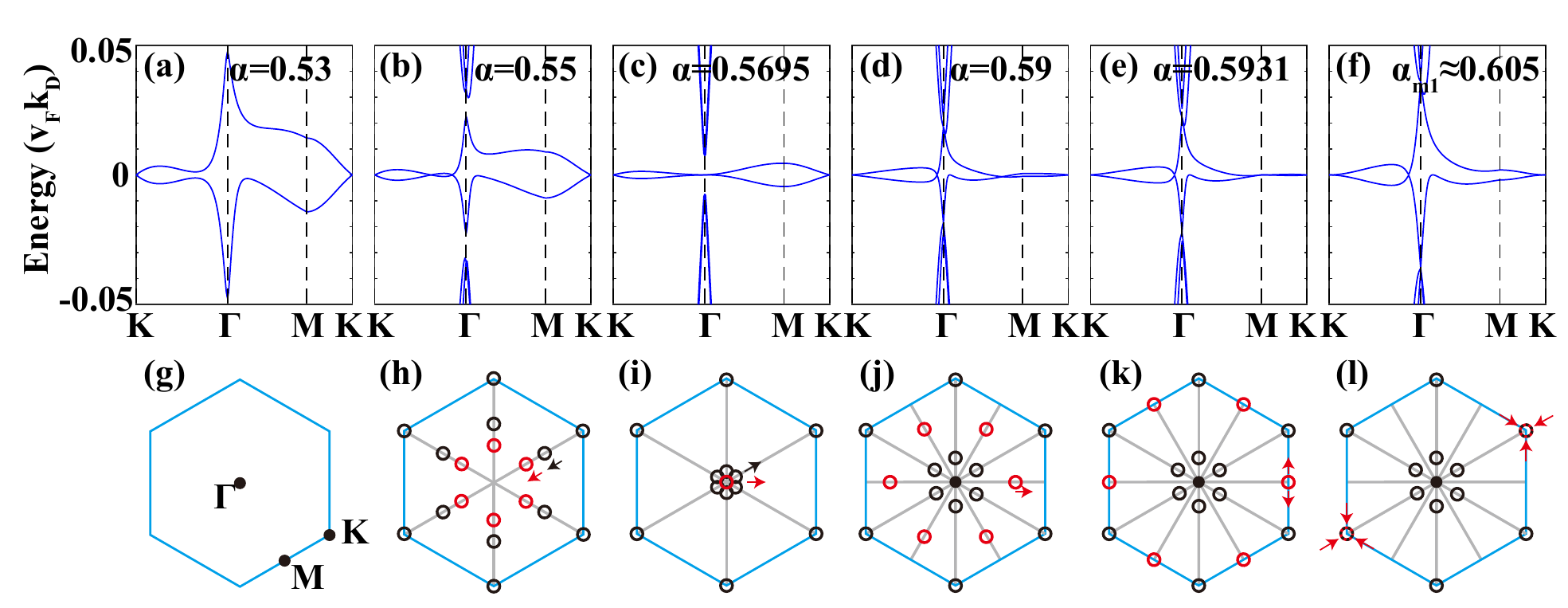}
\par\end{centering}

\protect\caption{\label{fig:magic} Generation of the first magic $\alpha$. Here $\alpha=w/\left(2v_{F}\left|\mathbf{K}\right|\sin\frac{\theta}{2}\right)$,
and $\theta$ is the twisting angle. In (a)-(f) we plot the band structures
at several $\alpha$, wherein (f) corresponds to the first magic $\alpha$.
In (g)-(l) we show the distribution of Dirac nodes between +1 and
-1 bands. The black and red circles represent Dirac nodes with different
vorticities, respectively.}
\end{figure}

\begin{figure}
\begin{centering}
\includegraphics[width=1\linewidth]{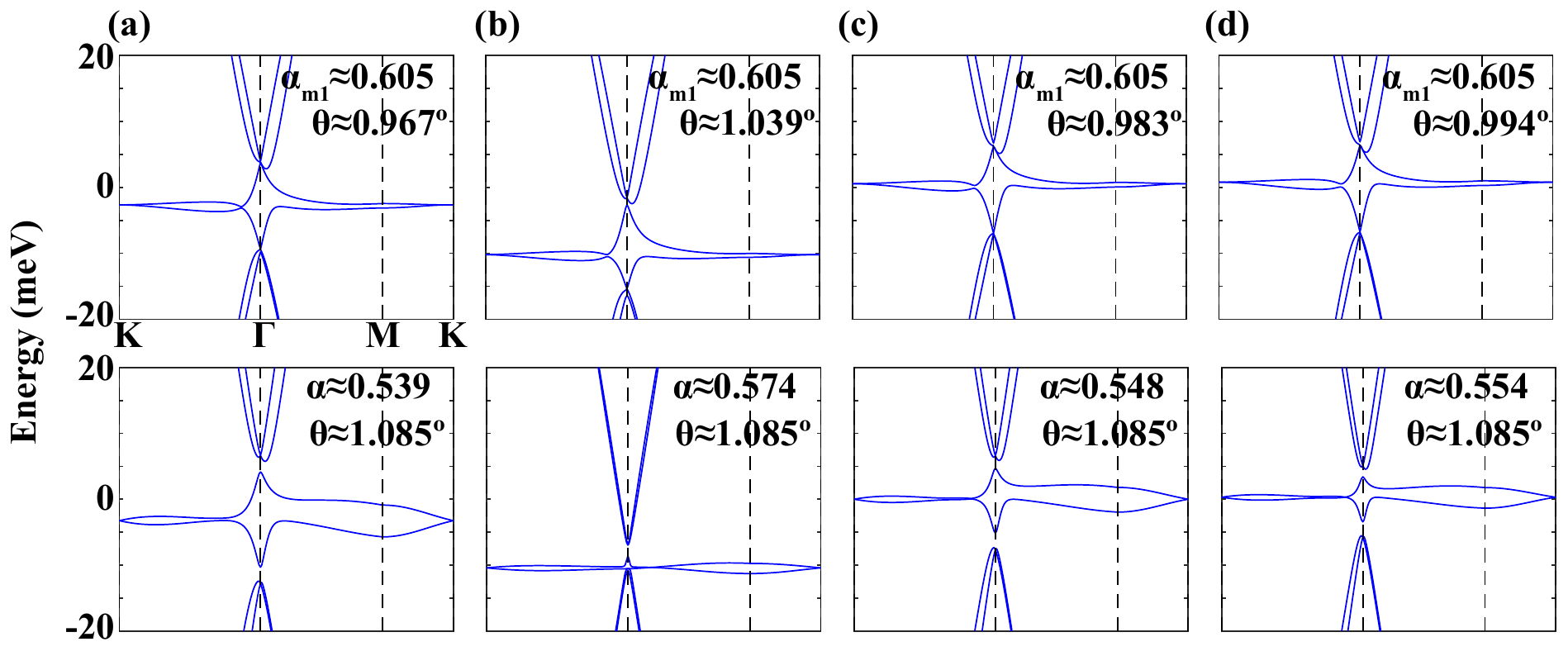}
\par\end{centering}

\protect\caption{\label{fig:PHbreaking} Particle-hole symmetry breaking. Here $\alpha=w/\left(2v_{F}\left|\mathbf{K}\right|\sin\frac{\theta}{2}\right)$, which equals to $w/\left(\frac{4\pi}{\sqrt{3}}t\sin\frac{\theta}{2}\right)$ for the model in appendix \ref{sub:PHbreaking}. Here $t$ is the nearest tight-binding hopping parameter, and $\theta$
is the twisting angle. In the top panel band structures at fixed $\alpha$,
which is chosen to be $\alpha_{m1}$ such that the $K$ point is a
quadratic touching, is plotted. In the bottom panel band structures
at fixed $\theta$, which is close to the experimental angle where
superconductor is observed \cite{cao_TBG2}, is plotted. In (a)-(d)
we use four sets of parameters: (a) $t=-2.97\mathrm{eV}$, $t^{\prime}=-0.073\mathrm{eV}$
\cite{TB_graphene}, (b) $t=-2.79\mathrm{eV}$, $t^{\prime}=-0.68\mathrm{eV}$
\cite{TB_graphene}, (c) $t=-2.92\mathrm{eV}$, $t^{\prime}=0.223\mathrm{eV}$
\cite{WTB_graphene}, (d) $t=-2.89\mathrm{eV}$, $t^{\prime}=0.242\mathrm{eV}$
\cite{TB_TBG}.}
\end{figure}

\subsection{Particle hole symmetry breaking \label{sub:PHbreaking}}

The PH symmetry is an approximate symmetry in MBM. It can be broken
by either introducing the $\theta$-dependence of $H^{\left(BB\right)}$ (Eq.
(\ref{eq:H(BB)})) or the $\mathbf{k}^{2}$ terms in $H^{\left(TT\right)}$
and $H^{\left(BB\right)}$ - the latter corresponding to breaking PH symmetry in the Dirac Hamiltonian of each single layer of graphene. In this subsection, we discuss the effect
of these PH symmetry breaking terms. 

First, we derive the $H^{\left(TT\right)}$ Hamiltonian with
$\mathbf{k}^{2}$ terms. We start with the tight-binding Hamiltonian of Graphene
\begin{equation}
H^{(Graphene)}\left(\mathbf{k}\right)=\begin{bmatrix}2t^{\prime}\sum_{i}\cos\left(\mathbf{d}_{i}\cdot\mathbf{k}\right) & t\sum_{i}e^{i\boldsymbol{\delta}_{i}\cdot\mathbf{k}}\\
t\sum_{i}e^{-i\boldsymbol{\delta}_{i}\cdot\mathbf{k}} & 2t^{\prime}\sum_{i}\cos\left(\mathbf{d}_{i}\cdot\mathbf{k}\right)
\end{bmatrix}
\end{equation}
where $\boldsymbol{\delta}_{i=1,2,3}$ are the three nearest-neighbor
vectors, $\mathbf{d}_{i=1\cdots6}$ are the six second-nearest-neighbor
vectors, $t$ is the nearest hopping, and $t^{\prime}$ is the second-nearest
hopping \cite{Graphene_RMP}. Expanding this Hamiltonian around the
$K$ point to second order $\mathbf{k}$, we get the $H^{\left(TT\right)}$
Hamiltonian as
\begin{equation}
H^{\left(TT\right)}\approx\frac{3t}{2}a\mathbf{k}\cdot\boldsymbol{\sigma}+\frac{3t}{8}a^{2}\left(\left(k_{x}^{2}-k_{y}^{2}\right)\sigma_{x}-2k_{x}k_{y}\sigma_{y}\right)+\frac{9t^{\prime}}{4}a^{2}\mathbf{k}^{2}\sigma_{0}\label{eq:HTT-noPH}
\end{equation}
Then the $\theta$-dependent $H^{\left(BB\right)}$ can be easily obtained from Eq. (\ref{eq:HBB-HTT}) as
\begin{equation}
H^{\left(BB\right)}\approx\frac{3t}{2}a\tilde{\mathbf{k}}\cdot\boldsymbol{\sigma}+\frac{3t}{8}a^{2}\left(\left(\tilde{k}_{x}^{2}-\tilde{k}_{y}^{2}\right)\sigma_{x}-2\tilde{k}_{x}\tilde{k}_{y}\sigma_{y}\right)+\frac{9t^{\prime}}{4}a^{2}\tilde{\mathbf{k}}^{2}\sigma_{0}\label{eq:HBB-noPH}
\end{equation}
where $\tilde{\mathbf{k}}=M_{\theta}^{-1}\mathbf{k}$. Replacing the
diagonal ($\mathbf{Q}=\mathbf{Q}^{\prime}$) Hamiltonian in Eq. (\ref{eq:M-model})
with Eq. (\ref{eq:HTT-noPH})-(\ref{eq:HBB-noPH}), we get a Moir{\'e} model breaking PH symmetry. 
We calculate the band structures of this Hamiltonian at fixed $\alpha$'s or fixed $\theta$'s with four different sets of $t$ and $t^\prime$: set-(i) parameters are fitted to the {\it ab-initio} bands along the high symmetry lines \cite{TB_graphene}, set-(ii) parameters are fitted to the {\it ab-initio} bands at high symmetry momenta \cite{TB_graphene}, set-(iii) parameters are obtained by ab-initio Wannier functions \cite{WTB_graphene}, set-(iv) parameters are also obtained by {\it ab-initio} Wannier functions \cite{TB_TBG}.

Fig. \ref{fig:PHbreaking}(a)-(d), we plot the band structure at the first magic angle (fixed $\alpha=0.605$ or fixted $\theta=1.085^\circ$) using parameter sets (i)-(iv), respectively.

We have also checked the band structures of the four gapped phases in Fig. \ref{fig:evolution}.
In doing this, for each gapped phase we choose a representative angle. 
Specifically, $\alpha=0.5695$ for A-phase, $\alpha=0.65$ for B-phase, $\alpha=1.83$ for C-phase, and $\alpha=3.1$ for D-phase. 
It turns out that in three of the four angles, \ie $\alpha=0.5695, 0.65, 1.83$, the middle two bands (called +1 and -1 bands by us) are disconnected from upper and lower bands for all the four sets of parameters, and all of them have the same irreps with A-phase in Fig. \ref{fig:evolution}.
On the other hand, at $\alpha=3.1$, for set-(i), -(iii), -(iv) parameters the middle two bands have the same irreps with A-phase, whereas for set-(ii) parameters the middle two bands are no longer seperated from other bands: the $\Gamma_2$ level in the original middle two bands interchanges with a $\Gamma_3$ level, and the $K_2K_3$ level in the original middle two bands interchanges with a $K_1$ level.

\section{Fragile topology of Moir{\'e} bands} \label{sec:fragile}

\subsection{Fragile topology protected by $C_{2z}T$\label{sub:fragile-C2zT}}

The fragile topology is an obstruction for constructing symmetric
Wannier functions for a given set of bands, with the particular characteristic that this obstruction can be \textit{removed} by adding other topologically \textit{trivial}
bands into the band set \cite{Ashvin_fragile, cano2017}. In more detail, suppose
we have two bands that have no  Wannier representation respecting the symmetries of the lattice plus time-reversal. Then they must have some kind of topology. This topology is called fragile if,
once we add to the problem a set of completely trivial bands,  and once the topological bands are coupled to these trivial bands, all of the bands together do have a symmetric Wannier representation. In this
appendix we show that the anti-unitary symmetry $C_{2z}T$, which
squares to $1$, can protect fragile topology with a $\mathbb{Z}$-classification 
in a two-band system. 

Let us denote the $n$-th eigenstate of $H(\mathbf{k})$ as $|u_{n,\mathbf{k}}\rangle$, represented by a one-column vector (in an orbital basis), and denote the collection of all the occupied states as a matrix $U_\mathbf{k} = [|u_{1\mathbf{k}}\rangle, \cdots, |u_{n_{occ}\mathbf{k}}\rangle]$, where $n_{occ}$ is the number of occupied bands. 
Taking the parametrization $\mathbf{k}=k_1 \mathbf{b}_1 + k_2\mathbf{b}_2$, where $\mathbf{b}_{1}$ and $\mathbf{b}_{2}$ are reciprocal lattices shown in Fig. \ref{fig:M-model}(b), we define the Wilson loop in integrated along $k_{2}$ as \cite{alexandradinata2016}
\begin{equation}
W\left(k_{1}\right)=U_{k_{1},0}^{\dagger}U_{k_{1},\frac{2\pi}{N}}U_{k_{1},\frac{2\pi}{N}}^{\dagger}U_{k_{1},\frac{2\times2\pi}{N}}\cdots U_{k_{1},\frac{\left(N-1\right)\times2\pi}{N}}^{\dagger}U_{k_{1},2\pi}
\end{equation}
Here $N$ is a large enough integer to describe the infinite lattice limit $N\rightarrow \infty$.
In general $H(\mathbf{k})$ can be non-periodic, \ie $H(\mathbf{k+G})=V^\mathbf{G}H(\mathbf{k})V^{\mathbf{G}\dagger}$, thus $U_{\mathbf{k+G}}=V^{\mathbf{G}}U_{\mathbf{k}}$. 
Therefore, the Wilson loop can be rewritten as
\begin{equation}
    W\left(k_{1}\right)=U_{k_{1},0}^{\dagger}U_{k_{1},\frac{2\pi}{N}}U_{k_{1},\frac{2\pi}{N}}^{\dagger}U_{k_{1},\frac{2\times2\pi}{N}}\cdots U_{k_{1},\frac{\left(N-1\right)\times2\pi}{N}}^{\dagger} V^{(0,2\pi)} U_{k_{1},0} \label{eq:Wilson}
\end{equation}

We now want to understand the constraints that $C_{2z}T$ symmetry imposes on the Wilson loops. 
For a gapped system, we can choose a gauge where the representation matrix of $C_{2z}T$  in occupied bands is identity \cite{Bernevig-C2zT}, \ie $C_{2z}T U_{\mathbf{k}} \equiv D(C_{2z}T) U_\mathbf{k} ^* = U_{\mathrm{k}} \mathbb{I}_2$, such that the Wilson loop matrix is real
\begin{equation}
W\left(k_{1}\right)=W^{*}\left(k_{1}\right)\label{eq:WL-C2zT}.
\end{equation}
(See Eq. (\ref{eq:D-C2T}) for definition of $D(C_{2z}T)$.)
To see the constraint on Wilson loop eigenvalues, we define the ``Wilson
loop Hamiltonian'' as 
\begin{equation}
\mathcal{H}\left(k_{1}\right) = -i\ln W\left(k_{1}\right), \label{eq:WL-H}
\end{equation}
$\mathcal{H}\left(k_{1}\right)$ is an Hermitian matrix whose eigenvalues
give the Wilson loop spectrum. Then Eq. (\ref{eq:WL-C2zT}) yields a ``particle-hole symmetry'' of $\mathcal{H}$
\begin{equation}
\mathcal{C}\mathcal{H}\left(k_{1}\right)\mathcal{C}^{-1}=-\mathcal{H}\left(k_{1}\right)\mod 2\pi, \label{eq:HWL-C2zT}
\end{equation}
where $\mathcal{C}=K$ is just complex conjugation. 
For two-occupied bands, the Wilson loop is a $2\times2$ matrix,  and we can always expand $\mathcal{H}\left(k_{1}\right)$
in Pauli matrices as
\[
\mathcal{H}\left(k_{1}\right)=d_{0}\left(k_{1}\right)\sigma_{0}+d_{x}\left(k_{1}\right)\sigma_{x}+d_{y}\left(k_{1}\right)\sigma_{y}+d_{z}\left(k_{1}\right)\sigma_{z}
\]
Eq. (\ref{eq:HWL-C2zT}) enforces $d_{x}\left(k_{1}\right)=d_{z}\left(k_{1}\right)=0$,
and $d_{0}\left(k_{1}\right)=0$ or $\pi$. For a gapped system $d_{0}\left(k_{1}\right)$
must be a constant, and so the only k-dependent component is $d_{y}\left(k_{1}\right)$.
The winding of $d_{y}\left(k_{1}\right)$ can be thought as a topological
invariant.  The above proof depends on the Pauli matrix representation of the two band Wilson loop Hamiltonian. Thus it cannot be applied for multi-band system, and in these systems we lose the $\mathbb{Z}$-classification of the phase. This is why this topology is fragile. However, the topology maintains a subtle stable part that we will present in next subsection. We hence need to re-asses the role played by fragile topology.

\subsection{Proof of stable topology of 2V-1B via homotopy group}
\label{sec:homotopy}
In the main text, we have shown that in the MBM-1V, particle-hole symmetry assumed, for any twist angle where the middle two bands are separated from the rest by direct energy gap, the middle 2B-1V has fragile topology, featuring nonzero winding in the Wilson loop spectrum.
Here we argue that when this winding number is odd, one can further prove that the fragile topology becomes stable, robust against superposition of trivial sets of bands; this points to a possible re-assessment of fragile topology.
We need to find a topological invariant that is (i) well defined when there are more than two occupied bands and (ii) takes nonzero value when the system is a superposition of a trivial sets of bands and one set of two bands having odd winding.

For an arbitrary system with $C_2T$ composite symmetry, we can always find a gauge such that $C_2T=K$ where $K$ is complex conjugation.
In this gauge, the Hamiltonian is real at each $\mathbf{k}$, but at the cost that $H(\mathbf{k})$ is not necessarily periodic at the BZ boundary.
In fact, we generally have
\begin{eqnarray}
H(-\pi,k_y)&=&O^T_x(k_y)H(\pi,k_y)O_x(k_y),\\
\nonumber
H(k_x,-\pi)&=&O^T_y(k_x)H(k_x,\pi)O_y(k_x).
\end{eqnarray}.

In order to have a unitary Wilson loop along $k_y$, we have to insert $O_y(k_x)$ matrix at the end, that is,
\begin{align}
W_{mn}(k_x)\equiv & \lim_{N\rightarrow\infty} \sum_{i_1 \cdots i_N}  \langle{u}_m(k_x,-\pi)|u_{i_1}(k_x,-\pi+2\pi/N)\rangle\langle{u}_{i_1}(k_x,-\pi+2\pi/N)|\cdots\\
\nonumber & |u_{i_{N-1}}(k_x,-\pi+2\pi(N-1)/N)\rangle\langle{u}_{i_{N-1}}(k_x,-\pi+2\pi(N-1)/N)|u_{i_N}(k_x,\pi)\rangle\langle{}u_{i_N}(k_x,\pi)|O^T_y(k_x)|u_n(k_x,-\pi)\rangle.
\end{align}
We next \textit{assume} that
\begin{equation}\label{eq:3}
\langle{u}_m(-\pi,k_y)|O^T_x(k_y)O_x(k_y+\delta{k})|u_n(-\pi,k_y+\delta{k})\rangle\rightarrow\langle{u}_m(-\pi,k_y)|u_n(-\pi,k_y+\delta{k})\rangle
\end{equation}
as $\delta{k}\rightarrow0$. This assumption is easy to prove for the embedding matrices $V^G$ which are $k$-independent. In general, it is not clear to us how to prove it. 
Using this assumption, it is easy to show that 
\begin{equation}
W(-\pi)=W(\pi).
\end{equation}

 $W(k_x)$ is not only unitary but also real, as all its terms are real.
Therefore $W(k_x)\in{O}(n)$, where $n\ge2$.
We define the $\mathbb{Z}_2$-invariant $\delta$ as a functional of $W(k_x)$ such that $\delta[W]=0$ when $W(k_x)$ can be continuously deformed to identity matrix, and $\delta[W]=1$ if such continuous path does not exist.
This definition is consistent, because mathematically we know that $\pi_1[O(n)]=\mathbb{Z}_2$.

To find the explicit value of $\delta$ for an arbitrary $W(k_x)$, we follow a recursive process.\\
(i) At every $k_x$, we can perform and operate on a matrix $f^n(k_x)\in{O}(n)$  (which in the end is the $W(k_x)$) in two parts: we first choose a direction in $S^{n-1}$ as an invariant direction and express  $f^n(k_x)\in{O}(n)$  as a combination of this direction and a rotation $f^{n-1}(k_x) \in{O(n-1)}$ on the remaining components, that is, 
\begin{equation}
f^n(k_x)=f^{n-1}(k_x)) \oplus s^{n-1}(k_x).
\end{equation}
(ii) Since $\pi_1(S^{n-1})=1$ for $n>2$, we can then adiabatically send $s^{n-1}(k_x)$ to identity, so that $f^n\rightarrow{f}^{n-1}\oplus\mathbb{I}_1$.\\
(iii) In the previous steps we have deformed $f^n(k_x)$ so that it has a fixed component (identity) for all $k_x$. Without loss of generality, we take it to be the $n$-th component.\\\\
(iv) Repeat the steps (i),(ii), (iii) until $n=2$, at which point we have reduced $f^n(k_x)$ to $f^2(k_x)\oplus\mathbb{I}_{n-2}$.

$f^2(k_x)\oplus\mathbb{I}_{n-2}$ is a rotation in $n$-dimensional real space which leaves all but the first two components invariant. We write it explicitly as:
\begin{equation}
f^n(k_x)=f^2(k_x)\oplus\mathbb{I}_{n-2}=[\cos(\theta(k_x))\sigma_0+i\sin(\theta(k_x))\sigma_y]\oplus\mathbb{I}_{n-2}.
\end{equation}
where $\theta(k)$ is a smooth function of $k_x$.
From the parametrization of $O(2)$ matrices, we know that any $\theta(k)$ can be smoothly sent to $\theta(k)=wk$, where $w\in{\mathbb{Z}}$ is the winding of the $O(2)$ matrix.  We have hence deformed any $f^n(k_x)\in{O}(n)$ to the standard form
\begin{equation}
g_w(k_x)\equiv[\cos(wk_x)\sigma_0+i\sin(wk_x)\sigma_y]\oplus\mathbb{I}_{n-2}.
\end{equation}
We prove that $\delta[g_w]=w$ mod 2 by contradiction. If $\delta[g_1]=0$, then from the fact that $g_w$ for arbitrary $w$ can shrink to identity, any $f^n(k_x)$ can  then shrink to identity, contradicting the fact that $\pi_1(O(n))=\mathbb{Z}_2$.
With $\delta[g_1]=1$ and using the $\mathbb{Z}_2$ group structure of $\pi_1(O(n))$, we have $\delta[g_{2k}]=0$ and $\delta[g_{2k+1}]=1$.

For 2B-1V, the odd winding cases correspond to $f^2(k_x)$'s that can be transformed transform into $g_w(k_x)$ with $w\in{odd}$.
Adding trivial sets of bands is equivalent to superimposing identity matrices to $f^2(k_x)$ and then turning on hybridization, and the end result is some $f^n(k_x)$.
However, an $f^n(k_x)$ thus obtained can smoothly deformed back to the case with zero hybridization, \ie $f^2(k_x)\oplus\mathbb{I}_{n-2}$.
Following the definition of $\delta[f^n]$, we can conclude that $\delta[f^n]=1$ if and only if $f^2(k_x)$ has odd winding.

We {\it conjecture} that the $\mathbb{Z}_2$ homotopy group can be revealed by determinant of the nested Wilson loop introduced in Ref. \cite{Benalcazar2017}.
This conjecture is based on the following observation: in the case where $C_2T=K$ and the embeding matrices are identities, or where orbitals defining the Hamiltonian sit on lattices, the determinant of nested Wilson loop is quantized to either $1$ or $-1$, corresponding to the trivial and nontrivial states, respectively.
Using the notation in appendix \ref{sec:notation}, the Bl\"och bases in this case are defined as $|\phi_{\alpha \mathbf{k}}\rangle = \frac{1}{\sqrt{N}}\sum_{\mathbf{R}} e^{i\mathbf{k}\cdot\mathbf{R}}|\alpha\mathbf{R}\rangle$, which always have an identity nested Wilson loop.
Therefore such atomic insulators always have determinant $1$, and $-1$ determinant implies some nontrivial topology, which should be the homotopy group topology discussed above.

\subsection{Particle-hole-symmetry-enforced  topology in Moir{\'e} model} \label{sub:fragile-Moire}

Having laid out the fundamentals of the topology of systems with $C_2T$ symmetry, we now show that the Moir{\'e} bands exhibit at least fragile and possibly stable (as per the arguments in the previous paragraph) topology at \textit{all} small twist angles. We \textit{stress} that our proof uses particle-hole symmetry. Without using this symmetry, one \textit{cannot} prove that the bands around the charge neutrality point, when separated from others, are topological. 

We calculate the Wilson loops of the two bands around the charge neutrality point, integrated along $\mathbf{b}_2$ and plot their spectra along $\mathbf{b}_1$,  where $\mathbf{b}_{1}$ and $\mathbf{b}_{2}$ are reciprocal lattices shown in Fig. \ref{fig:M-model}(b), for the four gapped phases in MBM-1V (Fig. \ref{fig:WL}).
All of them have a winding of 1, thereby preventing a localized Wannier description of these bands; by the arguments in the previous section, the topology could be stable. 

We now provide an understanding of the fragile topology from the viewpoint of Elementary Band Representations (EBR's) \cite{Bernevig_TQC}.
As proved in appendix \ref{sub:Irreps}, as long as the PH symmetry
is present, the gapped middle two bands always have the irreps $\Gamma_{1}+\Gamma_{2}$,
$M_{1}+M_{2}$, and $K_{2}K_{3}$. From table \ref{tab:EBR-magSG}
we find that these irreps  \textit{cannot} be written as a positive
number of EBR's. 
Instead, they can be written as a difference of EBR's.  For example they can be written as
\begin{equation}
    G^{2c}_{A_1} + G^{1a}_{A2} - G^{1a}_{A1} \label{eq:fragileEBR}
\end{equation}
This alone suggests at least a  fragile topology. If we only take into account their representation content (which alone cannot tell us if the Wilson loop winding is even or odd, see appendix \ref{sub:WLfromIrrep}, and hence whether it is fragile or stable as per appendix \ref{sec:homotopy}), we can add a band which is an EBR $G^{1a}_{A_{1}}$  and couple  it to the middle two bands of the MBM-1V such that the three bands in total can be understood as a sum of two EBR's, and hence completely trivial. Therefore  \textit{(crucially)}  the PH symmetry in MBM-1V and the representation content of the bands guarantees the non trivial fragile topology.

Slight breaking of PH symmetry will clearly not change the topology of the 2 bands around the charge neutrality point. We have
checked the gapped phases in Fig. \ref{fig:PHbreaking}(a), (b), (d) numerically 
and find that they all have same type of the Wilson loops shown in
Fig. \ref{fig:WL}. In appendix \ref{sub:WLfromIrrep} we show how the Wilson loop winding is related to the irreps of the EBR's.

\begin{figure}
\begin{centering}
\includegraphics[width=1\linewidth]{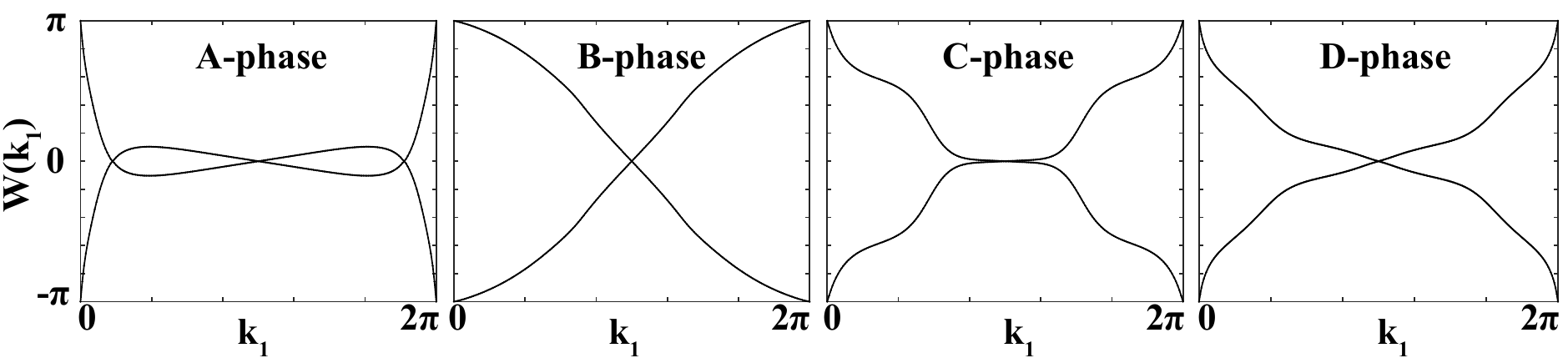}
\par\end{centering}

\protect\caption{\label{fig:WL}The Wilson loop spectra of the four gapped phases in MBM-1V. Wilson loops are integrated along $\mathbf{b}_2$ and plotted along $\mathbf{b}_1$, where $\mathbf{b}_1$ and $\mathbf{b}_2$ are reciprocal lattices shown in Fig. \ref{fig:M-model}. Due to the $C_{2x}$ symmetry, the Wilson loop spectrum is symmetric for $k_{1}\to2\pi-k_{1}$.}
\end{figure}

\subsection{Topology of higher bands in Moir\'e model} \label{sub:higherbands}

To show why PH symmetry is essential in proving topology in graphene, we compute the topology of some other low-lying energy bands. Without PH symmetry, any of these higher bands can be, in principle, deformed to be close to the Fermi level.  

We apply the criteria of fragile topology for higher bands in the Moir\'e model. For exemplification, here we take $\alpha=0.56$ (A-phase).
In doing this, we (i) group the higher bands into many sets, each of which satisfies compatibility relations \cite{Bernevig_TQC,Ashvin_indicator_2017} and so is ``gapped'' from other bands along the high symmetry lines, (ii) try to decompose each set to EBR's and diagnosing whether it has fragile topology or not. We obtain the following results.\\
(1) The +2 to +5 bands form an EBR, $G^{2c}_{E}$.\\
(2) The +6 to +7 bands form $\Gamma_1+\Gamma_2$, $M_1+M_2$, $K_2K_3$, which are identical to irreps formed by the middle two bands and can only be written as a difference of BRs, such as $G^{2c}_{A_1}+G^{1a}_{A_2}-G^{1a}_{A_1}$. \\
(3) The +8 to +15 bands form a BR, $3G^{1a}_E + G^{1a}_{A_1} + G^{1a}_{A_2}$. \\
(4) The +16 to +17 bands, forming $\Gamma_3$, $M_1+M_2$, $2K_1$, can only be written as a difference of BRs, such as $G^{2c}_E-G^{1a}_{E}$. \\
(5) The +18 to +22 bands form a BR, $G^{1a}_E + G^{1a}_{A_1} + G^{2c}_{A_2}$. \\
(6) The +23 to +24 bands form a BR, $G^{1a}_{E}$.\\
(7) The +25 to +27 bands form a BR, $G^{1a}_{A_1} + G^{1a}_E$.

If the two bands in case-(2) (or (4)) are disconnected from other bands in the whole BZ, then they must be fragile topological phase. In the other cases, the bands can be trivial.


\subsection{Diagnosing fragile topology from symmetry eigenvalues} \label{sub:WLfromIrrep}

In this subsection we prove that the irreps of the two bands around the Fermi level, which we proved to be $\Gamma_{1}+\Gamma_{2}$, $M_{1}+M_{2}$, and $K_{2}K_{3}$ at \textit{all} angles if PH symmetry is preserved, must correspond to a nonzero Wilson loop winding.  As the winding number is a topological invariant, it remains unchanged if the Wilson loop path is deformed continuously. 
Here we choose the $N_1$ bent Wilson loops shown in Fig. \ref{fig:BWL}(a), where $N_1$ is a large enough even integer to describe the infinite limit.
For $0 \le i \le N_1/2$, the $i$-th Wilson loop $W(i)$ is defined on the path
\begin{equation}
W(i) = W_{K^{(3)} \to \boldsymbol{\kappa}_i \to K^{(2)}},
\end{equation}
where
\begin{equation}
\boldsymbol{\kappa}_i = -\frac{N_1/2-i}{N_1/2}K^{(1)},
\end{equation}
and $K^{(1,2,3)}$ are the momenta translated from the Moir{\'e} $K$ point by Moir{\'e} reciprocal lattice vectors, as shown in Fig. \ref{fig:BWL}(a). For $N_1/2 < i \le N_1$, the $i$-th Wilson loop $W(i)$ is defined on the path
\begin{equation}
    W(i) = W_{ K^{(3)} \to \boldsymbol{\kappa}_i \to K^{(1)}\to \boldsymbol{\kappa}_i^\prime \to K^{(2)}},
\end{equation}
where
\begin{equation}
    \boldsymbol{\kappa}_i =-\frac{i-N_1/2}{N_1/2}K^{(2)}, \qquad
    \boldsymbol{\kappa}_i^\prime =-\frac{i-N_1/2}{N_1/2}K^{(3)}.
\end{equation}
The $N_1$-th Wilson loop is equivalently defined on the path $-K^{(2)}\to K^{(1)} \to -K^{(3)}$ since the two sections, $K^{(3)}\to -K^{(2)}$ and $-K^{(3)}\to K^{(2)}$, at the beginning and the ending cancel each other (the product of the Wilson line matrices on these two lines is exactly the identity matrix). 
Therefore $W(N_1)$ is defined on the same path (up to reciprocal lattice vectors) as $W(0)$, and the Wilson loop spectra of $W(0)$ to $W(N_1)$ are identical due to this periodicity.

\begin{figure}
\begin{centering}
\includegraphics[width=1\linewidth]{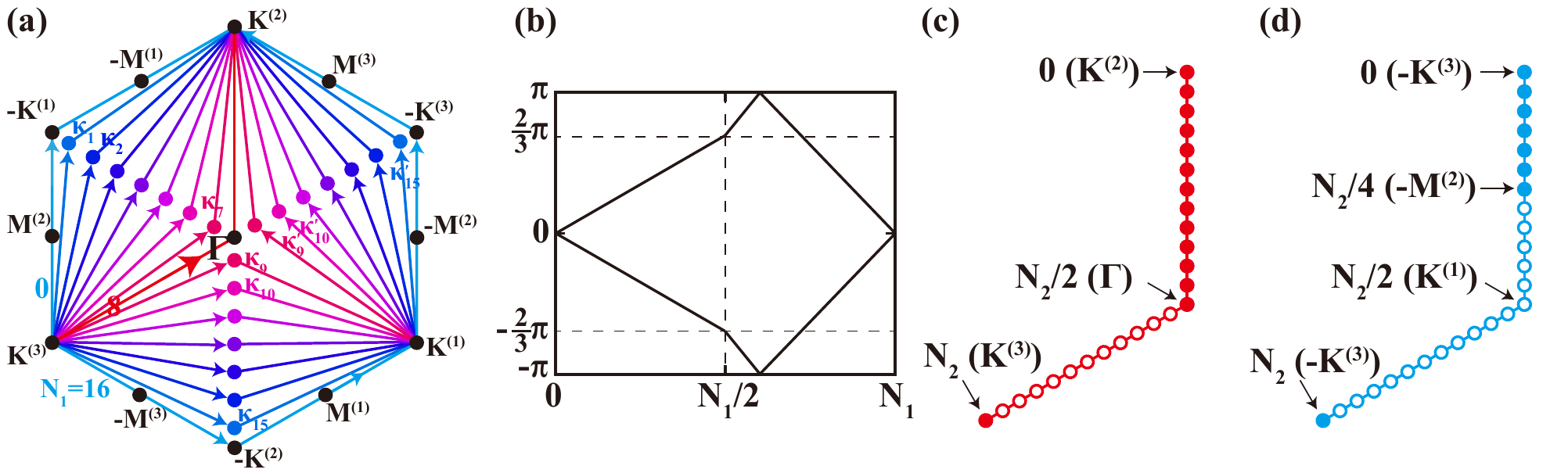}
\par\end{centering}    
\protect\caption{\label{fig:BWL}Bent Wilson loops used to prove the relation between winding number and irreps. 
(a) The bent Wilson loops. There are in total $N_{1}$ Winson loops. In this figure we set $N_1=16$ for illustration. For the first $N_1/2+1$ Wilson loops, the integration path goes as $\mathbf{K}^{(3)}\to\boldsymbol{\kappa}_i\to\mathbf{K}^{(2)}$, where $0\le i\le N_1/2$, whereas for the next $N_1/2$ Wilson loops, the integration path goes as $\mathbf{K}^{(3)} \to \boldsymbol{\kappa}^{i} \to \mathbf{K}^{(1)} \to \boldsymbol{\kappa}_i^\prime \to\mathbf{K}^{(2)}$, where $N_1/2 < i\le N_1$. 
In the $N_1$-th Wilson loop, the first section cancels with the last section such that we can calculate the Wilson loop eigenvalues just from the path $-\mathbf{K}^{(2)}\to\mathbf{K}^{(1)}\to-\mathbf{K}^{(3)}$.
(b) The spectra of the bent Wilson loops with winding number $\pm1$. 
(c) The wave function indices in the $N_{1}/2$-th Wilson loop. 
(d) The wave function indices in the $N_{1}$-th Wilson loop.
The indices in (c) and (d) are used in the proof in text. }
\end{figure}

In the first step, we prove that the eigenvalues of $W(N_{1}/2)$ are quantized as $e^{\pm i\frac23\pi}$ for bands with representations  $\Gamma_1+\Gamma_2$, $M_1+M_2$, and $K_2K_3$ at $\Gamma, M$ and $K$ respectively.
Since the eigenvalues of Wilson loops are gauge-independent, we can choose the gauge freely in our proof.  We denote by  $i=1\ldots N_2-1$  indices of the wave function  forming  the $N_{1}/2$-th Wilson loop (see Fig \ref{fig:BWL}). 
To make the proof simple and concise, we take the gauge where (i) the wave functions at indices $i=N_2/2+1,\cdots,N_2-1$ are transformed from the wave functions at indices $i=1,\cdots, N_2/2-1$ as
\begin{equation}
C_{3z}U_{i}=U_{N_{2}-i}\qquad\text{for}\quad i=1\cdots N_{2}/2-1,
\end{equation}
and (ii) wave functions at $C_{3z}$-invariant points, \ie the $0$-th point $K^{(2)}$ and the $N_2/2$-th point $\Gamma$, transform under $C_3$ as
\begin{equation}
V^{\mathbf{b}_2-\mathbf{b}_1}C_{3z}U_{0}=U_{0}S_{K}\qquad C_{3z}U_{N_{2}/2}=U_{N_{2}/2}S_{\Gamma} \label{euqationc3zfor2points}
\end{equation}
where $S_{K}=e^{i\frac{2\pi}{3}\sigma_{y}}$ (or $e^{-i\frac{2\pi}{3}\sigma_{y}}$) and $S_{\Gamma}=\sigma_{0}$.

At $K$ point, the bases are chosen as $\frac{1}{\sqrt{2}}|1\rangle-\frac{i}{\sqrt{2}}|-1\rangle$ and $-\frac{i}{\sqrt{2}}|1\rangle+\frac{1}{\sqrt{2}}|-1\rangle$ (or $\frac{1}{\sqrt{2}}|1\rangle+\frac{i}{\sqrt{2}}|-1\rangle$ and $\frac{i}{\sqrt{2}}|1\rangle+\frac{1}{\sqrt{2}}|-1\rangle$), where $|\pm1\rangle$ are the eigenstates $C_{3z}$ with eigenvalues $e^{\pm i\frac{2\pi}{3}}$, such that $C_{2z}T=K$.
At $\Gamma$, as both the $\Gamma_1$ and $\Gamma_2$ irreps have $C_3$-eigenvalue $+1$, in any gauge, including the $C_{2z}T=K$ gauge, the $C_{3z}$ representation matrix must be identity.
Therefore
\begin{eqnarray}
W\left(N_{1}/2\right) & = & U_{0}^{\dagger} V^{\mathbf{b}_1-\mathbf{b}_2\dagger} U_{N_{2}-1}U_{N_{2}-1}^{\dagger}U_{N_{2}-2}\cdots U_{N_{2}/2+1}^{\dagger}U_{N_{2}/2}U_{N_{2}/2}^{\dagger}U_{N_{2}/2-1}\cdots U_{1}^{\dagger}U_{0}\nonumber \\
 & = & U_{0}^{\dagger}V^{\mathbf{b}_1-\mathbf{b}_2\dagger}C_{3z}U_{1}U_{1}^{\dagger}C_{3z}^{\dagger}C_{3z}U_{2}\cdots U_{N_{2}/2-1}^{\dagger}C_{3z}^{\dagger}U_{N_{2}/2}U_{N_{2}/2}^{\dagger}U_{N_{2}/2-1}\cdots U_{1}^{\dagger}U_{0}\nonumber \\
 & = & S_{K}U_{0}^{\dagger}U_{1}U_{1}^{\dagger}U_{2}\cdots U_{N_{2}/2-1}^{\dagger}U_{N_{2}/2}S_{\Gamma}^{\dagger}U_{N_{2}/2}^{\dagger}U_{N_{2}/2-1}\cdots U_{1}^{\dagger} U_{0}\nonumber \\
 & = & S_{K}W^{\prime\dagger}\left(N_{1}/2\right)S_{\Gamma}^{\dagger}W^{\prime}\left(N_{1}/2\right)
\end{eqnarray}
Here $W^{\prime}\left(N_{1}/2\right)=U_{N_{2}/2}^{\dagger}U_{N_{2}/2-1}\cdots U_{1}^{\dagger}U_{0}$
is a unitary matrix. Since $S_\Gamma= \sigma_0$, we have that  $W^{\prime\dagger}\left(N_{1}/2\right)S_{\Gamma}^{\dagger}W^{\prime}\left(N_{1}/2\right) = W^{\prime\dagger}\left(N_{1}/2\right) W^{\prime}\left(N_{1}/2\right)= 1$. (Alternatively, in the $C_{2z}T=K$ gauge $W^{\prime}\left(N_{1}/2\right)=W^{\prime *}\left(N_{1}/2\right)$ (Eq. (\ref{eq:WL-C2zT})), thus $W^{\prime}\left(N_{1}/2\right)$ takes the form $\exp\left(i\theta^{\prime}\sigma_{0}+i\theta\sigma_{y}\right)$, where $\theta^{\prime}=0$ or $\pi$). Then it follows
\begin{equation}
W\left(N_{1}/2\right)=S_{K}S_{\Gamma}^{\dagger}=e^{i\frac{2\pi}{3}\sigma_{y}}\qquad(\text{or}\quad e^{-i\frac{2\pi}{3}\sigma_{y}})
\end{equation} 
We secondly look at the Wilson loop on the $N_{1}$-th path, shown in Fig. \ref{fig:BWL}.
This path passes through the $M$ point. This suggests that we can use the symmetry eigenvalues at this point, which can be read from the irreps at the $M$ point, to find out the constraints on the corresponding Wilson loop.
We take the gauge that wave functions at indices $i=N_{2}/4+1,\cdots,N_{2}-1$ are rotated from
the wave functions at $i=1\cdots N_{2}/4-1$ (Fig. \ref{fig:BWL}(c)),
\ie
\begin{equation}
C_{2x}U_{i}=U_{N_{2}/2-i}\qquad\text{for}\quad i=0\cdots N_{2}/4-1
\end{equation}
\begin{equation}
V^{\mathbf{b}_1}C_{3z}U_{i}=U_{N_{2}-i}\qquad\text{for}\quad i=1\cdots N_{2}/2-1
\end{equation}
For the $C_{2x}$-invariant point  (the $N_2/4$-th point),  $-M^{(2)}$, we set
\begin{equation}
C_{2x}U_{N_{2}/4}=U_{N_{2}/4}S_{M}.
\end{equation}
And for the $C_{3z}$-invariant point, \ie the $N_2/2$-th point $K^{(1)}$, we set
\begin{equation}
V^{\mathbf{b}_1}C_{3z}U_{N_{2}/2}=U_{N_{2}/2}S_{K},
\end{equation}
We use the same $S_K$ matrix with Eq. (\ref{euqationc3zfor2points}) because $K^{(1)}$ here differs from $K^{(2)}$ in Eq. (\ref{euqationc3zfor2points}) by only a reciprocal lattice.
The $C_3$ representation matrix of the other $C_3$-invariant point, \ie the $0$-th point $-K^{(3)}$, can be derived from the one of $K^{(1)}$
\begin{equation}
V^{\mathbf{b}_2}C_{3z}U_{0}=V^{\mathbf{b}_2}C_{3z}C_{2x}U_{N_{2}/2}=C_{2x}V^{\mathbf{b}_2}C_{3z}^{\dagger}U_{N_{2}/2}=C_{2x}U_{N_2/2}S_{K}^{\dagger}=U_{0}S_{K}^{\dagger}
\end{equation} 
For the gapped middle two bands, the irreps at $K$ and $M$ are always
$K_{2}K_{3}$ and $M_{1}+M_{2}$, thus we take $S_{K}=e^{i\frac{2\pi}{3}\sigma_{y}}$
(or $e^{-i\frac{2\pi}{3}\sigma_{y}}$) and $S_{M}=\sigma_{z}$.
The Wilson loop
\begin{eqnarray}
W\left(N_{1}\right) & = & U_{0}^{\dagger}V^{\mathbf{b}_1-\mathbf{b}_2\dagger}U_{N_{2}-1}U_{N_{2}-1}^{\dagger}U_{N_{2}-2}\cdots U_{N_{2}/2+1}^{\dagger}U_{N_{2}/2}U_{N_{2}/2}^{\dagger}U_{N_{2}/2-1}\cdots U_{1}^{\dagger}U_{0}\nonumber \\
 & = & U_{0}^{\dagger}V^{-\mathbf{b}_2\dagger}C_{3z}U_{1}U_{1}^{\dagger}(V^{\mathbf{b}_1}C_{3z})^{\dagger}(V^{\mathbf{b}_1}C_{3z})U_{2}\cdots U_{N_{2}/2-1}^{\dagger} (V^{\mathbf{b}_1}C_{3z})^{\dagger}U_{N_{2}/2}U_{N_{2}/2}^{\dagger}U_{N_{2}/2-1}\cdots U_{1}^{\dagger}U_{0}\nonumber \\
 & = & S_{K}^{\dagger}W^{\prime\dagger}\left(N_{1}\right)S_{K}^{\dagger}W^{\prime}\left(N_{1}\right)\label{eq:WN1-tmp}
\end{eqnarray}
where the Wilson line can be separated to take advantage of the known $M$-point eigenvalues.
\begin{eqnarray}
W^{\prime}\left(N_{1}\right) & = & U_{N_{2}/2}^{\dagger}U_{N_{2}/2-1}\cdots U_{N_{2}/4+1}^{\dagger}U_{N_{2}/4}U_{N_{2}/4}^{\dagger}U_{N_{2}/4-1}\cdots U_{1}^{\dagger}U_{0}\nonumber \\
 & = & U_{0}^{\dagger}C_{2x}^{\dagger}C_{2x}U_{1}\cdots U_{N_{2}/4-1}^{\dagger}C_{2x}^{\dagger}U_{N_{2}/4}U_{N_{2}/4}^{\dagger}U_{N_{2}/4-1}\cdots U_{1}^{\dagger}U_{0}\nonumber \\
 & = & W^{\prime\prime\dagger}\left(N_{1}\right)S_{M}W^{\prime\prime}\left(N_{1}\right)
\end{eqnarray}
where we have defined the Wilson line:
\begin{equation} W^{\prime\prime}\left(N_{1}\right)=U_{N_{2}/4}^{\dagger}U_{N_{2}/2-1}\cdots U_{1}^{\dagger}U_{0}\end{equation}
is a unitary matrix Wilson line. In the $C_{2z}T=K$ gauge $W^{\prime\prime}\left(N_{1}\right)=W^{\prime\prime *}\left(N_{1}\right)$ (Eq. (\ref{eq:WL-C2zT})), thus $W^{\prime\prime}\left(N_{1}\right)$ takes the form $\exp\left(i\theta^{\prime}\sigma_{0}+i\theta\sigma_{y}\right)$, where $\theta^{\prime}=0$ or $\pi$. Substituting this form into $W^{\prime}\left(N_{1}\right)$
and then substituting  $W^{\prime\prime}\left(N_{1}\right)$ back into
Eq. (\ref{eq:WN1-tmp}) we get
\begin{equation}
W\left(N_{1}\right)=S_{K}^{\dagger}S_{M}S_{K}^{\dagger}S_{M}=\sigma_{0}
\end{equation}

In the second step, we prove that the eigenvalues from $W\left(0\right)$
to $W\left(N_{1}\right)$ must wind. As discussed in subsection \ref{sub:fragile-C2zT},
in the $C_{2z}T=K$ gauge the Wilson loops take the form $W\left(i\right)=\exp\left(i\phi^{\prime}\sigma_{0}+i\phi\left(i\right)\sigma_{y}\right)$.
Here $\phi^{\prime}=0$ because the $W\left(N_{1}\right)=\sigma_{0}$.
From the analysis above, we have $\phi\left(N_{1}/2\right)=\pm\frac{2\pi}{3}\mod2\pi$,
$\phi\left(N_{1}\right)=0\mod2\pi$.  
Because they are defined on the paths  that differ from each other by a reciprocal lattice, $\phi\left(0\right)$ must equal to $\phi\left(N_{1}\right)$ (Fig. \ref{fig:BWL}(a)). In order to  figure out how the eigenvalues are  connected, we
notice that $\phi$$\left(N_{1}/2-i\right)$ and $\phi\left(N_{1}/2+i\right)$
are indeed not independent. 
As shown in Fig. \ref{fig:BWL}, the $N_1/2+i$-th Wilson loop can be divided into two sections, \ie the section on path $K^{(3)}\to \boldsymbol{\kappa}_{N_1/2+i} \to K^{(1)}$ and the section on path $K^{(1)}\to \boldsymbol{\kappa}_{N_1/2+i}^\prime \to K^{(2)}$.
Each of the two section can be rotated from the Wilson loop defined on path $K^{(2)}\to\boldsymbol{\kappa}_{N_1/2-i}\to K^{(3)}$, which is just the hermitian conjugate of the $N_1/2-i$-th Wilson loop.
Accordingly, it is immediate that $W\left(N_{1}/2+i\right)= W^{\dagger}\left(N_{1}/2-i\right)W^{\dagger}\left(N_{1}/2-i\right)$,
which yields
\begin{equation}
\phi\left(N_{1}/2+i\right)=-2\phi\left(N_{1}/2-i\right)\mod2\pi
\end{equation}
Therefore, if from $i=0$ to $i=N_{1}/2$ the phase $\phi\left(i\right)$
goes from $0$ to $\left(n\pm\frac{1}{3}\right)2\pi$, where $n$
is an arbitrary integer, then from $i=N_{1}/2$ to $i=N_{1}$ the
phase $\phi\left(i\right)$ goes from $\left(n\pm\frac{1}{3}\right)2\pi$
to $\left(3n\pm1\right)2\pi$. Therefore the total winding number
is $3n\pm1$, which must be nonzero. In Fig. \ref{fig:BWL}(b) we
sketch the Wilson loop spectrum for $n=0$.
In conclusion, the Wilson loop must be winding for any given two bands, disconnected from other bands, and of irreps $\Gamma_{1}+\Gamma_{2}$, $M_{1}+M_{2}$, and $K_{2}K_{3}$. By replacing the representation matrices in the proof, Wilson loops compatible with other possible irreps can also be derived.
In table \ref{tab:winding} we tabulate all the compatibility-relation-allowed irreps \cite{Bernevig_TQC,Ashvin_indicator_2017} and the corresponding compatible winding numbers. 

\begin{table}
\begin{tabular}{|c|c|}
\hline 
Irreps & Compatible winding number \tabularnewline
\hline 
\hline 
$2\Gamma_{1}$, $2M_{1}$, $2K_{1}$ & $3n$\tabularnewline
\hline 
$2\Gamma_{1}$, $2M_{1}$, $K_{2}K_{3}$ & $3n$\tabularnewline
\hline 
$2\Gamma_{2}$, $2M_{2}$, $2K_{1}$ & $3n$\tabularnewline
\hline 
$2\Gamma_{2}$, $2M_{2}$, $K_{2}K_{3}$ & $3n$\tabularnewline
\hline 
$\Gamma_{1}+\Gamma_{2}$, $M_{1}+M_{2}$, $2K_{1}$, & $3n$\tabularnewline
\hline 
$\Gamma_{1}+\Gamma_{2}$, $M_{1}+M_{2}$, $K_{2}K_{3}$ & $3n\pm1$\tabularnewline
\hline 
$\Gamma_{3}$, $M_{1}+M_{2}$, $2K_{1}$ & $3n\pm1$\tabularnewline
\hline 
$\Gamma_{3}$, $M_{1}+M_{2}$, $K_{2}K_{3}$ & $3n$ or $3n\pm1$\tabularnewline
\hline 
\end{tabular}

\caption{\label{tab:winding}All the compatibility-relation-allowed two-band
irreps in magnetic space group $P6^{\prime}2^{\prime}2$ and the corresponding
constraints on Wilson loop winding.}

\end{table}

\section{Tight-binding models}
\subsection{Four-band model for one valley (TB4-1V)}\label{sub:TB4-1V}
In appendix \ref{sec:fragile} we showed that the middle two bands in MBM-1V, provided that they are disconnected from other bands, must have fragile topology due to the PH symmetry.
If the PH symmetry is broken, as long as the gap between the middle two bands and other bands remains finite (so the irreps keep the same), the topology still exists.
Thus there is a Wannier obstruction for these two bands. In this subsection, we provide a four-band tight-binding model (referred as TB4-1V in the following), whose bands form two separated ``valence and conduction'' branches having the $+1$ and $-1$ winding, respectively.


First let us try to obtain the symmetry content of a TB4-1V model from the viewpoint of EBR's.
As discussed in appendix \ref{sub:fragile-Moire}, the irreps of the middle two bands in MBM-1V, \ie $\Gamma_1+\Gamma_2$, $M_1+M_2$, and $K_2K_3$, cannot be written as a positive number of EBR's. Instead, they can be written as a difference of EBR's, $G^{2c}_{A_1}+G^{1a}_{A_1}-G^{1a}_{A_2}$ (Eq. (\ref{eq:fragileEBR})).  Here, we interpret these irreps in another way: they can be thought as forming one disconnected, topological branch of the composite band representation (BR), formed by the sum of EBR's $G^{2c}_{A_1}$ and $G^{2c}_{A_2}$ (table \ref{tab:EBR-magSG}). 
This understanding gives us an ansatz for building the TB4-1V model. 
We start with two independent EBR's, which give the irreps $2\Gamma_1$, $2M_1$, $K_2K_3$, and $2\Gamma_2$, $2M_2$, $K_2K_3$, respectively. We then mix the two EBR's and decompose the bands into two new branches, each of which forms the irreps $\Gamma_1+\Gamma_2$, $M_1+M_2$, and $K_2K_3$, such that each branch has a Wilson loop winding $3n\pm1$ according to discussion in appendix \ref{sub:WLfromIrrep}. 
By this method, the TB4-1V model is able to reproduce the irreps of the middle two bands. What is left  to check is whether it can reproduce the correct Wilson loop. (In principle it may wind $3n\pm 1$ times, even or odd.)

The orbitals and hoppings are shown in Fig. \ref{fig:TB4-1V}. There are six parameters in total: $\Delta$ is the energy splitting between $s$ and $p_z$ orbitals, $t_{s,p}$ are the nearest hoppings, $\lambda$ is the second-nearest hopping, and $t_{s,p}^\prime$ are the third-nearest hoppings. Only $\Delta$ is real; all the other parameters can be complex in general. For simplicity, here we set $t_s=t_p=t$, $t_s^\prime=t_p^\prime=t^\prime$ and $\lambda$ all real numbers. Then the Hamiltonian in momentum space can be written as
\begin{eqnarray}
    H^{(TB4-1V)}\left(\mathbf{k}\right) & = & \Delta\mu_{z}\sigma_{0}+\mu_{0}\sigma_{x}\sum_{i=1}^3 \left[t\cos\left(\boldsymbol{\delta}_{i}\cdot\mathbf{k}\right)+t^{\prime}\cos\left(-2\boldsymbol{\delta}_{i}\cdot\mathbf{k}\right)\right]\nonumber \\
     &  & -\mu_{0}\sigma_{y}\sum_{i=1}^3\left[t\sin\left(\boldsymbol{\delta}_{i}\cdot\mathbf{k}\right)+t^{\prime}\sin\left(-2\boldsymbol{\delta}_{i}\cdot\mathbf{k}\right)\right]-2\lambda\mu_{y}\sigma_{z}\sum_{i=1}^3\sin\left(\mathbf{d}_{i}\cdot\mathbf{k}\right)
\end{eqnarray}
Here $\boldsymbol{\delta}_1=\frac{1}{3}\mathbf{a}_1 +\frac{2}{3}\mathbf{a}_2$, $\boldsymbol{\delta}_2=-\frac{2}{3}\mathbf{a}_1-\frac{1}{3}\mathbf{a}_2$, and $\boldsymbol{\delta}_2=\frac{2}{3}\mathbf{a}_1 +\frac{1}{3}\mathbf{a}_2$ are the nearest vectors, $\mathbf{d}_1=\mathbf{a}_1$, $\mathbf{d}_2=\mathbf{a}_2$, and  $\mathbf{d}_3=-\mathbf{a}_1-\mathbf{a}_2$ are the second-nearest vectors, $\mu_0$ and $\sigma_0$ are 2 by 2 identities, $\mu_{x,y,z}$ and $\sigma_{x,y,z}$ are Pauli matrices.
The symmetry operators for this model is 
\begin{equation}
    C_{2z}T=\sigma_{x}K\qquad C_{2x}=\mu_{z}\qquad C_{3z}=1
\end{equation} 
The energy levels at high symmetry momenta can be solved analytically. At $\Gamma$, we have
\begin{equation}
    E\left(\Gamma_{1}\right)=\Delta\pm3\left(t+t^{\prime}\right)\qquad E\left(\Gamma_{2}\right)=-\Delta\pm3\left(t+t^{\prime}\right)
\end{equation}
at $M$ we have
\begin{equation}
    E\left(M_{1}\right)=\Delta\pm\left(t-3t^{\prime}\right)\qquad E\left(M_{2}\right)=-\Delta\pm\left(t-3t^{\prime}\right)
\end{equation}
and at $K$ we have
\begin{equation}
    E\left(K_{2}K_{3}\right)=\pm\sqrt{\Delta^{2}+27\lambda^{2}}
\end{equation}
The parameters should satisfy $3|t+t^\prime|>|\Delta|$ and $|3t-3t^\prime|>|\Delta|$ for the lower two bands to form the irreps $\Gamma_1+\Gamma_2$, $M_1+M_2$, $K_2K_3$. To make the bands as flat as possible, we further ask the averaged energy of the lower two bands at $\Gamma$, $M$, $K$ equal to each other, which requires $t^\prime=-\frac{1}{3}t$, $\lambda=\frac{2}{\sqrt{27}}t$. The band structure and Wilson loop with parameters satisfying these conditions are plotted in Fig. \ref{fig:TB4-1V}(b) and (c). It turns out that the TB4-1V model not only reproduces the correct irreps but also the correct Wilson loop winding and flat dispersion. 

As shown in Fig. \ref{fig:TB4-1V}(b), the upper two bands are far away from the neutrality point and hence will not enter the low energy physics. These two bands are only used to cancel the Wannier obstruction of the lower two bands.

\subsection{Eight-band model for two valleys (TB8-2V)} \label{sub:TB8-2V}
The two-valley tight-binding preserving valley symmetry is just a stacking of the TB4-1V model and its time-reversal counterpart. We refer to this model as TB8-2V. Following the convention that $t_1=t_2=t$, $t_1^\prime=t_2^\prime=t^\prime$ and $\lambda$ are all real, the TB8-2V model can be written as
\begin{eqnarray} \label{eq:TB8-2V}
    H^{(TB8-2V)}\left(\mathbf{k}\right) & = & \Delta\tau_{0}\mu_{z}\sigma_{0}+\tau_{0}\mu_{0}\sigma_{x}\sum_{i}\left[t\cos\left(\boldsymbol{\delta}_{i}\cdot\mathbf{k}\right)+t^{\prime}\cos\left(-2\boldsymbol{\delta}_{i}\cdot\mathbf{k}\right)\right]\nonumber \\
     &  & -\tau_{0}\mu_{0}\sigma_{y}\sum_{i}\left[t\sin\left(\boldsymbol{\delta}_{i}\cdot\mathbf{k}\right)+t^{\prime}\sin\left(-2\boldsymbol{\delta}_{i}\cdot\mathbf{k}\right)\right]-2\lambda\tau_{0}\mu_{y}\sigma_{z}\sum_{i}\sin\left(\mathbf{d}_{i}\cdot\mathbf{k}\right)
\end{eqnarray}
Here $\tau_0$ is a 2 by 2 identity representing the valley degree. 
The time-reversal and spatial symmetry operators are
\begin{equation}
    T=\tau_{x}K\qquad C_{2z}=\tau_{x}\sigma_{x}\qquad C_{2x}=\mu_{z}\qquad C_{3z}=1
\end{equation}
and the valley symmetry operator is $\mathcal{V}=\tau_z$.

\begin{figure}
\begin{centering}
\includegraphics[width=0.6\linewidth]{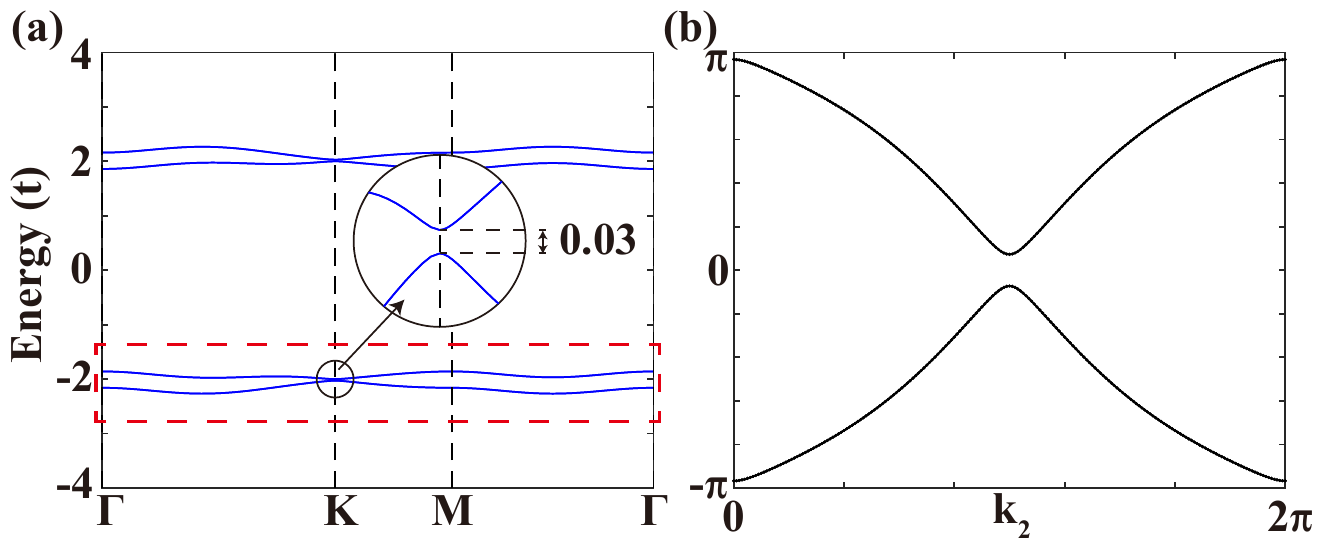}
\par\end{centering}
\protect\caption{\label{fig:TB8-2V} The eight-band tight-binding model for two valleys (TB8-2V). (a) The band structure. The inter-valley coupling $\zeta$ opens a gap at $K$. (b) The Wilson loop of the lower four bands. (a) and (b) are calculated with the parameters $t^\prime=-\frac{1}{3}t$, $\lambda=\frac{2}{\sqrt{27}}t$, $\Delta=0.15t$, and $\zeta=0.2t$. }
\end{figure}

As shown by first principle calculations in appendix \ref{sec:DFT}, the inter-valley coupling is small but not zero (around 0.2meV) at small twisting angle (Fig. \ref{fig:DFT-gap}). Such effect can be simulated in the TB8-2V model by adding valley symmetry breaking terms such as $\zeta \tau_y \mu_z \sigma_z$. 
We find that the $\zeta\tau_y\mu_z\sigma_z$ term opens a gap at $K$ point and trivializes the Wilson loop, as shown in Fig. \ref{fig:TB8-2V}.
Therefore, in absence of valley symmetry, the Wilson loop winding is no longer protected.

\subsection{Four-band model for two valleys (TB4-2V)} \label{sub:TB4-2V}

For the TB8-2V model, the Wannier obstruction can be removed by breaking the valley symmetry. This valley symmetry breaking is observed (and increasing) at small angles in  first principle calculation (appendix \ref{sec:DFT}). Thus it is feasible to build a tight-binding model for the lower four bands in TB8-2V by breaking the valley symmetry. In this subsection, we build such a model by explicitly constructing Wannier functions for the lower four bands in TB8-2V.

\begin{figure}
\begin{centering}
\includegraphics[width=1\linewidth]{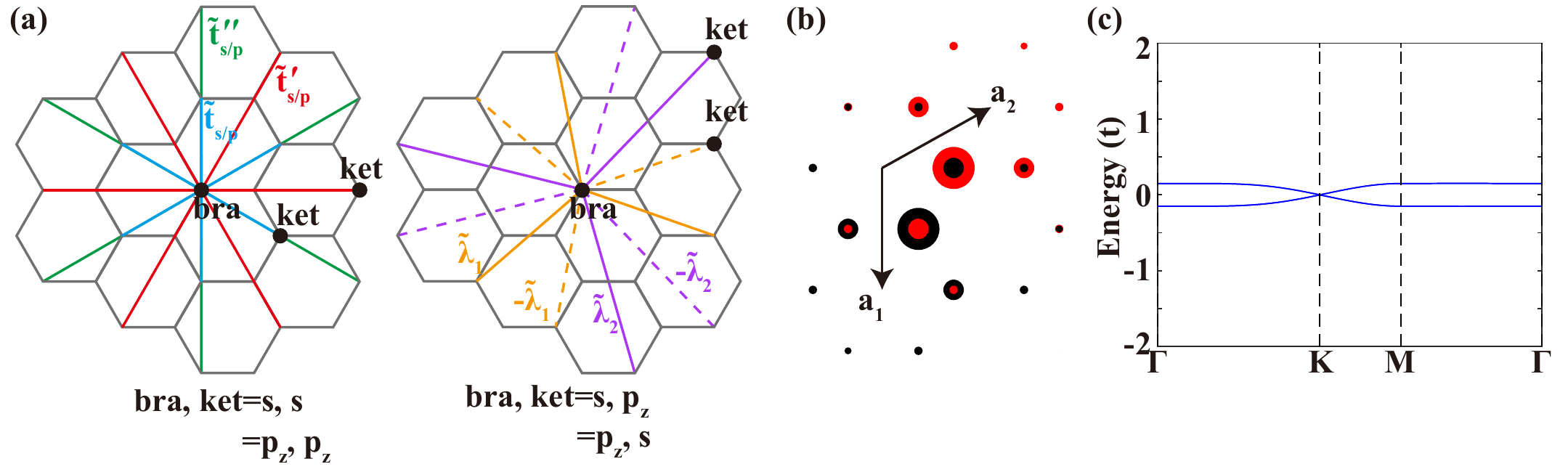}
\par\end{centering}
\protect\caption{\label{fig:TB4-2V} The four-band tight-binding (TB4-2V) model for two valleys got by constructing Wannier functions for the lower four bands in TB8-2V model. The parameters in TB8-2V are set as $t^\prime=-\frac{1}{3}t$, $\lambda=\frac{2}{\sqrt{27}}t$, $\Delta=0.15t$, and $\zeta=0$. (a) Some largest hopping parameters ($>0.001t$). $\tilde{t}_s=0.00238t$, $\tilde{t}_p=-0.01960t$, $\tilde{t}_s^\prime=-0.00252t$, $\tilde{t}_p^\prime=0.01971t$, $\tilde{t}_s^{\prime\prime}=0.01456t$, $\tilde{t}_p^{\prime\prime}=0.00382t$, $\tilde{\lambda}_1=0.01842t$, $\tilde{\lambda}_2=0.00509t$, the onsite energies $\tilde{\epsilon}_s=-1.9382t$, $\tilde{\epsilon}_p=-2.1731t$. (b) The shape of Wannier functions. Size of red circles represents the density of $|s\mathbf{t}_1\rangle$ orbital and size of black circles represent the density of $|p\mathbf{t}_2\rangle$ orbital. (c) The band structure of Eq. (\ref{eq:TB4-2V}), where the identity term $\Upsilon_+(\mathbf{k})$ is omitted, with tuned parameters. The tuned parameters are set as $\tilde{\Delta}_-=0.1174t$, $\tilde{t}_-=0.011t$, $\tilde{t}^\prime_-=-\tilde{t}_-$, $\tilde{t}^{\prime\prime}_-=\frac{1}{3}(\tilde{\Delta}_- -3\tilde{t}_- +6\tilde{t}_-^\prime)$, $\tilde{\lambda}_1=0.01842t$, and $\tilde{\lambda}_2=0.00509t$. }
\end{figure}

We start with a TB8-2V model \textit{preserving} the valley symmetry, \ie $\zeta=0$, but take a symmetry-breaking gauge for the Wannier functions.
We follow the projection procedure described in Ref. \cite{Vanderbilt_Wannier_Z2}.
The lower four bands of TB8-2V form irreps $\Gamma_1+\Gamma_2+\Gamma_3+\Gamma_4$, $M_1+M_2+M_3+M_4$, and $2 K_3$, which can be induced by the EBR's $G^{2c}_{A_1}$ and $G^{2c}_{A_2}$ (table \ref{tab:EBR-SG}). Thus the Wannier functions for the lower four bands should be the $s$ and $p_z$ orbitals $2c$, which form the $A_1$ and $A_2$ representations, respectively. 
For convenience, in the following, we label the eight bases in the TB8-2V model (Eq. (\ref{eq:TB8-2V})) as 
$|s^K \mathbf{t}_1+\mathbf{R}\rangle$, 
$|s^K \mathbf{t}_2+\mathbf{R}\rangle$, 
$|p_z^K\mathbf{t}_1+\mathbf{R}\rangle$, 
$|p_z^K\mathbf{t}_2+\mathbf{R}\rangle$, 
$|s^{K^\prime}\mathbf{t}_1+\mathbf{R}\rangle=T|s^{K}\mathbf{t}_1+\mathbf{R}\rangle$, 
$|s^{K^\prime} \mathbf{t}_2+\mathbf{R}\rangle=T|s^K \mathbf{t}_2+\mathbf{R}\rangle$, 
$|p_z^{K^\prime}\mathbf{t}_1+\mathbf{R}\rangle=T|p_z^K\mathbf{t}_1+\mathbf{R}\rangle$, 
$|p_z^{K^\prime}\mathbf{t}_2+\mathbf{R}\rangle=T|p_z^K\mathbf{t}_2+\mathbf{R}\rangle$. 
Here $\mathbf{t}_1=\frac13\mathbf{a}_1+\frac23\mathbf{a}_2$ and $\mathbf{t}_2=\frac23\mathbf{a}_1+\frac13\mathbf{a}_2$ are sublattice vectors (Fig. \ref{fig:TB4-1V}(a)), $\mathbf{R}$ is lattice vector, and superscript $K$ and $K^\prime$ represent the belonging valley of the corresponding orbitals.
We then define the four projection orbitals for Wannier functions as 
\begin{equation} \label{eq:proj-1}
    |s,\mathbf{t}_{1,2}+\mathbf{R}\rangle =c_{1}|s^K\mathbf{t}_{1,2}+\mathbf{R}\rangle +c_{1}^{*}|s^{K^\prime}\mathbf{t}_{1,2}+\mathbf{R}\rangle 
\end{equation}
\begin{equation} \label{eq:proj-2}
    |p_z,\mathbf{t}_{1,2}+\mathbf{R}\rangle =c_{2}|p_z^K\mathbf{t}_{1,2}+\mathbf{R}\rangle +c_{2}^{*}|p_z^{K^\prime}\mathbf{t}_{1,2}+\mathbf{R}\rangle  
\end{equation}
such that they respect the symmetry of space group $P622$ plus time-reversal but break valley symmetry.
To be specific, $|s,\mathbf{t}_{1,2}+\mathbf{R}\rangle$ form the $A_1$ irrep at $2c$ and $|p_z,\mathbf{t}_{1,2}+\mathbf{R}\rangle$ form the $A_2$ irrep at $2c$.
The breaking of valley symmetry is obvious since bases from the two valleys are mixed in the projection orbitals.
The projected Bl{\"o}ch wave-functions are
\begin{equation}
    \left|\bar{\psi}_{\alpha,s,\mathbf{k}}\right\rangle =\sum_{n=1}^{4}\left|\psi_{n\mathbf{k}}\right\rangle \left\langle \psi_{n\mathbf{k}}\right|\alpha,\mathbf{t}_s \rangle e^{i\mathbf{k}\cdot\mathbf{t}_s} \label{barpsialphask}
\end{equation}
where $\alpha=s,p_z$ is the orbital index and $s=1,2$ is the sublattice index, and $n=1,\ldots, 4$ runs over the lower energy bands of the TB8-2V model. The overlap matrix above, \ie $\left\langle \psi_{n\mathbf{k}}\right|\alpha,\mathbf{t}_s \rangle$, can be calculated as
\begin{equation}
    \left\langle \psi_{n\mathbf{k}}\right|\alpha,\mathbf{t}_s\rangle e^{i\mathbf{k}\cdot\mathbf{t}_s}=\frac{1}{\sqrt{N}}\sum_{a}u_{a,n}^{*}\left(\mathbf{k}\right)\left\langle a,\mathbf{t}_s\right|\alpha,\mathbf{t}_s\rangle
\end{equation}
Here $a=s^K,s^{K^\prime},p_z^K,p_z^{K^\prime}$ is the orbital index in TB8-2V model, $u_{a,n}(\mathbf{k})$ is the periodic part of Bl{\"o}ch wave-function, and $\langle a,\mathbf{t}_s| \alpha \mathbf{t}_s \rangle$ is given by Eq. (\ref{eq:proj-1}) and (\ref{eq:proj-2}). 
Definning the overlap matrix $S_{\beta s^\prime,\alpha s}(\mathbf{k})=\langle\bar{\psi}_{\beta,s^\prime,\mathbf{k}}|\bar{\psi}_{\alpha,s,\mathbf{k}}\rangle$ (with the use of Eq. (\ref{barpsialphask}), the orthonormal Bl{\"o}ch-like wave-functions are then given by 
\begin{equation}
    \left|\tilde{\psi}_{\alpha,s,\mathbf{k}}\right\rangle  = \sum_{\beta s^\prime}S_{\beta s^\prime,\alpha s}^{-\frac{1}{2}}\left(\mathbf{k}\right)\left|\bar{\psi}_{\beta,s^\prime,\mathbf{k}}\right\rangle  = \sum_{n=1}^{4}\left|\psi_{n\mathbf{k}}\right\rangle B_{n,\alpha s}\left(\mathbf{k}\right)\label{eq:psi-tilde}
\end{equation}
where $B_{n,\alpha s}(\mathbf{k}) = \sum_{\beta s^\prime} \left\langle \psi_{n\mathbf{k}}\right|\beta, \mathbf{t}_{s^\prime} \rangle e^{i\mathbf{k}\cdot\mathbf{t}_{s^\prime}} S_{\beta s^\prime, \alpha s}^{-\frac{1}{2}}\left(\mathbf{k}\right)$. The projected Wannier functions are the Fourier transformation of the orthonormal Bl{\"o}ch-like wave-functions
\begin{eqnarray}
    \left|\alpha,\mathbf{t}_s+\mathbf{R}\right\rangle  & = & \frac{1}{\sqrt{N}}\sum_{\mathbf{k}}e^{-i\mathbf{k}\cdot\left(\mathbf{R}+\mathbf{t}_s\right)}\left|\tilde{\psi}_{\alpha,s,\mathbf{k}}\right\rangle 
\end{eqnarray}
As long as the $\det S(\mathbf{k})$ is nonzero in the whole BZ, the projected Wannier functions are exponentially localized. For our calculations, we use the parameters $t^\prime=-\frac{1}{3}t$, $\lambda=\frac{2}{\sqrt{27}}t$, $\Delta=0.15t$, and $\zeta=0$ and choose $c_1=1+i$ and $c_2=1-i$ for the projection orbitals (Eq. (\ref{eq:proj-1})-(\ref{eq:proj-2})). 
We find that $\det S(\mathbf{k})$ is quite flat in the BZ ($15.8\lesssim \det S(\mathbf{k})\lesssim 16$), implying that the projected Wannier functions are well localized. 
In Fig. \ref{fig:TB4-2V}(b) the shape (density) of the projected Wannier functions are plotted.
We emphasize that the obtained Wannier functions respect the $P622$ plus time-reversal symmetries, since they are constructed from simple projection \cite{Vanderbilt_Wannier_Z2} without maximallizing \cite{Marzari1997}.
But they do break the valley symmetry in TB8-2V, indeed that is why they can be exponentially localized, since we mix the orbitals from two valleys (Eq. (\ref{eq:proj-1}) and (\ref{eq:proj-2})). 
One would find that, if the valley-symmetry is enforced, then $S(\mathbf{k})$ must be singular somewhere in the BZ such that the resulting Wannier functions are power law decayed.

The effective tight-binding model on the projected Wannier functions can be obtained as 
\begin{eqnarray} 
    \left\langle \mathbf{O}\alpha\right|\hat{H}\left|\mathbf{R}\beta\right\rangle  & = & \frac{1}{N}\sum_{\mathbf{k}}\left\langle \tilde{\psi}_{\alpha\mathbf{k}}\right|e^{i\mathbf{k}\cdot\left(\boldsymbol{\tau}_{\alpha}\right)}\hat{H}e^{-i\mathbf{k}\cdot\left(\mathbf{R}+\boldsymbol{\tau}_{\beta}\right)}\left|\tilde{\psi}_{\beta\mathbf{k}}\right\rangle \nonumber \\
     & = & \frac{1}{N}\sum_{\mathbf{k}}\sum_{n=1}^{4}e^{-i\mathbf{k}\cdot\left(\mathbf{R}+\boldsymbol{\tau}_{\beta}-\boldsymbol{\tau}_{\alpha}\right)}B_{n\alpha}^{*}\left(\mathbf{k}\right)\epsilon_{n\mathbf{k}}B_{n\beta}\left(\mathbf{k}\right) \label{eq:HR}
\end{eqnarray}
The largest hoppings  obtained from this method are illustrated in Fig. \ref{fig:TB4-2V}(a). With these parameters, the Hamiltonian in momentum space can be written as
\begin{equation} \label{eq:TB4-2V}
    H^{(TB4-2V)}\left(\mathbf{k}\right)=\Upsilon_+(\mathbf{k})\tau_0\sigma_0 + \Upsilon_-\left(\mathbf{k}\right)\tau_{z}\sigma_{0}+\Re\left(\Lambda\left(\mathbf{k}\right)\right)\tau_{x}\sigma_{x}-\Im\left(\Lambda\left(\mathbf{k}\right)\right)\tau_{x}\sigma_{y}
\end{equation} 
where 
\begin{equation}
    \Upsilon_\pm\left(\mathbf{k}\right)=\tilde{\Delta}_\pm+2\tilde{t}_\pm\sum_{i=1}^3\cos\left(\mathbf{d}_{i}\cdot\mathbf{k}\right)+2\tilde{t}^{\prime}_\pm\sum_{i=1}^3\cos\left(3\boldsymbol{\delta}_{i}\cdot\mathbf{k}\right)+2\tilde{t}^{\prime\prime}_\pm\sum_{i=1}^3\cos\left(2\mathbf{d}_{i}\cdot\mathbf{k}\right)
\end{equation}
\begin{equation}
    \Lambda\left(\mathbf{k}\right)=-\tilde{\lambda}_{1}\sum_{i=1}^{3}\left(e^{i\mathbf{k}\cdot\mathbf{D}_{i}}-e^{i\mathbf{k}\cdot C_{2x}\mathbf{D}_{i}}\right)+\tilde{\lambda}_{2}\sum_{i=1}^{3}\left(e^{i\mathbf{k}\cdot\mathbf{D}_{i}^{\prime}}-e^{i\mathbf{k}\cdot C_{2x}\mathbf{D}_{i}^{\prime}}\right)
\end{equation}
\begin{equation}
    \tilde{\Delta}_\pm=\frac{\epsilon_{s}\pm\epsilon_{p}}{2}\qquad 
    \tilde{t}_\pm=\frac{\tilde{t}_{s}\pm\tilde{t}_{p}}{2}\qquad 
    \tilde{t}^{\prime}_\pm=\frac{t_{s}^{\prime}\pm t_{p}^{\prime}}{2}\qquad 
    \tilde{t}^{\prime\prime}_\pm=\frac{t_{s}^{\prime\prime}\pm t_{p}^{\prime\prime}}{2}
\end{equation}
\begin{equation}
    \mathbf{D}_{i}=2\mathbf{d}_{i}+\boldsymbol{\delta}_{i+1} \qquad
    \mathbf{D}_{i}^{\prime}=2\mathbf{d}_{i}+\boldsymbol{\delta}_{i}
\end{equation}
In the following, we will omit the identity term $\Upsilon_+(\mathbf{k})$.  It should be noticed that, although the TB8-2V model we start with is gapless at K point,
the $K$ point becomes gapped after a cutoff in the hoppings in Eq. (\ref{eq:HR}). This gap can be changed by tuning the parameters. At high symmetry momenta the Hamiltonian can be solved analytically
\begin{equation}
    \Upsilon_-(\Gamma) = \tilde{\Delta}_- +6\tilde{t}_- +6\tilde{t}_-^\prime +6\tilde{t}^{\prime\prime}_- \qquad \Lambda(\Gamma) = 0
\end{equation}
\begin{equation}
    \Upsilon_-(M) = \tilde{\Delta}_- -2\tilde{t}_- -2\tilde{t}_-^\prime +6\tilde{t}^{\prime\prime}_- \qquad \Lambda(M) = 0
\end{equation}
\begin{equation}
    \Upsilon_-(K) = \tilde{\Delta}_- -3\tilde{t}_- +6\tilde{t}_-^\prime -3\tilde{t}^{\prime\prime}_- \qquad \Lambda(K) = 0
\end{equation}
We tune the parameters to satisfy (i) $\tilde{t}^{\prime}_- = - \tilde{t}_-$ such that $\Gamma$ and $M$ have same energy eigenvalues, (ii) $\tilde{t}^{\prime\prime}_-=\frac{1}{3}(\tilde{\Delta}_- -3\tilde{t}_- +6\tilde{t}_-^\prime)$ such that the $K$ point is gapless.
It should be clear that the valley symmetry is broken regardless of that $K$ point is tuned to gapless or not, since the Wannier gauge, Eq. (\ref{eq:proj-1}) and (\ref{eq:proj-2}), explicitly mixes two valleys.
The band structure with these tuned parameters is shown in Fig. \ref{fig:TB4-2V}(c).

\section{Large-scale ab-initio calculations}\label{sec:DFT}

Our first-principle calculations are based on density-functional theory (DFT) as implemented in Vienna ab initio simulation package (VASP)~\cite{vasp1996,vasp1999} and with the projector augmented wave (PAW) method~\cite{paw1994}. 
The local density approximation (LDA) was adopted for the exchange-correlation potential~\cite{lda}. 
The kinetic energy cutoff of the plane wave basis was set to 300 eV. Due to computational difficulty, only the gamma point was considered for the smallest-angle (around 1 degree) twisted bilayer graphene. 
The calculations were carried out at the HPC Platform of Shanghaitech University Library and Information Services, and School of Physical Science and Technology. A node of HPC has 72 cups and its total memory is about 2 Terabyte. The computing time for TBG's depends on the size of the Moir\'e unit cells. 
It takes 6 hours to finish the calculation for each k point in the TBG structure with $i=10$, it grows to about 1 day, 3 days, and to 30 days for $i=16$, $i=23$, and  $i=30$, respectively. The original lattice parameter of graphene $a_0=2.456\text{\AA}$  was employed. The distance ($z$) between two layers was initially set to be $z=d_0=3.35{\text{\AA}}$ but was tuned away from this value in some calculations to enhance symmetry-unprotected crossings. 
``A bilayer''-structures were generated by taking AA graphene bilayers and rotating the top layer by an angle $\theta$ with no shift $\dd=0$. The rotating center was choose to be the center of the Carbon hexagon. No relaxation was used. 

\begin{figure}
\begin{centering}
\includegraphics[width=0.8\linewidth]{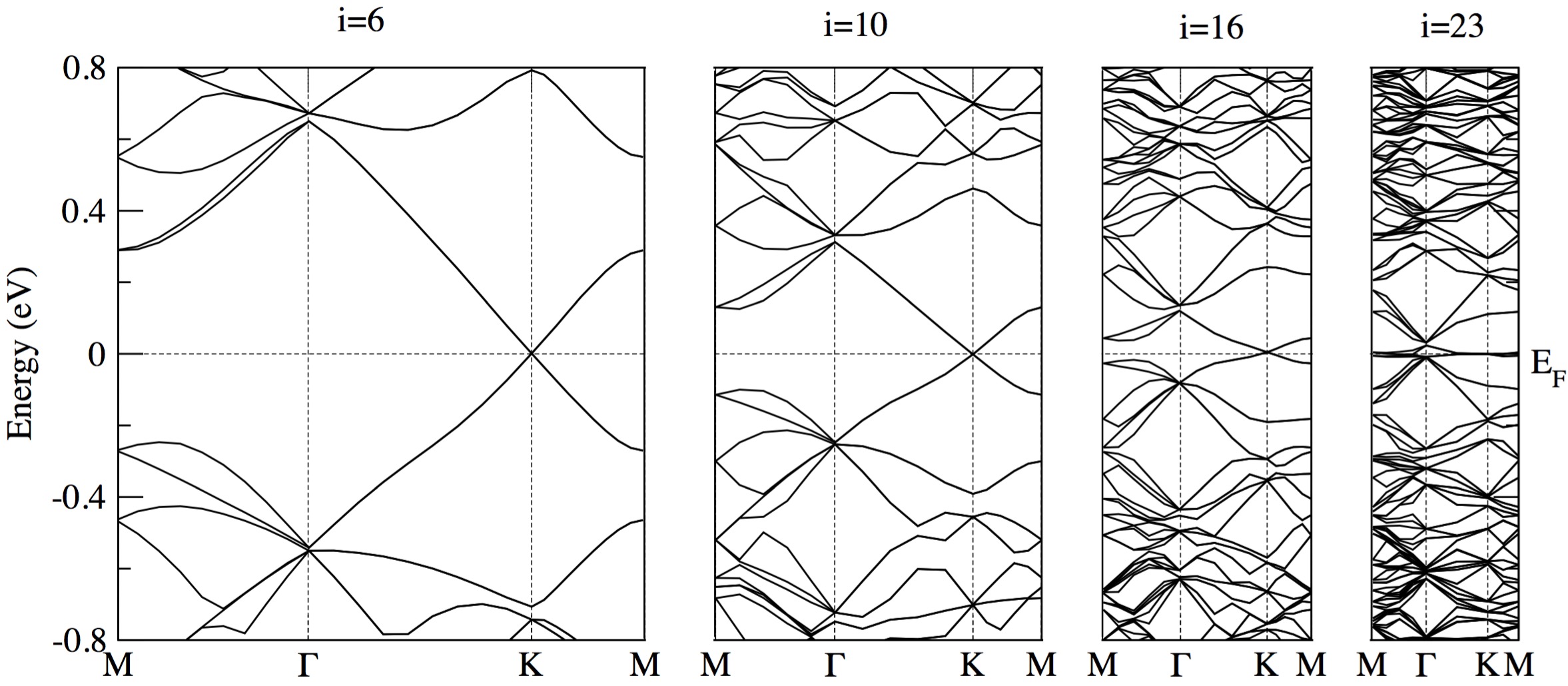}
\par\end{centering}
\caption{Electronic band structures for $i=$ 6, 10, 16, and 13, respectively.}
\label{dftbands1}
\end{figure}

For the AA stacking, the shift $\dd$ affects electronic band structures and properties. However, for the small-angle TBG, it only shifts the Moir\'e pattern in large, but doesn't change the electronic band structures at all. By using the {\it ab-initio} calculations, we have systematically investigated the electronic band structures up to the first ``magic'' angle ($\theta_{30}=1.08^\circ$) and $\dd=0$ in the commensurate twisted bilayer structures. We adopt the commensurate relation: $cos(\theta_i)=\frac{3i^2+3i+0.5}{3i^2+3i+1}$~\cite{TBGstr}. The structure of $i=0$ corresponds to AB stacking with $\theta_0=60^\circ$. We have performed the calculations systematically for $i=$ 6, 10, 16, 23, 27 and 30. The band structures for $i=$ 6, 10, 16, and 23 are presented in Fig.~\ref{dftbands1}. The middle Dirac bands becomes flatter as increasing the $i$ value. To show the evolution of the energy bands at $\Gamma$, we computed the irreps of those Bl{\"o}ch eigenstates and tracked their energy levels as a function of $i$ or $\theta_i$. For $i=6$, we concluded that the 4 low-energy bands near E$_F$ are $\Gamma_2+\Gamma_3$ (lower) and $\Gamma_1+\Gamma_4$ (higher); two ``4-fold'' bands closest to the four bands are $\Gamma_5+\Gamma_6$. The evolution of those 4 branches of energy bands at $\Gamma$ are illustrated in Fig.~\ref{dftgaps}, which we call ``low56'', ``23'', ``14'' and ``high56'' for short, respectively. We clear saw that the gap between ``low56'' (blue) and ``23'' (gray) is much smaller than that between ``14'' (red) and ``high56'' (blue). The closing of the former gap happens at $i=16$, while the gap closing happens at $i=30$ for the latter one. This results in a ``metallic'' phase for middle 4 bands for $16<i<30$, which means those 4 bands are not separated from all others. This ``metallic'' phase can only occur when the PHS is broken, which is also consistent with the results from the MBM by breaking the PHS. Another significant difference from the MBM, is that the gap at K point grows as increasing $i$ (or decreasing $\theta_i$). The K gaps obtained in {\it ab-initio} calculations are indicated in Fig.~\ref{dftgaps}. We conjecture that the K gap due to the inter-valley coupling is not negligible for an extremely small angle. As an extremely small angle means a huge supperunit cell with tons of atoms, which is hard to deal with in {\it ab-initio}, we plan to double-check the tendency of the K gap by decreasing the distance $z$ (increasing the inter-layer hopping $w$) for a relatively larger angle $\theta_{10}$. Both features of the flattening of middle bands and increasing of the K gap are obtained and presented in Fig.~\ref{fig:DFT-gap}.

\begin{figure}
\begin{centering}
\includegraphics[width=0.16\linewidth]{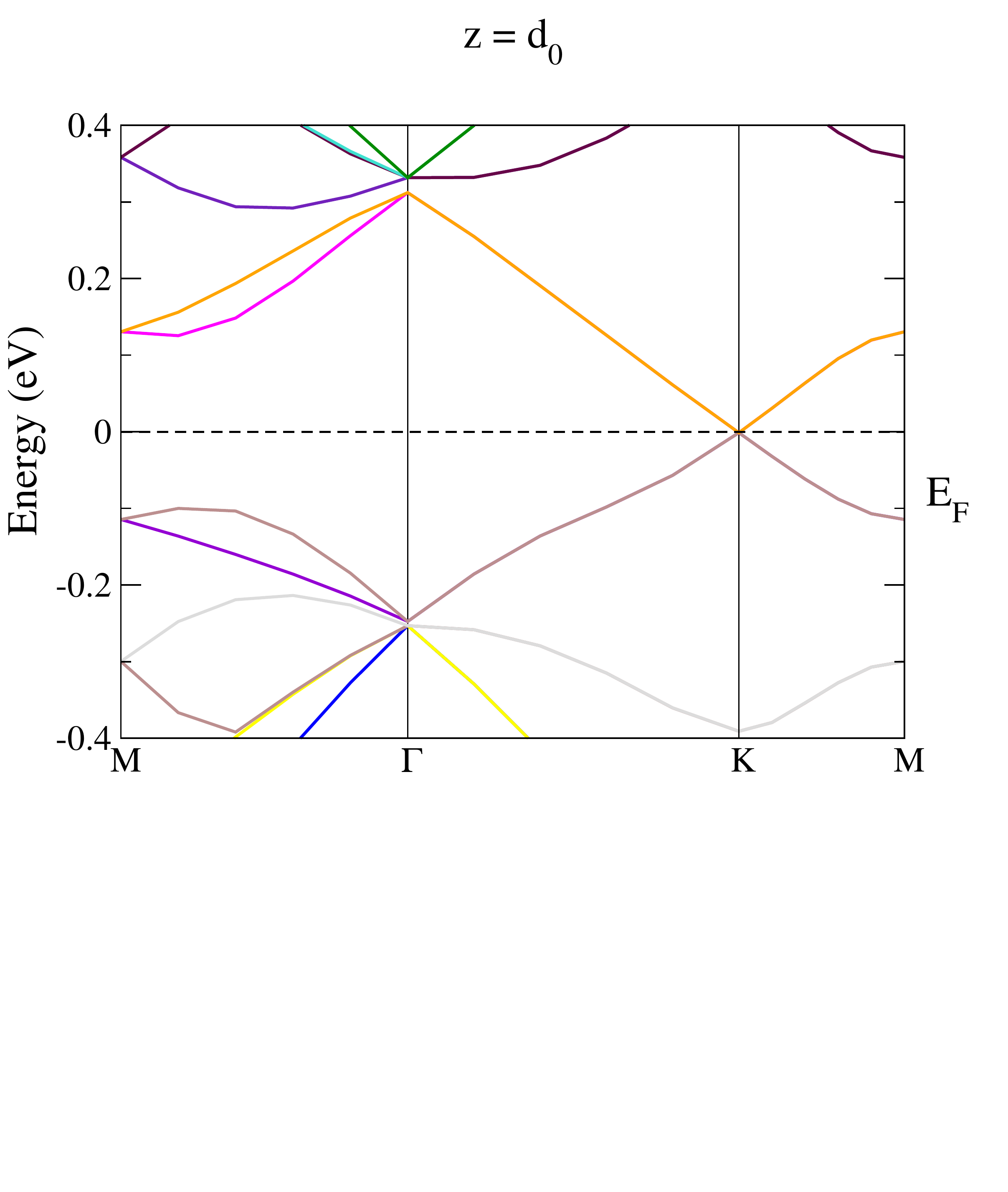}
\includegraphics[width=0.16\linewidth]{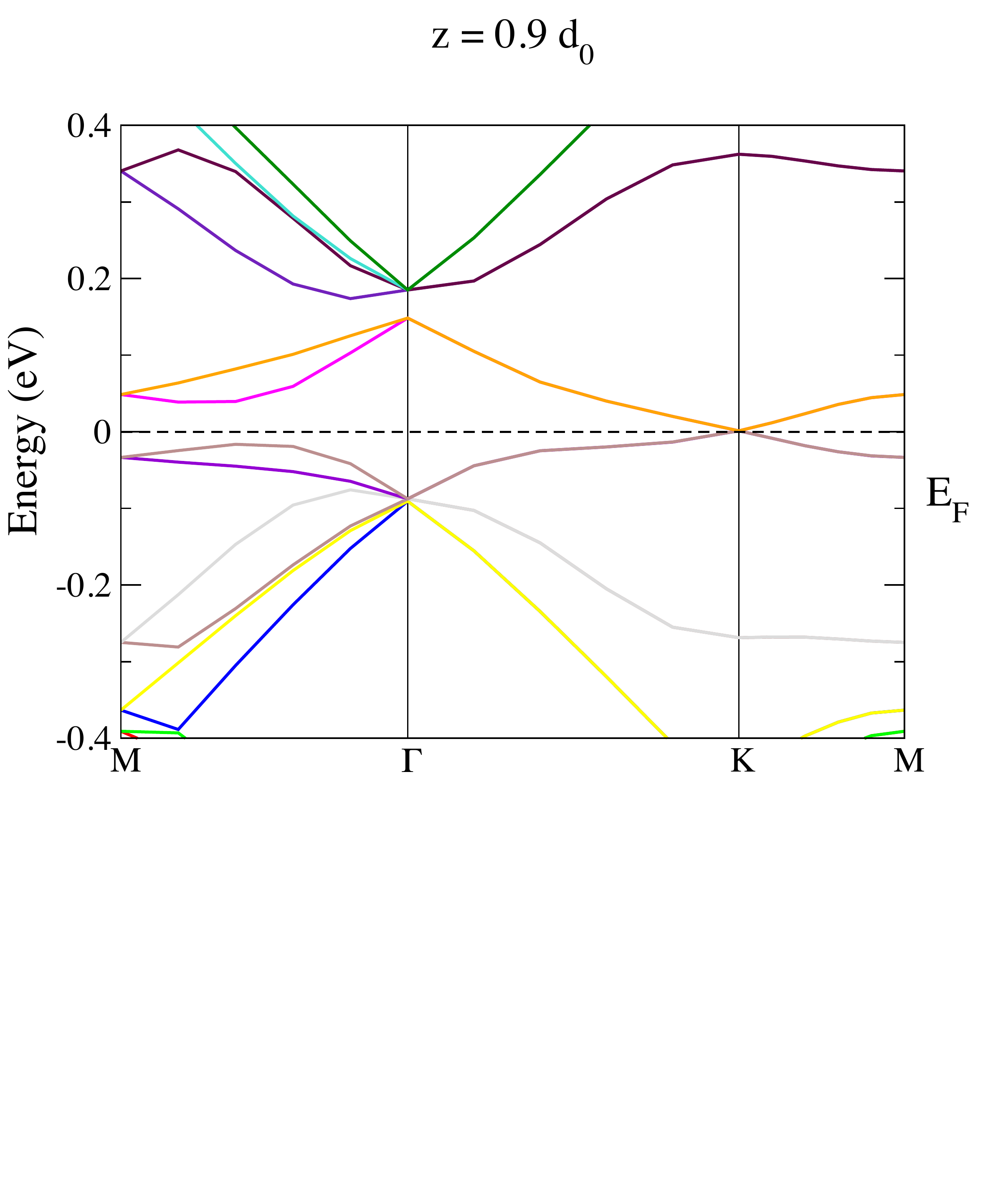}
\includegraphics[width=0.16\linewidth]{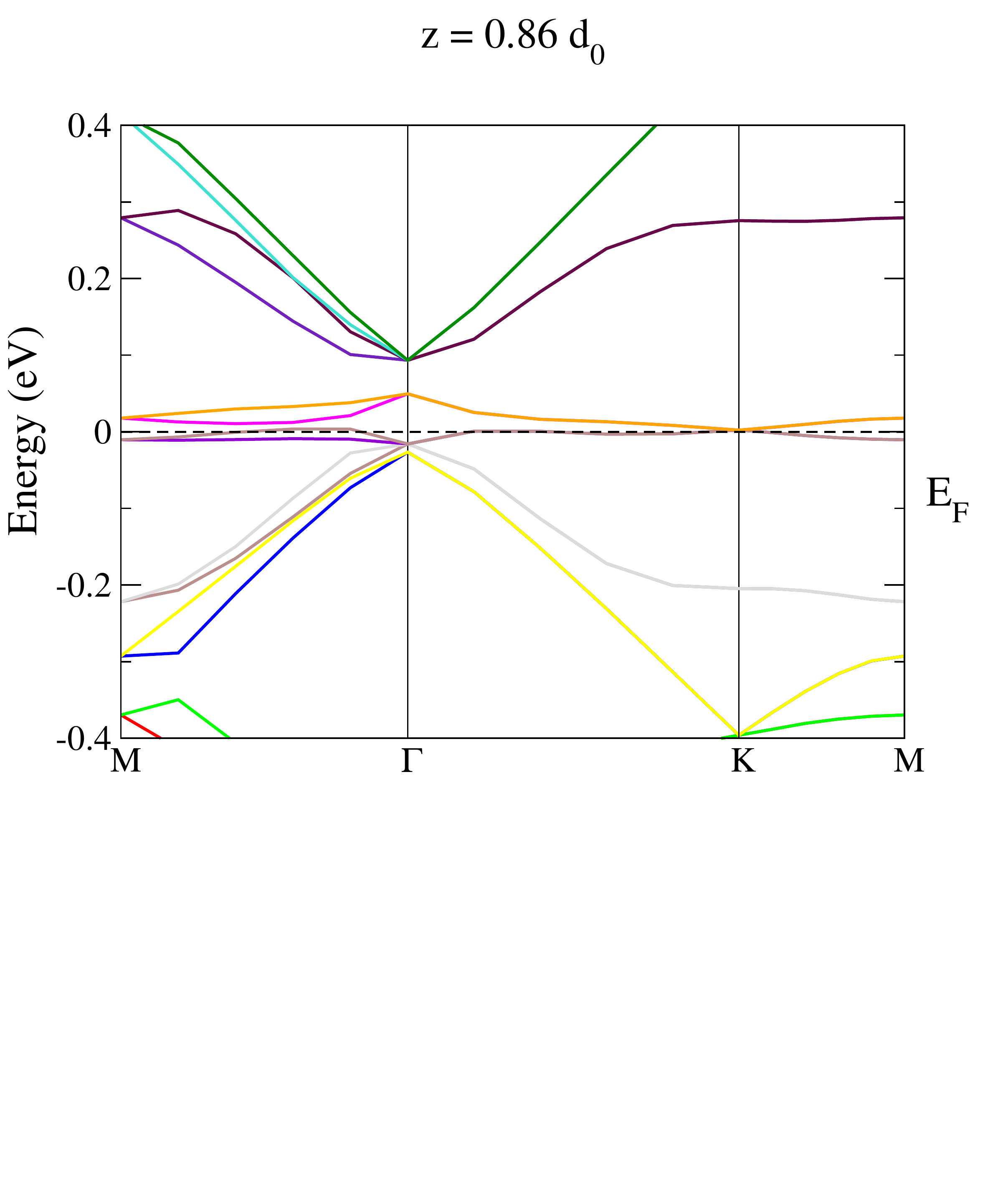}
\includegraphics[width=0.16\linewidth]{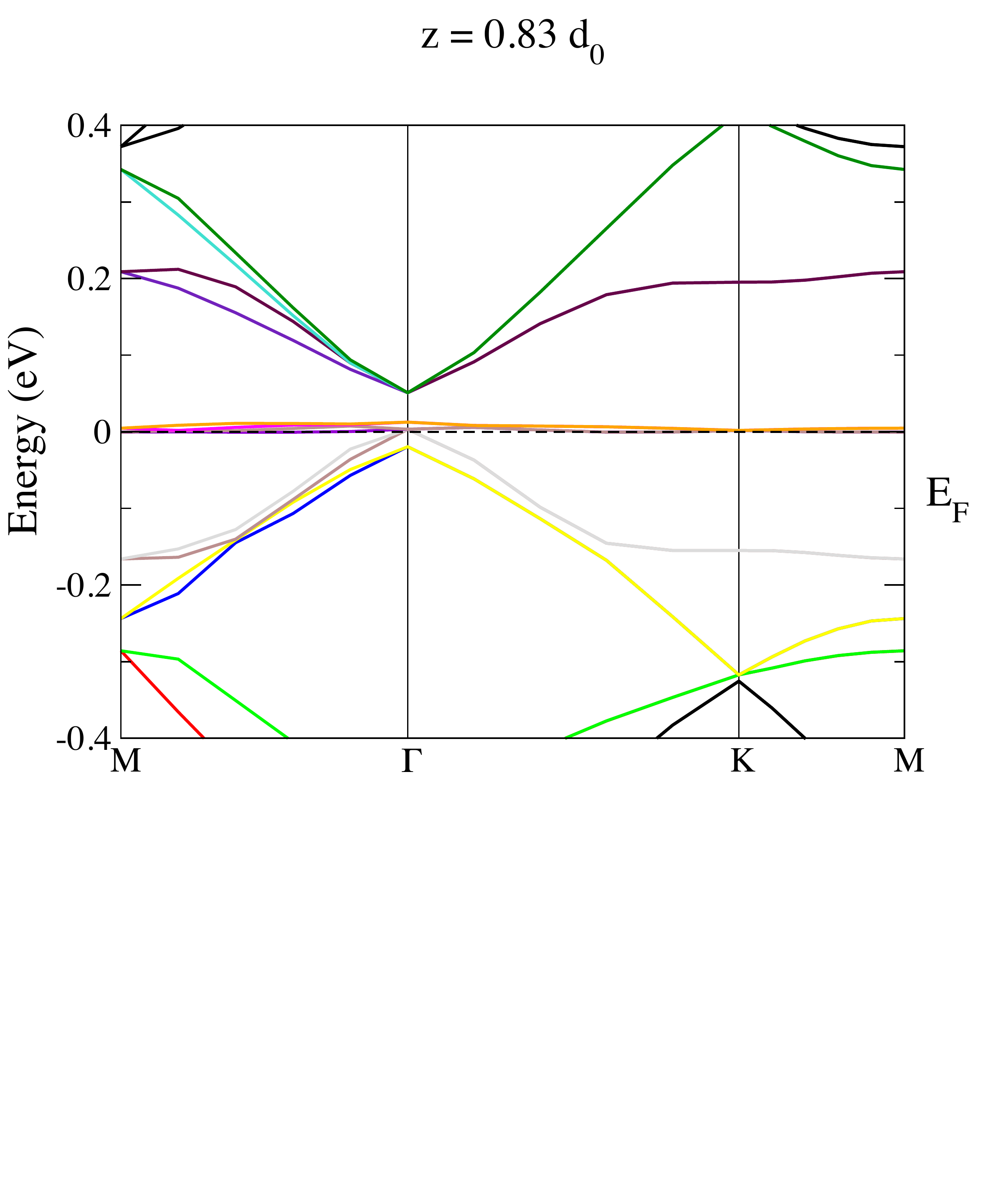}
\includegraphics[width=0.16\linewidth]{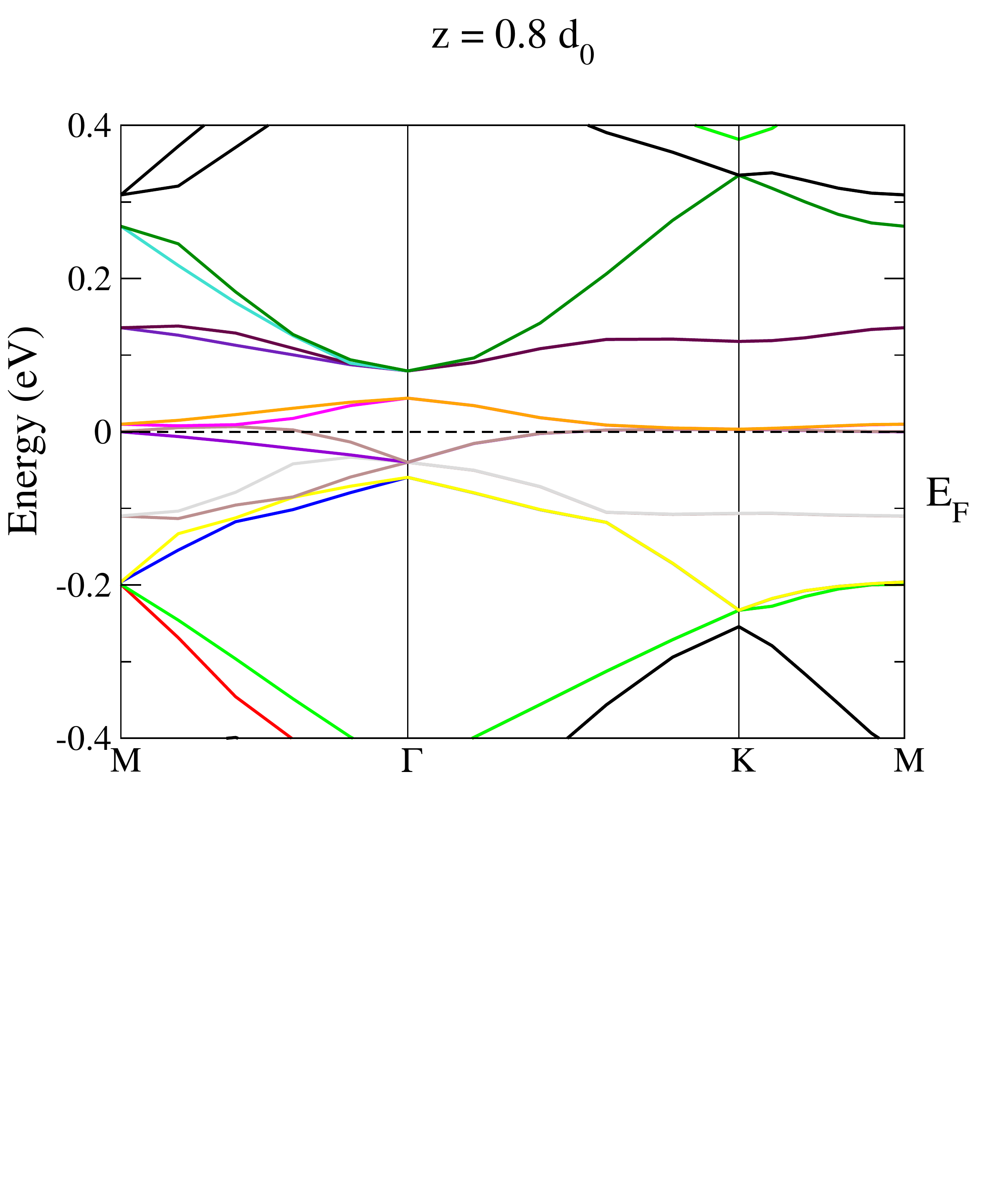}
\includegraphics[width=0.8\linewidth]{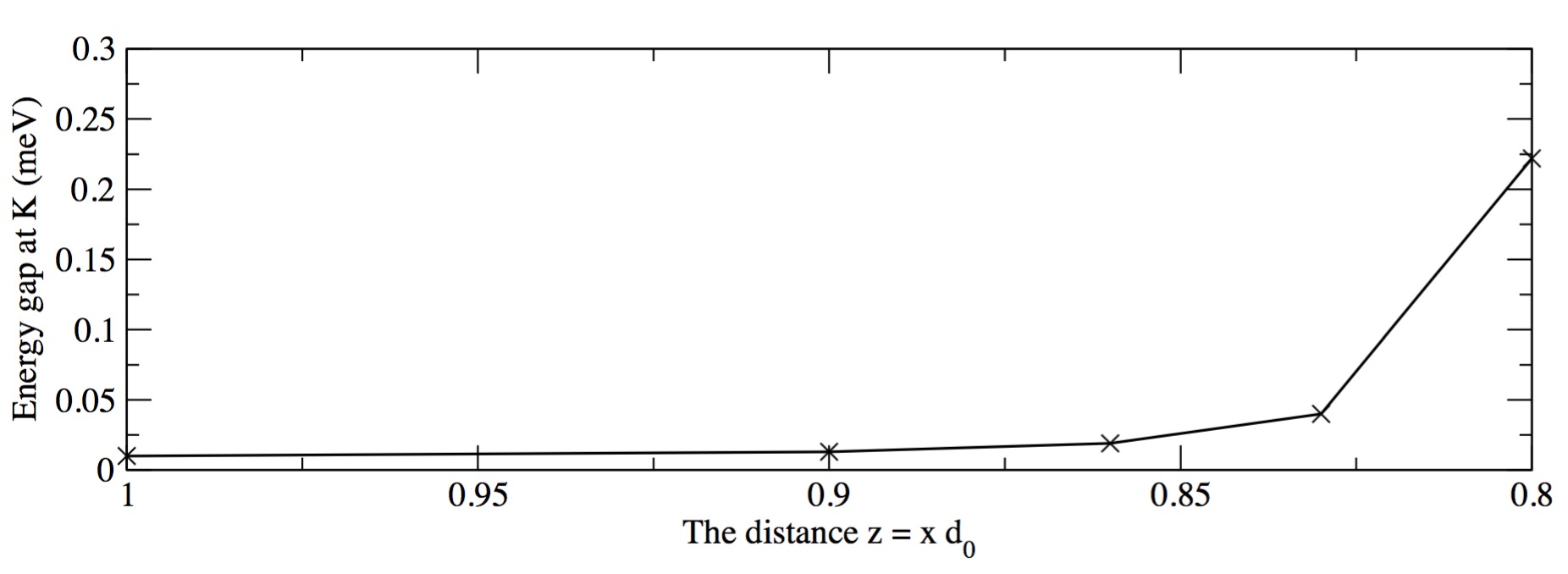}
\par\end{centering}
\caption{Electronic band structures for $i=10$ with varying distance $z$.  We find that the tendency of decreasing the distance $z$ is as same as that of decreasing the twist angle $\theta$. $d_0=3.35$\AA~is the distance between two layers in graphite. We find that the gap at K grows as decreasing $z$. \label{fig:DFT-gap}}
\end{figure}
\newpage

\end{widetext}

\end{document}